\documentclass[12pt,draftclsnofoot,onecolumn]{IEEEtran}
\usepackage{amsfonts}
\usepackage{amssymb}
\usepackage{amsmath}
\usepackage{graphicx}
\usepackage{algorithm}

\usepackage{enumerate}
\usepackage{subfigure}
\usepackage{cite}
\allowdisplaybreaks[4]
\usepackage{stfloats}
\graphicspath{{figures/}}

\usepackage{algorithm}
\allowdisplaybreaks[4]
\usepackage{dsfont}
\usepackage{algorithm}
\usepackage{algorithmic}
\usepackage{multirow}
\usepackage{hyperref}
\usepackage{xcolor}
\usepackage{datetime}
\newcommand{\mb}[1]{\mathbf{#1}}
\newtheorem{proposition}{{Proposition}} 
\normalsize

	\author{Xiaojun~Yuan,~\IEEEmembership{Senior Member,~IEEE,}
	        Haiyang~Xin,
	        Soung-Chang~Liew,~\IEEEmembership{Fellow,~IEEE}
	        and~Yong~Li

	\thanks{	
	This work was supported by the Natural Science Foundation of China
	under Grant No. 61471241. The work in this paper has been partially presented at the IEEE Global Communications Conference, Singapore 2017. 
	 X. Yuan is with the Center for Intelligent Networking and Communications (CINC), University of Electronic
	Science and Technology of China, Chengdu, China. H. Xin and S.C. Liew
	are with the Department of Information Engineering, The Chinese University
	of Hong Kong, Hong Kong SAR, China. Y. Li is with School of Information
	Science and Technology, ShanghaiTech University, Shanghai, China.}}
		\ifCLASSINFOpdf
		\else
		\fi
		\title{Capacity of   the Gaussian  Two-Pair Two-Way Relay Channel to Within ${1\over 2} $ Bit }
		
\begin{document}
			\maketitle
					\begin{abstract}
		This paper studies the transceiver design of the Gaussian two-pair two-way relay channel (TWRC), where two pairs of users exchange information through a common relay in
		a pairwise manner. Our main contribution is to show that the
		capacity of the Gaussian two-pair TWRC is achievable to within
		$1\over 2$
		bit for arbitrary channel conditions. In the proof, we develop
		a hybrid coding scheme involving Gaussian random coding,
		nested lattice coding, superposition coding, and network-coded
		decoding. Further, we present a message-reassembling strategy
		to decouple the coding design for the user-to-relay and relay-to-user links, so as to provide flexibility to fully exploit the channel
		randomness. Finally, judicious power allocation at the relay
		is necessary to approach the channel capacity under various
		channel conditions.
		\end{abstract}

		\section{Introduction}
		Two-way relaying (TWR), in which two users exchange   information via a single relay, has attracted much research interest in the past decade.   The transmission scheme over a two-way relay channel (TWRC) consists of two links. In the user-to-relay link, the two users 
		transmit signals to the relay; in the relay-to-user link, the relay
		broadcasts  signals to the users. The main idea, termed  physical-layer network coding (PNC) \cite{Zhang06}, is to allow the relay to decode a linear function of incoming messages, and to allow each user to decode the message from the other user by exploiting the self-message. Compared with  conventional  relaying,  PNC-aided two-way relaying has the potential to double the network throughput \cite{Nam10}. More recent progress on TWRC and PNC has been reported in  \cite{Nazer11, Liew13}  and the references therein.

		A natural extension of TWR is multi-pair TWR that supports multiple pairs of users engaged in pair-wise data exchange. Multi-pair TWR finds applications in a variety of communications scenarios. For example, in satellite communications, a satellite can serve as a relay to enable multiple ground stations to exchange information simultaneously. Compared with the single-pair case, the transceiver design for multi-pair TWR is more intricate, since the latter needs to carefully deal with the inter-pair interference. It has been shown that the capacity of the two-pair TWRC can be achieved to within $\frac{3}{2}$ bits by the so-called divide-and-conquer relay strategy \cite{Sezgin12}. Multiple-input multiple-output (MIMO) techniques have also been introduced into the TWR systems for spatial multiplexing \cite{Liew13}. For example, \cite{Joung10,  Xin11, Ding13} studied the beamforming design for the two-pair MIMO TWRC; \cite{Xin16} investigated the capacity limits of the two-pair MIMO TWRC from the perspective of principal angles.

		The multi-way relay channel (MWRC) is a generalization of the multi-pair TWRC, where more advanced data exchange models are allowed for information delivery \cite{Lee10, Chaaban13, Ong12, Gunduz13, Yuan14,Liu17, Tian14, Wang16}. For example, the authors in \cite{Lee10} and \cite{ Chaaban13} studied pairwise data exchange, where any two users in the network are allowed to exchange data. The authors in \cite{Ong12} studied full data exchange, where each user transmits a common message to all the other users. Furthermore, the authors in \cite{Gunduz13} and \cite{Yuan14} studied more general models in which users are divided into groups, and the users in each group exchange data with each other. Existing works on MWRC are mostly focused on analyzing the degrees of freedom, or roughly speaking, the asymptotic slope of the network capacity in the high signal-to-noise ratio (SNR) regime. So far, there is limited understanding of the capacity limits of the multiway relay network, especially in the practical SNR regime.

		This paper studies the transceiver design for the two-pair
		TWRC in the finite SNR regime. The main contribution
		of the paper is to show that our scheme can achieve the
		capacity of the two-pair TWRC to within $1\over 2$ bit per user. Our result is  tighter than the state-of-the-art capacity gap
		developed in \cite{Sezgin12} by one bit per user. 	Compared with \cite{Sezgin12}, a more general channel model is considered in our work. In specific, we consider the Gaussian two-pair TWRC with an individual power budget and a different noise level at each user and at the relay, while in \cite{Sezgin12} a common power budget and a uniform noise power is assumed at every node. More importantly, to derive the capacity bounds, we employ a number of new techniques in the proof, as detailed below. 
		\begin{enumerate}[(i)]
			\item We derive a genie-aided outer bound for the Gaussian two-pair TWRC. This new bound is tighter than the cut-set outer bound used in \cite{Sezgin12}.
			\item For the user-to-relay link, we use the same encoding scheme as in \cite{Sezgin12}. But in our scheme, the relay appropriately scales its received signal for nested lattice decoding, so that our scheme achieves higher rates at the user-to-relay link than the scheme in \cite{Sezgin12}.
			\item We further present a message-reassembling strategy at the relay to decouple the coding design for the user-to-relay and relay-to-user transmissions. This provides more flexibility to the coding design for the relay-to-user link, so as to more efficiently exploit the channel randomness of the user-to-relay and relay-to-user links.
			\item Power allocations at the users and at the relay are carefully designed to adapt to various user-to-relay and relay-to-user channel conditions.
		\end{enumerate}		
		Roughly speaking, the use of the genie-aided outer bound in (i) accounts for one half bit reduction of the capacity gap, and the transceiver design in (ii)-(iv) accounts for the other half bit reduction. We show that, with a careful design of the power allocation strategy at the users and at the relay, every boundary point of the outer bound can be achieved to within ${1\over 2}$ bit under arbitrary channel conditions.

		The remainder of this paper is organized as follows. Section II describes the system model. In Section III, we propose our transceiver scheme. Our main result is introduced in Section IV. The proof of the main result is presented in Section V. Section VI concludes the paper.

			\begin{figure} 
				\centering
				\includegraphics[width= 3.25 in]{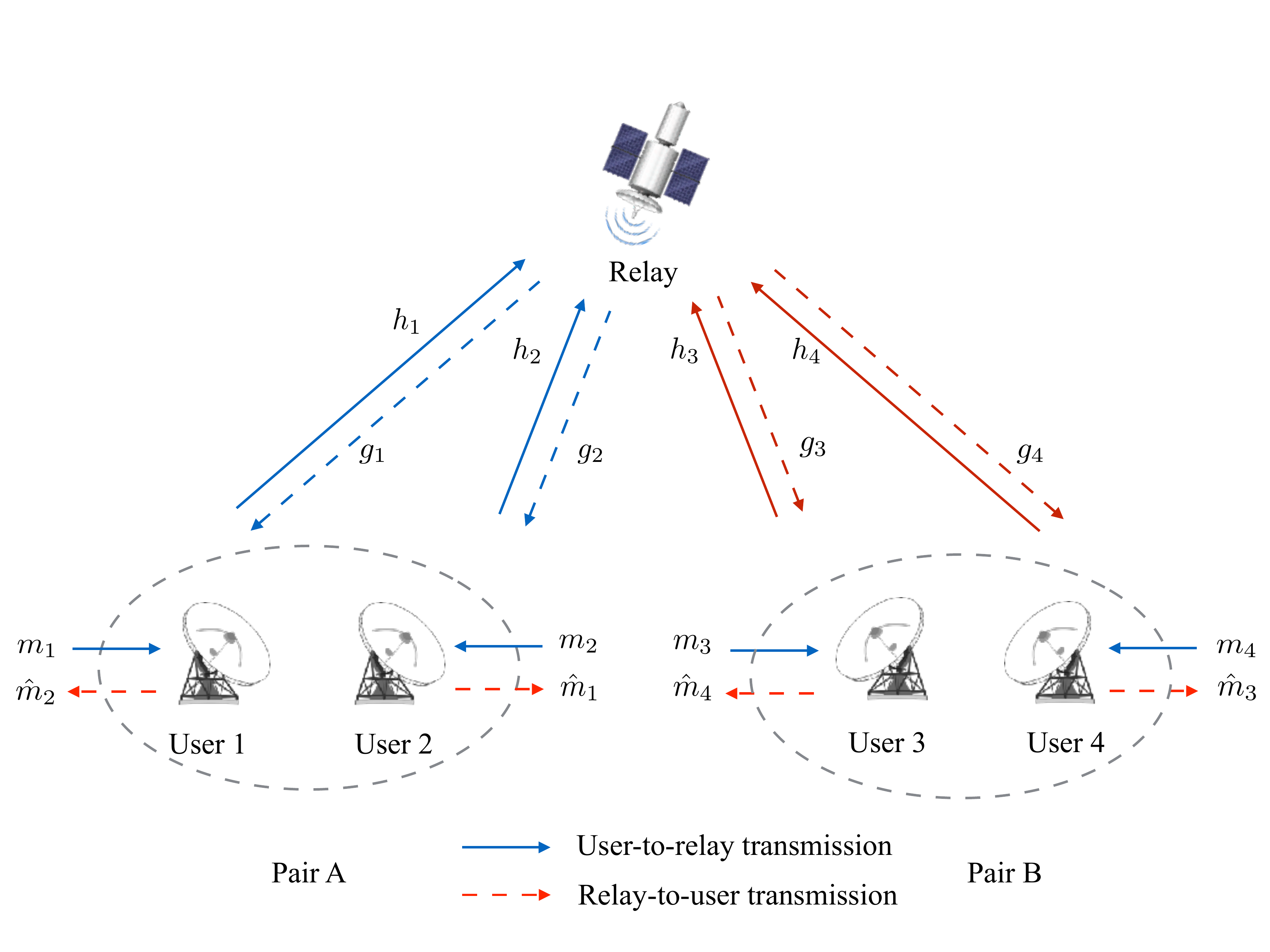}  \vspace{-2mm}
				\caption{The system model of a two-pair two-way relay channel, consisting of four users and a relay. \vspace{-5mm}}\vspace{-2mm}
			\end{figure}
				\section{System Model}
			\subsection{Channel Model}
As illustrated in Fig. 1, we consider a Gaussian two-pair TWRC  with  four  user nodes and one relay node.  The  users  exchange information  in a pairwise manner with the help of the relay. Specifically, users 1 and 2 form a pair, referred to as pair A; users 3 and 4 form the other pair, referred to as pair B. The channel consists of two links, namely, the user-to-relay link and the relay-to-user link. In the user-to-relay link, all the users simultaneously transmit signals to the relay; in the relay-to-user link, the relay broadcasts signals to all the users.
We assume full duplex transmission, in which each node transmits and receives signals simultaneously at different frequency bands.

Block transmission is assumed, i.e., each round of information exchange consists of two transmission blocks with equal duration, one for the user-to-relay link and the other for the relay-to-user link. The two transmission blocks are arranged without overlapping in time, with the user-to-relay block coming first. In this way, the relay is able to decide what to transmit after the reception of the whole transmission block for the user-to-relay link.

Denote $\mathcal{I} \!=\! \{1,2,3,4\}$, $\mathcal{I}_\textrm{A} \!=\!\{1,2\}$, and $\mathcal{I}_\textrm{B}\!=\!\{3,4\}$. User $i$, $i \in \mathcal{I}$,  has a message $m_i \in \{1,...,2^{n{ R}_i}\}$ for transmission, where $n$ is the number of channel uses, and $R_i$ is the rate of user $i$.  The message $m_i$ is encoded into a codeword  $\mb x_i$, where $\mb x_i \!=\!\left[x_i^{(1)},...,x_i^{(t)},...,x_i^{(n)}\right]  \! \in \!\mathbb{R}^n$.
 In the user-to-relay link, the four users transmit signals simultaneously.  The received signal at the relay in time slot $t$, $t\in \{1,...,n\}$,  is given by
\begin{align} \label{up}
			 y_\mathrm{R}^{(t)}=\sum_{i \in \mathcal{I} }  h_i  x_i^{(t)} +    n_\mathrm{R}^{(t)} 
\end{align}
where  $ x_i^{(t)}$ is power-constrained by    $p_i  = {\rm E} \left[ |x_i^{(t)}|^2\right] \leq P_i$,  $ y_\mathrm{R}^{(t)} \! \in \!\mathbb{R}$ is the received signal, $h_i\! \in \!\mathbb{R}  $  is the channel gain between user   $i$ and the relay,  and $n_\mathrm{R}^{(t)} \! \in \!\mathbb{R}$   is the white Gaussian noise   $ \sim \mathcal{N}(0, {\sigma_\mathrm{R}^2})$. The corresponding vector form of the user-to-relay link in (\ref{up}) is given by 
\begin{align} \label{up_v}
			 \mb y_\mathrm{R}=\sum_{i \in \mathcal{I} }  h_i  \mb x_i +    \mb n_\mathrm{R} 
\end{align}
where $\mb y_{\rm R}\!=\! \left[y_\mathrm{R}^{(1)},...,y_\mathrm{R}^{(n)}\right]$, $\mb x_i \!=\!\left[x_i^{(1)},...,x_i^{(n)}\right]  $, and $\mb n_{\rm R}\!=\! \left[n_\mathrm{R}^{(1)},...,n_\mathrm{R}^{(n)}\right]$.

Upon receiving $\mb y_{\rm R}$, the relay performs decoding and then re-encodes the decoded messages into a codeword $\mb x_{\rm R}$, where $\mb x_{\rm R} \!=\!\left[x_{\rm R}^{(1)},...,x_{\rm R}^{(t)},...,x_{\rm R}^{(n)}\right]  \! \in \!\mathbb{R}^n$. Then, in the relay-to-user link, the received signal at user $i$ in time slot $t$ is  
\begin{align} \label{down}
					  y_i^{(t)} =  g_i   x_\mathrm{R}^{(t)}  +   n_i^{(t)}, \quad  i \in \mathcal{I} 
\end{align}
where   $x_\mathrm{R}^{(t)}$  is power-constrained by $p_\mathrm{R} = {\rm E}\left[| x_\mathrm{R}^{(t)}|^2\right] \leq P_\mathrm{R}$,  $ y_i^{(t)} \! \in \!\mathbb{R} $ is the received signal at user $i$, $g_i \! \in \!\mathbb{R}$ is the channel gain between the relay and user  $i$, and $n_i^{(t)} \! \in \!\mathbb{R}$  is the white Gaussian noise   $\sim \mathcal{N}(0, {\sigma_i^2})$. The corresponding vector form of the relay-to-user channel in (\ref{down}) is given by 
\begin{align} \label{down_v}
			 \mb y_i =  g_i   \mb x_\mathrm{R}  +  \mb n_i, \quad  i \in \mathcal{I}
\end{align}
where $\mb y_i\!=\!\left[y_{i}^{(1)},...,y_{i}^{(n)}\right]$, $\mb x_{\rm R} \!=\!\left[x_{\rm R}^{(1)},...,x_{\rm R}^{(n)}\right]  $, and $\mb n_i\!=\!\left[n_{i}^{(1)},...,n_{i}^{(n)}\right]$. 
Following the convention in \cite{Nam10, Sezgin12}, we assume perfect    channel state information (CSI), i.e., $\{h_i, g_i|i\in \mathcal{I}\}$ are perfectly known by the nodes in the network.

With the help of self-message $m_i$, user  $i \in  \mathcal{I}_l$, $l \in \{\mathrm{A},\mathrm{B}\}$,   decodes $m_{\bar i}$ based on $\mb y_i$, yielding an estimated message  $\hat m_{\bar i}$,  where $\bar i$ is the complement of $i$ in $\mathcal{I}_l$, $l \in \{\rm A, \rm B\}$.  A rate tuple $(R_1,R_2,R_3,R_4)$ is said to be achievable if the probability of $\hat m_i \neq  m_i $, $i \in  \mathcal{I}$, vanishes as $n$ goes to infinity.  The capacity region  is   defined as the closure of all  achievable rate tuples.

\subsection{Outer Bound of the Capacity Region}
In this section, we present a genie-aided outer bound of the capacity region for the Gaussian two-pair TWRC. 

	\begin{proposition} 
 An outer bound of the capacity region is  given by
\begin{subequations}
	\label{outer}
\begin{align}
\label{outer_a}
R_1 +R_3 &\leq \min  (C_{13}, \max (D_2, D_4) ) \\
R_1 +R_4 &\leq \min  (C_{14}, \max (D_2, D_3)) \\
R_2 +R_3 &\leq \min  (C_{23}, \max  (D_1, D_4))  \\
R_2 +R_4& \leq \min  (C_{24}, \max (D_1, D_3)) \\
\label{outer_e}
R_1  &\leq \min (C_1, D_2) \\
R_2& \leq \min (C_2, D_1) \\
R_3 &\leq \min (C_3, D_4) \\
R_4 &\leq \min (C_4, D_3) \\
R_i & \geq 0, i \in \mathcal{I} 
\end{align}
\end{subequations}
where 
\begin{align}
\label{C_i}
C_i  & =   {1\over 2}\log\left(1+  {h_i^2 P_i\over \sigma_{\rm R}^2} \right), \ i \in \mathcal{I}  \\
\label{D_i}
D_i  & =   {1\over 2}\log\left(1+  {g_i^2 P_{\rm R}\over \sigma_i^2} \right), \ i \in \mathcal{I}  \\
\label{C_ij}
C_{ij}& =   {1\over 2}\log\left(1+  {h_i^2 P_i + h_j^2 P_j \over \sigma_{\rm R}^2} \right), \ i,j \in \mathcal{I} 
\end{align}
where  ``log" denotes the logarithm with base 2.
  \end{proposition}

\emph{Proof:} See Appendix A. 
  
We note that the above bound is   referred to as the {\it{restricted outer bound}} in \cite{Sezgin12}. As a contribution of this paper, we prove that it is indeed a capacity outer bound by using the genie technique. We emphasize that the genie-aided outer bound is tighter than the cut-set outer bound  in \cite{Sezgin12} by ${1 \over 2}$ bit per user. We will use this bound as the benchmark to analyze the capacity region of the TWRC.

\section{ Transceiver Design}
In this section, we propose a  transmission scheme to establish an inner bound of the capacity region.
		
\subsection{Preliminaries}
Throughout the paper, we assume $R_1\geq R_2$ and $R_3\geq R_4$. This does not lose generality since the discussions hereafter  apply to the cases of other rate orders by swapping the user indices in the two user pairs. For example, for the case of $R_1 \geq R_2$ and $R_3 \leq R_4$, the subsequent discussions hold literally, except that we need to swap the indices of users 3 and 4. We will design different coding and power allocation strategies for different  channel conditions. Clearly, there are 24 different   orders of the four user channels for the user-to-relay link,  and  the same amount of orders  for the relay-to-user link. It will be a formidable task to enumerate all the $24\times 24$ possibilities in system design. 

To alleviate this burden, we consider an auxiliary two-pair two-way relay system $(\{\hat h_i\}, \{\hat g_i\}, \{\hat \sigma_i^2\})$, where the user-to-relay channel coefficients of the auxiliary system are denoted by $\{\hat h_i\}$, the relay-to-user channel coefficients denoted by $\{\hat g_i\}$, and the noise powers denoted by $\{\hat \sigma_i^2\}$. We say that the new system $(\{\hat h_i\}, \{\hat g_i\}, \{\hat \sigma_i^2\})$ is an effective system to the original system $(\{h_i\}, \{g_i\}, \{\sigma_i^2\})$ if the new system has the same outer bound as the original one in (\ref{outer}). Then, we have the following proposition.
 \begin{proposition} With $R_1\geq R_2$ and $R_3\geq R_4$, for an arbitrary two-pair two-way relay system $(\{h_i\}, \{g_i\}, \{\sigma_i^2\})$, there always exists an effective system $(\{\hat h_i\}, \{\hat g_i\}, \{\hat \sigma_i^2\})$ satisfying the following conditions: (i)  $\hat h_i^2 P_i \leq  h_i^2P_i$ and  ${\hat g_i^2 \over \hat \sigma_i^2} \leq {  g_i^2 \over \sigma_i^2}$, $i\in \mathcal{I}$; (ii) $\hat h_1^2 P_1\geq \hat h_2^2 P_2$, $\hat h_3^2P_3 \geq \hat h_4^2P_4$, ${\hat g_2^2 \over \hat \sigma_2^2} \geq{ \hat g_1^2 \over \hat \sigma_1^2}$ and  ${\hat g_4^2 \over \hat \sigma_4^2}  \geq { \hat g_3^2 \over \hat \sigma_3^2 }$.
  \end{proposition}

The proof of Proposition 2 is similar to that of Lemma 2 in \cite{Sezgin12}. For completeness, we present the proof in the following.  

{\emph {Proof}}: 
We prove by construction. Specifically, consider an arbitrary rate tuple $\mb R=(R_1, R_2, R_3, R_4)$ with $R_1\geq R_2$ and $R_3 \geq R_4$. We need to construct an effective system $(\{\hat h_i\}, \{\hat g_i\}, \{\hat \sigma_i^2\})$ such that: first, conditions (i) and (ii) are met; second, if $\mb R$ is in the outer bound (\ref{outer}) then $\mb R$ is also in the outer bound of the effective system, and vice versa. Note that we set $\hat g_i=g_i$, $i \in \mathcal{I}$, for the effective system.

Now consider $\{\hat h_i\}$. By symmetry, it suffices to only consider pair A.  Recall that $R_1$ and $R_2$ satisfy the following inequalities in  (\ref{outer}):
\begin{subequations}
	\label{outer001}
	\begin{align} \label{7a}
	R_1 +R_3 &\leq \min  (C_{13}, \max (D_2, D_4) ) \\
		R_1 +R_4 &\leq \min  (C_{14}, \max (D_2, D_3)) \\
		R_2 +R_3 &\leq \min  (C_{23}, \max  (D_1, D_4))  \\
	R_2 +R_4& \leq \min  (C_{24}, \max (D_1, D_3)) \\
	R_1  &\leq \min (C_1, D_2) \\
	R_2& \leq \min (C_2, D_1). 
	\end{align}
\end{subequations}
We construct $\hat h_1$ and $\hat h_2$ for the user-to-relay link of the effective system with the same outer bound as follows. If $h_1^2P_1\geq h_2^2P_2$, setting $\hat h_1=h_1$ and $\hat h_2 = h_2$ meets the two conditions of Proposition 2 with respect to pair A.

Otherwise, we have $h_1^2P_1<h_2^2P_2$, implying $C_1<C_2$, $C_{13}<C_{23}$, and $C_{14}<C_{24}$. Note that $R_2+R_3 \leq R_1+R_3 \leq  C_{13}$, where the first inequality follows from $R_2 \leq R_1$, and the second inequality from (\ref{7a}). Thus, $R_2 +R_3 \leq \min  (C_{13}, \max  (D_1, D_4))$. Similarly, we have $R_2 +R_4 \leq R_1 +R_4 \leq C_{14}$ and $R_2 \leq R_1 \leq C_1$, and so $R_2 +R_4 \leq \min  (C_{14}, \max (D_1, D_3))$ and $R_2 \leq \min (C_1, D_1)$ hold.
  Then, (\ref{outer001}) can be rewritten as
\begin{subequations}
	\label{outer002}
	\begin{align}
	R_1 +R_3 &\leq \min  (C_{13}, \max (D_2, D_4) ) \\
	R_1 +R_4 &\leq \min  (C_{14}, \max (D_2, D_3)) \\
	R_2 +R_3 &\leq \min  (C_{13}, \max  (D_1, D_4))  \\
	R_2 +R_4& \leq \min  (C_{14}, \max (D_1, D_3)) \\
	R_1  &\leq \min (C_1, D_2) \\
	R_2& \leq \min (C_1, D_1). 
	\end{align}
\end{subequations}
By inspection, we see that (\ref{outer002}) gives the outer bound related to $R_1$ and $R_2$ for the effective system obtained by replacing $h_2$ with $h_1\sqrt{P_1\over P_2}$ (while all the other channel parameters remain unchanged). Therefore,  we can set $\hat h_1  =  h_1$ and $\hat h_2 = h_1  \sqrt {P_1 \over P_2}$  to meet conditions (i) and (ii) for the user-to-relay link.  

We now consider the settings of $\{\hat \sigma_i^2\}$ for the relay-to-user link. Again by symmetry, it suffices to focus on $\hat \sigma_1^2$ and $\hat \sigma_2^2$.
If ${g_2^2 \over \sigma_2^2}\geq { g_1^2 \over \sigma_1^2}$, then  setting ${\hat  \sigma_1^2} = { \sigma_1^2}$ and ${\hat  \sigma_2^2} = { \sigma_2^2}$ is enough to meet the conditions in Proposition 2. Otherwise, we have  ${g_2^2 \over \sigma_2^2}< { g_1^2 \over \sigma_1^2}$, implying $D_2<D_1$. 
Note that $R_2+R_3\leq R_1+R_3\leq \max (D_2, D_4)$, $R_2+R_4\leq R_1+R_4\leq \max (D_2, D_3)$, and $R_2\leq R_1\leq D_2$.
Then, we can rewrite (\ref{outer001}) as
\begin{subequations}
	\begin{align}	R_1 +R_3 &\leq \min  (C_{13}, \max (D_2, D_4) ) \\
	R_1 +R_4 &\leq \min  (C_{14}, \max (D_2, D_3)) \\
	R_2 +R_3 &\leq \min  (C_{23}, \max  (D_2, D_4))  \\
	R_2 +R_4& \leq \min  (C_{24}, \max (D_2, D_3)) \\
	R_1  &\leq \min (C_1, D_2) \\
	R_2& \leq \min (C_2, D_2).
	\end{align}
\end{subequations}
Therefore, to meet conditions (i) and (ii), it suffices to set $\hat \sigma_1^2$ and $\hat \sigma_2^2$ satisfying
   ${g_1^2 \over \hat \sigma_1^2  }=  {g_2^2    \over  \hat \sigma_2^2} ={  g_2^2   \over \sigma_2^2}$.  The  coefficients of the effective system for pair B can be constructed in a similar way, which concludes the proof. \hfill $\blacksquare$

 {\bf Remark 1}: In Proposition 2, condition (i) ensures that the effective system $(\{\hat h_i\},\,\{\hat g_i\},\,\{\hat \sigma_i^2\})$ is always worse than the original system. This implies that if a rate tuple is achievable in the effective system (with the hatted channel), then it is always achievable in the original system. Condition (ii) ensures that the channel coefficients of the effective system always satisfy certain orders. Therefore, for $R_1\geq R_2$ and $R_3\geq R_4$,  Proposition 2 allows us to only consider the following situation: $h_1^2P_1 \geq h_2^2P_2$ and $h_3^2P_3 \geq  h_4^2P_4$ for the user-to-relay link; ${g_2^2 \over \sigma_2^2} \geq{ g_1^2 \over \sigma_1^2}$ and  ${g_4^2 \over \sigma_4^2}  \geq {g_3^2 \over \sigma_3^2 }$ for the relay-to-user link. This simplifies the subsequent analysis.

\subsection{User-to-Relay Transmission}
In the user-to-relay link, users send signals to the relay. The transmission strategy follows the approach in \cite{Sezgin12}. Specifically, in each pair, the user with the stronger channel  transmits a superposition of a Gaussian codeword and a  lattice codeword\footnote{A Gaussian codeword is a vector with the entries independently and identically drawn from a Gaussian distribution. A lattice codeword is a vector selected from the codebook of a nested lattice code.}, and the other user only transmits a   lattice codeword. The  two  users in each pair share a common nested lattice code.  For each pair,  the relay decodes both the Gaussian codeword and a linear function of  the two  lattice codewords following the idea of network-coded decoding \cite{Nam10}. The details are as follows.

We first describe the encoding operations at pair A. Recall that $R_1\geq R_2$. We   split the message $m_1$ of user 1 into  $m_{10}$ and $m_{11}$, satisfying $m_{10} \in \{1,...,2^{n{ R}_{10}}\}$ and $m_{11} \in \{1,...,2^{n{ R}_{11}}\}$ with $R_{10} =R_2$ and $R_{11}=R_1-R_{2}$. 
We construct  a  nested lattice code following the Construction A method used in \cite{Nam10}. Specifically, let $ \Lambda_{\rm 1}^{c}$ and $\Lambda_{\rm 1}^{f}$  respectively    be a  coarse lattice and a fine lattice satisfying  $\Lambda_{\rm 1}^{c}\subseteq\Lambda_{\rm 1}^{f}$.
Denote by $\mathcal {V}(\Lambda)$ the fundamental Voronoi region of lattice $\Lambda$. A nested lattice codebook is then constructed as $\mathcal{C}_{\rm 10}= \Lambda_{\rm 1}^f\bigcap \mathcal{V}{(\Lambda_{\rm 1}^c)}$ with size $2^{nR_{\rm 10}}$.
Then we encode $m_{10}$  into $\mb x_{10}$ and $m_{2}$  into $\mb x_{2}$ using codebook $\mathcal{C}_{\rm 10}$.
The message $m_{11}$ is encoded  into a codeword  ${\bf x}_{{\rm 1}1}$ chosen from a   Gaussian codebook of size  $2^{nR_{{\rm 1}1}}$.

Similarly, in pair B, we   split the message $m_3$ into $m_{30}$ and $m_{31}$, satisfying $m_{30} \in \{1,...,2^{n{ R}_{30}}\}$ and $m_{31} \in \{1,...,2^{n{ R}_{31}}\}$ with $R_{30}=R_4$ and $R_{31}=R_3-R_4$.  We construct  nested lattices  $\Lambda_{\rm 3}^c\subseteq\Lambda_{\rm 3}^f$ with codebook $\mathcal{C}_{\rm 30}=\Lambda_{\rm 3}^f\bigcap\mathcal{V}(\Lambda_{\rm 3}^c)$ with size $2^{nR_{\rm 30}}$.
Then we   encode $m_{30}$  into $\mb x_{30}$,   $m_{4}$  into $\mb x_{4}$ using codebook $\mathcal{C}_{\rm 30}$. The message $m_{31}$ is encoded into a   codeword  ${\bf x}_{{\rm 3}1}$ chosen from a Gaussian codebook  of size $2^{nR_{{\rm 3}1}}$.

 Users 1 and  3 transmit $\mb x_1$ and $\mb x_3$, respectively:
\begin{subequations}
\begin{align}
&  {\bf  x}_{1}  =  {\bf  x}_{10} +  {\bf  x}_{11} \\
& {\bf  x}_{3} =  {\bf  x}_{30} +  {\bf  x}_{31} 
\end{align}
\end{subequations}
where  ${\bf  x}_{i0}$  and ${\bf  x}_{i1}$  are power-constrained  by $\alpha_{i0}P_i$ and $\alpha_{i1}P_i$, with $\alpha_{i0}  +  \alpha_{i1}    \leq 1$,  $i\in  \{1,3\}$.

Users 2 and  4    transmit $\mb x_2$ and $\mb x_4$,  power-constrained by $\alpha_2 P_2$ and $\alpha_4 P_4$, with  $\alpha_2 \leq 1$  and $\alpha_4 \leq 1$, respectively.  

The power factors   $\alpha_{10}$,    $\alpha_2$, $\alpha_{30}$  and    $\alpha_4$ are assigned  such that  the  nested lattice codewords of each pair arrive at the relay at the same power level.  That is,
\begin{subequations}
\begin{align}
p_{{\rm 1}0} &   = h_1^2\alpha_{10} P_1    =  h_2^2\alpha_{2} P_2    \\
p_{{\rm 3}0} & =  h_3^2  \alpha_{30} P_3   =  h_4^2\alpha_{4} P_4, 
\end{align}		
\end{subequations}
where $p_{{\rm 1}0}$ represents the signal power of $\mb x_{10}$ (or $\mb x_2$) received by the relay, and $p_{{\rm 3}0}$ represents that of $\mb x_{30}$ (or $\mb x_4$). This ensures that the two lattice codewords in each pair sit in the same fine lattice at the relay, so as to facilitate network-coded decoding. Furthermore, we  have
\begin{subequations}
	\begin{align}
p_{{\rm 1}1} &    = h_1^2 \alpha_{11}  P_1  \\
p_{{\rm 3}1} & =  h_3^2 \alpha_{31}    P_3,
\end{align}		
\end{subequations}
where $p_{11}$ is the   power of  $\mb x_{11}$ seen at the relay, and $p_{31}$ is that of $\mb x_{31}$.

 Upon receiving $\mb y_\mathrm{R}$,  the relay needs to decode $\mb x_{11}$ to obtain $m_{11}$, decode   $\mb x_{31}$ to  obtain  $m_{31}$, decode a combination of  $\mb x_{10}$ and $\mb x_{2}$ to obtain  a network-coded message $m_{\rm A}$, and decode a combination of $\mb x_{30}$ and $\mb x_{4}$ to obtain another network-coded message
 $m_{\rm B}$. 
 We now consider network-coded decoding. More specifically, for pair A, the relay computes $h_1  \mb x_{10}   \! +   \!  h_2  \mb x_{2}  
 $
to obtain the network-coded message $m_{\rm A}$ with rate $R_{\rm 10}$; for pair B, the relay computes $h_3 \mb x_{30}   \! +   \!  h_4  \mb x_{4}  
$ to obtain the network-coded message $m_{\rm B}$ with rate $R_{\rm 30}$.  
 There are in total $4!=24$ decoding orders.

We first consider decoding Gaussian codewords. For example, when the relay decodes Gaussian codeword  $\mb x_{11}$  to obtain $m_{11}$, the un-decoded codewords   are treated as interference. Generally, the signal model is given by
 \begin{align}
 \mb y_R = h_1   \mb x_{11}  + \mb s +\mb n_{\rm R} 
 \end{align}
 where   $\mb s$ is the interference. 
 Recall that $\mb{x}_{11}$ is drawn from a Gaussian distribution. If $\mb{s}$ is also Gaussian, then the capacity of the channel in (15) is given by 
 \begin{align} \label{14a} 
 R_{{\rm 1}1}  & \leq  {1\over 2} \log  \left(1+  { p_{{\rm 1}1}  \over { p_{\rm s} + \sigma_{\rm R}^2} } \right)  		
 \end{align}
 where  $p_{\rm s}$ is the power of $\mb s$.
 From the information theory, for Gaussian signaling, the worst noise distribution that minimizes the channel input output mutual information is the Gaussian distribution.  Therefore, the rate in (16) is always achievable for an arbitrary distribution of $\mb{s}$.
 
 We now consider network-coded decoding. 	
For example, if the relay decodes   the network-coded message $m_{\rm A}$,  the signal model is given by
\begin{align}
\label{network_decoding_y_r}
\mb y_R = h_1   \mb x_{10}  + h_2   \mb x_{2} + \mb s' +\mb n_{\rm R} 
\end{align}
where   $\mb s'$ is the interference. 
From Appendix B,  the decoding error probability goes to zero as $n\rightarrow \infty$,  provided
\begin{align}
\label{network_decoding_r_10}
R_{{\rm 1}0}  & \leq  {1\over 2} \left[\log  \left( {1\over 2} +  { p_{{\rm 1}0}  \over { p_{\rm s'}+ \sigma_{\rm R}^2} } \right)\right]^{+}  		
\end{align}
where  $p_{\rm s'}$ is the power of $\mb s'$, and $[x]^{+} = \max\, (x,\,0)$.

Suppose that  the relay first decode Gaussian codewords $\mb x_{11}$ and  $\mb x_{31}$,  and then decode the network-coded messages with the decoding order given by
 \begin{align}
 \mb x_{11} \rightarrow \mb x_{31} \rightarrow   \{\mb x_{10}, \mb x_{2}\}  \rightarrow  \{\mb x_{30}, \mb x_{4}\}.  
 \end{align}

  With successive interference cancellation,  the   following rates are achievable:
  \begin{subequations} \label{18}
\begin{align}
R_{11} &   =  {1\over 2} \log \left(  1  +  {p_{11} \over {2p_{10} + 2p_{30} + p_{31} + \sigma_{\rm R}^2} }\right) \\
R_{31} &      =  {1\over 2} \log \left(  1  +  {p_{31} \over {2p_{10} + 2p_{30}   + \sigma_{\rm R}^2} }\right)   \\
R_{10} &    =  {1\over 2} \left[\log \left(  {1\over 2} +  {p_{10} \over {  2p_{30}   + \sigma_{\rm R}^2} }\right)\right]^{+}  \\
R_{30} &    =  {1\over 2} \left[\log \left(  {1\over 2} +  {p_{30} \over {    \sigma_{\rm R}^2} }\right)\right]^{+}
\end{align}
 \end{subequations}
 where the power coefficients satisfy 
 \begin{subequations} 
 	\begin{align}
 	p_{{\rm 1}1}  +  p_{{\rm 1}0} &      \leq h_1^2  P_1   \\
 	p_{{\rm 1}0} &    \leq h_2^2   P_2 \\
 	p_{{\rm 3}1} + p_{{\rm 3}0} &  \leq h_3^2  P_3 \\
 	p_{{\rm 3}0} &    \leq h_4^2   P_4. 
 	\end{align}  
 \end{subequations} 
 
\subsection{Message Reassembling}
In the relay-to-user link, the relay forwards the  four decoded messages $\{m_{{\rm A}}, m_{{\rm B}}, m_{11}, m_{31}\}$ to users.  In the previous work  \cite{Sezgin12}, these four messages are   re-encoded into four Gaussian codewords, and a superposition of these codewords is transmitted. 
This implies that the rate-splitting pattern of  the user-to-relay link uniquely determines that of the relay-to-user link.
However, due to channel randomness, the rate-splitting pattern that fits the user-to-relay link may not be a good choice for the relay-to-user link. To increase flexibility, we introduce  a new message-reassembling strategy at the relay. The main purpose  is to decouple the rate pattern design for the user-to-relay and relay-to-user links, so as to fully exploit  the channel asymmetry.  

The message reassembling strategy consists of two  operations, namely, message splitting and message concatenating. Message splitting is to split a message into two parts. For example, we can split a binary sequence $``{\bf{10101}}11100"$   into $``{\bf{10101}}"$ and $``11100"$. Message concatenating is to concatenate  two messages into a new message.  For example, we can concatenate  two binary sequences $``{\bf{10101}}"$  and $``{{11111}}"$ into a new message $``{\bf{10101}}{{11111}}"$. The detailed operations of message reassembling depends on the channel conditions of the relay-to-user link, and will be elaborated in the following subsection.

\subsection{Relay-to-User Transmission}

 Let ${\bar \sigma_i^2  \!=\! {\sigma_i^2 \over  g_i^2} }$, $i\in  \mathcal{I}$ be the effective noise power seen by user $i$ in the relay-to-user link.
 From Remark 1 in Section III-A, we always have  $  \bar \sigma_1^2   \geq \bar \sigma_2^2 $ and $\bar \sigma_3^2 \geq \bar \sigma_4^2$. We further  assume $ \bar \sigma_4^2\geq \bar \sigma_2^2$. This assumption does not lose generality because for other cases we only need to change the pair order (i.e., to swap the indices of users 1 and 3, as well as the indices of users 2 and 4) in the subsequent discussions. 
Hence, we only need to consider the following three channel orders:
	\begin{subequations} \label{case}
	 \begin{align} \label{case1}
 \rm {Case \ I:} \quad \quad   &\bar \sigma_3^2   \geq \bar \sigma_4^2    \geq  \bar \sigma_1^2 \geq \bar \sigma_2^2  \\ \label{case2}
 \rm {Case \ II:} \quad \quad     & \bar \sigma_3^2   \geq \bar \sigma_1^2    \geq  \bar \sigma_4^2 \geq \bar \sigma_2^2\\ \label{case3}
 \rm {Case \ III:} \quad \quad 		&  \bar \sigma_1^2   \geq \bar \sigma_3^2    \geq  \bar \sigma_4^2 \geq \bar \sigma_2^2.   
		\end{align}
	\end{subequations}
In the following, we describe the relay-to-user transceiver design for each case in (\ref{case}).

\subsubsection{Case I}
In this case, the message reassembling at the relay is first to concatenate $m_{\rm A}$ and $m_{11}$ into  a single message, denoted by $\bar m_{\rm A}$,   and concatenate $m_{\rm B}$ and $m_{31}$ into  a single message $\bar m_{\rm B}$. 
 Then, the relay maps $\bar m_{\rm B}$ to   a codeword $ {\mb x}_{\rm R1}$   chosen from a Gaussian codebook of size $2^{n{R}_{\rm R1}}$,  and maps  $\bar m_{\rm A}$  to a codeword ${\mb x}_{{\rm R2}}$   chosen from a Gaussian codebook of size $2^{n{ R}_{{\rm R2}}}$, where
 	\begin{subequations} 
 \begin{align} \label{27a}
R_{\rm R1}&= R_{\rm 30} + R_{31}=R_3 \\   \label{27b}
R_{\rm R2}&= R_{\rm 10}+ R_{11} =R_1.
 \end{align}
 	\end{subequations} 
 The relay transmits 
 \begin{align}
 {\bf x}_{\rm R} = 	{\bf x}_{{\rm R1}} + {\bf x}_{{\rm R2}}
 \end{align}
 where $ {\mb x}_{{\rm R}i}$ is power-constrained by  ${1\over n}||{\mb x}_{{\rm R}i}||^2 \leq  p_{{\rm R}i}$, $i\in\{1,2\}$, and
\begin{align}
p_{\rm R1}+p_{\rm R2} =P_{\rm R}.
\end{align}
The encoding operation at the relay is shown in Fig. 2.

 \begin{figure*}
 	\centering
 	\includegraphics[width= 5.5 in]{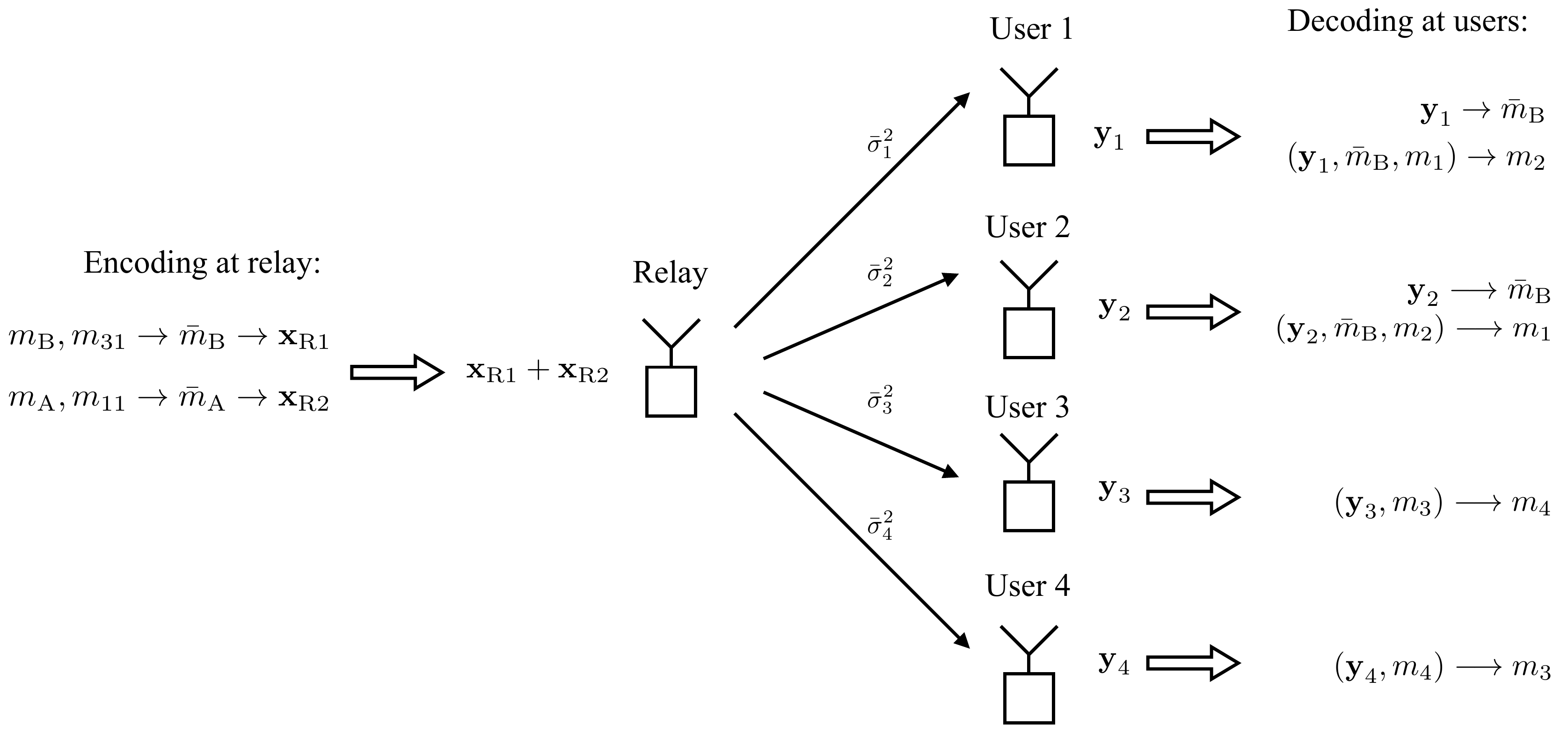}  \vspace{-2mm}
 	\caption{{The message flow table  for Case I: $\bar \sigma_3^2   \geq \bar \sigma_4^2    \geq  \bar \sigma_1^2 \geq \bar \sigma_2^2$}. \vspace{-2mm}}\vspace{-2mm}
 \end{figure*}

The received signal  of user $i$ in pair A is given by
\begin{align}
\mb y_i = g_i \mb x_{\rm R1} + g_i\mb x_{\rm R2} + \mb n_i, \ i \in \mathcal{I}_{\rm A}.
\end{align}
Each user  in pair A  first decodes  $	{\bf x}_{{\rm R1}}$ to obtain $\bar m_{\rm B}$ by treating $	{\bf x}_{{\rm R2}}$ as noise. The decoding error probability goes to zero as $n\rightarrow \infty$, provided
\begin{align} \label{27}
R_{\rm R1}    &\leq  {1\over 2} \log \left(1+ {p_{\rm R1}\over {p_{\rm R2}+\bar \sigma_i^2}}\right), \ i\in\mathcal{I}_{\rm A}. 
\end{align}
After removing ${\bf x}_{{\rm R1}}$ from the received signal, each user $i$ further decodes ${\bf x}_{{\rm R2}}$ with the help of its self-message.  
With their specific self-message, the sizes of the decoding codebooks of users 1 and 2 are respectively given by $2^{nR_2}$ and $2^{nR_1}$, where  $R_1= R_{\rm R_2}\geq R_2$.
The decoding operation at pair A is shown in Fig. 2. 
From [2, Theorem 1], the decoding error probability goes to zero as $n\rightarrow \infty$,  provided
\begin{subequations}  \label{28}
\begin{align} \label{32a}
 R_2 &  \leq {1\over 2} \log \left(1+ {p_{\rm R2}\over \bar \sigma_1^2}\right)\\  \label{32b}
 R_1 &  \leq {1\over 2} \log \left(1+ {p_{\rm R2}\over \bar \sigma_2^2}\right). 
\end{align}
\end{subequations}

The received signal  of user $i$ in pair B is given by
\begin{align}
\mb y_i = g_i \mb x_{\rm R1} + g_i\mb x_{\rm R2} + \mb n_i, \ i \in \mathcal{I}_{\rm B}.
\end{align}
By treating $ 	{\bf x}_{{\rm R2}}$ as noise,  users in pair B  decode $	{\bf x}_{{\rm R1}}$ with the help of their self-message  to obtain   the partner's message.  The decoding error probability goes to zero as $n\rightarrow \infty$, provided
\begin{subequations} \label{30}
	\begin{align} \label{34a}
	R_{4} &\leq {1\over 2} \log \left(1+ { p_{\rm R1}\over {p_{\rm R2}+\bar \sigma_3^2}}\right)\\   \label{34b}
	 R_{3} &\leq {1\over 2} \log \left(1+ { p_{\rm R1}\over {p_{\rm R2}+\bar \sigma_4^2}}\right). 
	\end{align}
\end{subequations}

We are now ready to present the following proposition.

\emph {Proposition 4.1:}   For Case I in (\ref{case1}), an achievable rate tuple for the relay-to-user link  is given by

\begin{subequations} \label{Pro4.1}
\begin{align} \label{Pro4.1_1}
R_{ 1}&   =  {1\over 2} \log \left(1+ { p_{\rm R2}\over \bar \sigma_2^2}\right) \\ \label{Pro4.1_2}
R_{ 2}&  = {1\over 2} \log \left(1+ {p_{\rm R2}\over \bar \sigma_1^2}\right) \\ \label{Pro4.1_3}
 R_{3} & = {1\over 2} \log \left(1+ {p_{\rm R1}\over {p_{\rm R2}+\bar \sigma_4^2}}\right) \\ \label{Pro4.1_4}
  R_{4} & ={1\over 2} \log \left(1+ {p_{\rm R1}\over {p_{\rm R2}+\bar \sigma_3^2}}\right). 
\end{align}
\end{subequations}
\emph {Proof:}  It suffices to show that the rates in (\ref{Pro4.1}) meet the conditions in (\ref{27}), (\ref{28}),  and (\ref{30}). Note that (\ref{Pro4.1_1}), (\ref{Pro4.1_2}), and (\ref{Pro4.1_4}) are straightforward from (\ref{28}) and (\ref{30}). For (\ref{Pro4.1_3}), we first see from (\ref{27a}) that $R_3 = R_{\rm R1}$. Together with $\bar \sigma_4^2 \geq \bar \sigma_1^2\geq \bar \sigma_2^2$, we see that  (\ref{Pro4.1_3})  satisfies both (\ref{27}) and (\ref{34b}), which completes the proof.

\subsubsection{Case II}

We propose two achievable schemes for Case II. In the first scheme, the relay carries out the following  message reassembling. The relay first splits the message $m_{11}$ into  $m_{11}^{(0)}$ and $m_{11}^{(1)}$, and  concatenates $m_{\rm A}$ and $m_{11}^{(0)}$ into  $\bar m_{\rm A}$. The relay  splits the message $m_{31}$ into  $m_{31}^{(0)}$ and $m_{31}^{(1)}$, and  concatenates $m_{\rm B}$ and $m_{31}^{(0)}$ into  $\bar m_{\rm B}$.
 Then, the relay maps $\bar m_{\rm B}$ to   $ {\mb x}_{\rm R1}$ using a Gaussian codebook of size $2^{n{R}_{\rm R1}}$,  maps $\bar m_{\rm A}$ to   $ {\mb x}_{\rm R2}$ using a Gaussian codebook of size $2^{n{R}_{\rm R2}}$, maps $m_{31}^{(1)}$ to  ${\mb x}_{{\rm R}3}$ using a Gaussian codebook of size $2^{n{R}_{{\rm R}3}}$, and maps $m_{11}^{(1)}$ to  ${\mb x}_{{\rm R}4}$ using a Gaussian codebook of size $2^{n{R}_{{\rm R}4}}$. Finally, the relay transmits a superposition of these four codewords as
 \begin{align}
 {\bf x}_{\rm R} =  {\mb x}_{{\rm R}1} +  {\mb x}_{{\rm R2}} +   {\mb x}_{{\rm R3}}+ {\mb x}_{{\rm R}4}
 \end{align}
 where    ${1\over n}|| {\mb x}_{{\rm R}i}||^2 \leq  p_{{\rm R}i}$,  $i \in \mathcal{I}$, and 
 \begin{align}
  p_{\rm R1}+p_{\rm R2}+p_{\rm R3}+p_{\rm R4}=\!P_{\rm R}.  
 \end{align}
 Also, form the above construction, we have  
 \begin{subequations}\label{ine1} 
\begin{align}
R_1 & = R_{\rm R2}  + R_{\rm R4} \\
R_2& \leq  R_{\rm R2}  \\
R_3 & = R_{\rm R1}  + R_{\rm R3} \\
R_4& \leq  R_{\rm R1}.  
\end{align} 
\end{subequations} 
 The encoding operation at the relay is shown in Fig. 3.

	\begin{figure*}
		\centering
	\includegraphics[width= 6 in]{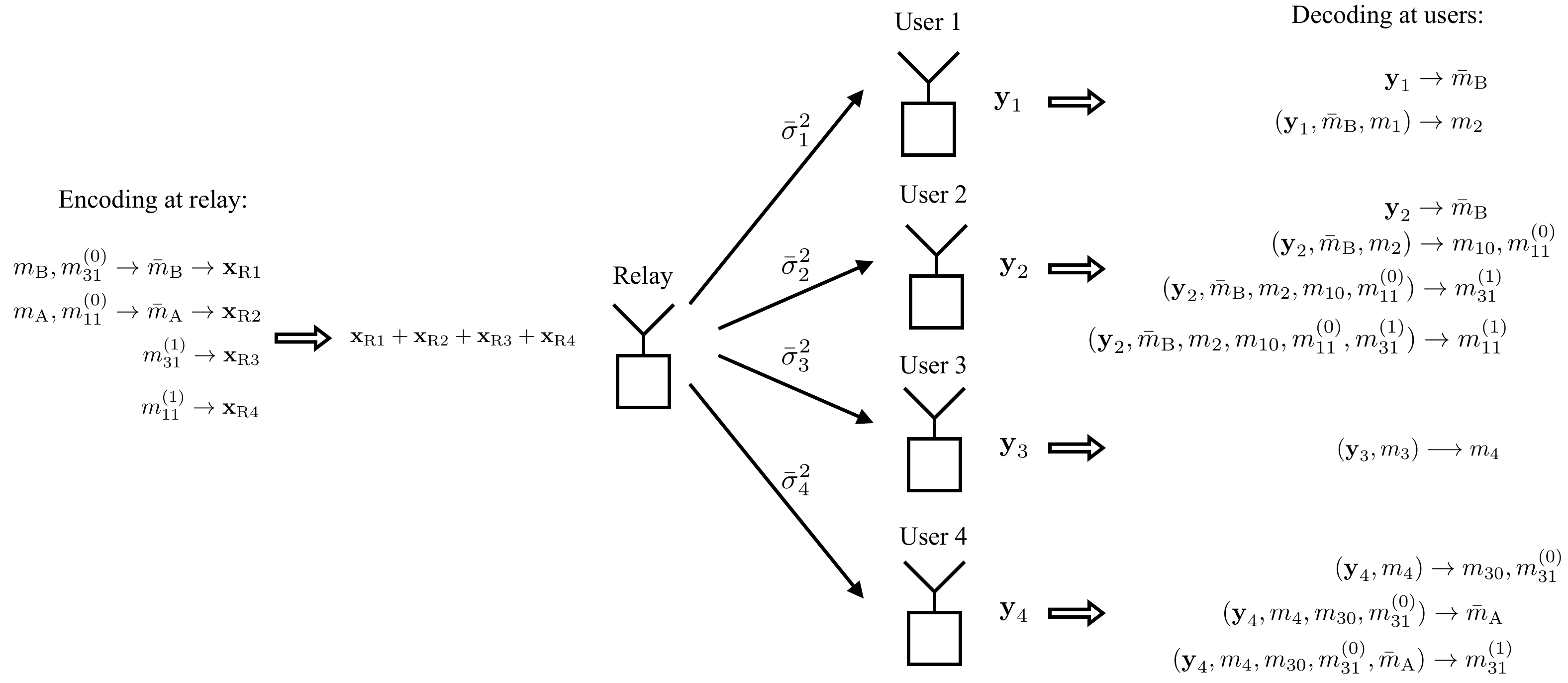}  \vspace{-2mm}
	\caption{ The message flow table of the first scheme for Case II: $\bar \sigma_3^2   \geq \bar \sigma_1^2    \geq  \bar \sigma_4^2 \geq \bar \sigma_2^2$.     \vspace{-2mm}}\vspace{-2mm}
	\end{figure*}

The received signal  at user 1 is 
\begin{align} 
\mb y_1 = g_1{\mb x}_{{\rm R}1} +  g_1{\mb x}_{{\rm R2}} +   g_1{\mb x}_{{\rm R3}}+ g_1{\mb x}_{{\rm R}4} + \mb n_1.
\end{align}
User 1 first decodes ${\mb x}_{{\rm R1}}$  to obtain $\bar m_{\rm B}$ of rate $R_{{\rm R1}}$ by treating the other signals as noise. The decoding error probability goes to zero as $n\rightarrow \infty$, provided
\begin{align}  \label{33}
	R_{{\rm R1}}  &  \leq    {1\over 2} \log \left( 1+ {  p_{{\rm R1}}\over   p_{{\rm R2}} +   p_{{\rm R3}} +  p_{{\rm R4}}+ \bar  \sigma_1^2}\right).  
\end{align}
After removing  ${\mb x}_{{\rm R1}}$  from $\mb y_1$, user 1   decodes ${\mb x}_{{\rm R2}}$ to obtain $m_{2}$ of rate $R_2$ via  the help of  self-message $m_{1}$ by treating $\mb x_{\rm R3}$ and $\mb x_{\rm R4}$ as interference.  The decoding operation at user 1 is shown in Fig. 3.  
From [2, Theorem 1],  the decoding error probability goes to zero as $n\rightarrow \infty$, provided
\begin{align}   \label{34}
 R_{2}  &  \leq    {1\over 2} \log \left(1+  {  p_{{\rm R2}}\over   p_{{\rm R3}} +   p_{{\rm R}4} + \bar  \sigma_1^2}\right).
\end{align}

The received signal  at user 2 is 
\begin{align}
\mb y_2 = g_2{\mb x}_{{\rm R}1} +  g_2{\mb x}_{{\rm R2}} +   g_2{\mb x}_{{\rm R3}}+ g_2{\mb x}_{{\rm R}4} + \mb n_2.
\end{align}	
 User 2 decodes ${\mb x}_{{\rm R1}}$, ${\mb x}_{{\rm R2}}$, ${\mb x}_{{\rm R3}}$,   and ${\mb x}_{{\rm R4}}$ sequentially to obtain $\bar m_{\rm B}$,  $\{m_{\rm A}, m_{11}^{(0)}\}$,  $m_{31}^{(1)}$ and  $m_{11}^{(1)}$  of rates  $R_{\rm R1}$, $R_{\rm R2}$, $R_{\rm R3}$ and $R_{\rm R4}$, where $R_{\rm R2} + R_{\rm R4} = R_1$.     The decoding operation at user 2 is shown in Fig. 3.  
With successive interference cancellation, the decoding error probability goes to zero as $n\rightarrow \infty$  provided
\begin{subequations}\label{36}
\begin{align} {\label{41a}}
R_{{\rm R1}}  &  \leq    {1\over 2} \log \left( 1+ {  p_{{\rm R1}}\over   p_{{\rm R2}}+    p_{{\rm R}3} +   p_{{\rm R4}} +\bar  \sigma_2^2}\right)   \\ {\label{41b}}
R_{{\rm R2}}  &  \leq    {1\over 2} \log \left(1+  {  p_{{\rm R2}}\over   p_{{\rm R3}} +   p_{{\rm R4}} + \bar  \sigma_2^2}\right)   \\ {\label{41c}}
R_{{\rm R3}} & \leq {1\over 2} \log \left( 1+ {  p_{{\rm R3}}\over   p_{{\rm R4}} +\bar  \sigma_2^2}\right)  \\ {\label{41d}}
R_{{\rm R4}}  & \leq   {1\over 2} \log \left(  1+ {  p_{{\rm R4}}\over\bar  \sigma_2^2}\right).
\end{align}	
\end{subequations}

The received signal  at user 3 is  given by
\begin{align}
\mb y_3 = g_3{\mb x}_{{\rm R}1} +  g_3{\mb x}_{{\rm R2}} +   g_3{\mb x}_{{\rm R3}}+ g_3{\mb x}_{{\rm R}4} + \mb n_3.
\end{align}	
User 3 decodes ${\mb x}_{{\rm R1}}$ (so as to acquire $m_4$) with the help of $m_3$, by treating the other signals as interference. The decoding operation at user 3 is shown in Fig. 3.  The decoding error probability goes to zero as $n\rightarrow \infty$, provided
\begin{align} \label{38}
 R_{4}  &  \leq    {1\over 2} \log \left( 1+ {p_{{\rm R1}}\over  p_{{\rm R2}} +  p_{{\rm R4}} + \bar  \sigma_3^2}\right).
\end{align}
Note that $p_{{\rm R3}}$ does not appear in (\ref{38}) since ${\mb x}_{{\rm R3}}$ is known to User 3.

The received signal  at user 4 is given by
\begin{align}
\mb y_4 = g_4{\mb x}_{{\rm R}1} +  g_4{\mb x}_{{\rm R2}} +   g_4{\mb x}_{{\rm R3}}+ g_4{\mb x}_{{\rm R}4} + \mb n_4.
\end{align}	
User 4 decodes ${\mb x}_{{\rm R1}}$, ${\mb x}_{{\rm R2}}$, and ${\mb x}_{{\rm R3}}$ at rates $R_{\rm R1}$, $R_{\rm R2}$, and $R_{\rm R3}$, respectively, with successive interference cancellation, where $R_{\rm R1}  +\!R_{\rm R3} = R_3$.  As shown in Fig. 3, user 4 obtains $\{m_{\rm B},m_{31}^{(0)}\}$, $\bar m_{\rm A}$, and $m_{31}^{(1)}$ from ${\mb x}_{{\rm R1}}$, ${\mb x}_{{\rm R2}}$, and ${\mb x}_{{\rm R3}}$, respectively. With successive interference cancellation, the decoding error probability goes to zero as $n\rightarrow \infty$, provided
\begin{subequations} \label{40}
\begin{align} {\label{45a}}
  R_{\rm R1}    &  \leq    {1\over 2} \log \left( 1+ {  p_{{\rm R1}}\over   p_{{\rm R2}}+    p_{{\rm R3}} +   p_{{\rm R4}} +\bar  \sigma_4^2}\right)   \\ {\label{45b}}
  R_{{\rm R2}}  &  \leq    {1\over 2} \log \left(1+  {  p_{{\rm R2}}\over   p_{{\rm R3}} +   p_{{\rm R4}} + \bar  \sigma_4^2}\right)  \\ {\label{45c}}
   R_{\rm R3}  & \leq {1\over 2} \log \left( 1+ {  p_{{\rm R3}}\over   p_{{\rm R4}} +\bar  \sigma_4^2}\right).
\end{align}	
\end{subequations}

We are now ready to present the following proposition.

\emph {Proposition 4.2:} For Case II in (\ref{case2}), an achievable rate tuple for the relay-to-user link is given by
	 \begin{subequations}  \label{Pro4.2}
	 	\begin{align} 	\label{Pro4.2_1}
	 	R_1 &  \! = \! {1\over 2} \log \left(  1\!+\! {  p_{{\rm R4}}\over\bar  \sigma_2^2}\right)  \!   + \!  {1\over 2} \log \left(1\!+ \! {  p_{{\rm R2}}\over   p_{{\rm R3}} \!+\!   p_{{\rm R4}} \!+\! \bar  \sigma_4^2}\right)  \\  \label{Pro4.2_2}
	 	R_2 &   \! =  \!  {1\over 2} \log \left(1+  {  p_{{\rm R2}}\over   p_{{\rm R3}} +   p_{{\rm R}4} + \bar  \sigma_1^2}\right)  \\  \label{Pro4.2_3}
	 	R_3 &   \! =  \!  {1\over 2} \log \left( 1+ {  p_{{\rm R1}}\over   p_{{\rm R2}} +   p_{{\rm R3}} +  p_{{\rm R4}}+ \bar  \sigma_1^2}\right)  \cr & \ \ \ \ \ \ +  {1\over 2} \log \left( 1+ {  p_{{\rm R3}}\over   p_{{\rm R4}} +\bar  \sigma_4^2}\right) \\  \label{Pro4.2_4}
	 	R_4 &  \! =  \!   \min\left\{{1\over 2} \log \left( 1+ {p_{{\rm R1}}\over  p_{{\rm R2}} +  p_{{\rm R4}} + \bar  \sigma_3^2}\right), \right .\nonumber\\
	 	& \ \ \ \ \ \ \left . \frac{1}{2}\log\left(1+\frac{p_{\rm R1}}{p_{\rm R2}+p_{\rm R3}+p_{\rm R4}+\bar\sigma_1^2}\right)\right\}.
	 	\end{align}
	 \end{subequations}

 \emph {Proof:} To prove Proposition 4.2, we need to show that the rate tuple in (\ref{Pro4.2}) satisfies (\ref{ine1}), (\ref{33}), (\ref{34}), (\ref{36}), (\ref{38}), and  (\ref{40}). To this end, we first combine (\ref{33}), (\ref{36}), and (\ref{40}) using the fact of $\bar \sigma_3^2   \geq \bar \sigma_1^2    \geq  \bar \sigma_4^2 \geq \bar \sigma_2^2$ for Case II, yielding
 \begin{subequations}\label{36x}
\begin{align} {\label{41ax}}
R_{{\rm R1}}  &  \leq    {1\over 2} \log \left( 1+ {  p_{{\rm R1}}\over   p_{{\rm R2}}+    p_{{\rm R}3} +   p_{{\rm R4}} +\bar  \sigma_1^2}\right)   \\ {\label{41bx}}
R_{{\rm R2}}  &  \leq    {1\over 2} \log \left(1+  {  p_{{\rm R2}}\over   p_{{\rm R3}} +   p_{{\rm R4}} + \bar  \sigma_4^2}\right)   \\ {\label{41cx}}
R_{{\rm R3}} & \leq {1\over 2} \log \left( 1+ {  p_{{\rm R3}}\over   p_{{\rm R4}} +\bar  \sigma_4^2}\right)  \\ {\label{41dx}}
R_{{\rm R4}}  & \leq   {1\over 2} \log \left(  1+ {  p_{{\rm R4}}\over\bar  \sigma_2^2}\right).
\end{align}	
\end{subequations} Then, substituting (\ref{36x}) into (\ref{ine1}), together with (\ref{34}) and (\ref{38}), we obtain Proposition 4.2.  \hfill $\blacksquare$

 We now propose the second scheme for Case II. The message reassembling at the relay is first to concatenate $m_{\rm B}$ and $m_{31}$ into  a single message,  mapped to   a codeword $ {\mb x}_{\rm R1}$   chosen from a Gaussian codebook of size $2^{n{R}_{\rm R1}}$.  Then   concatenate $m_{\rm A}$ and $m_{11}$ into  a single message,  mapped  to a codeword ${\mb x}_{{\rm R2}}$   chosen from a Gaussian codebook of size $2^{n{ R}_{{\rm R2}}}$. Thus
  	\begin{subequations} 
  \begin{align} \label{Pro4.3_0_R3}
 R_{\rm R1}&= R_{\rm 30} + R_{31}=R_3 \\   
 R_{\rm R2}&= R_{\rm 10}+ R_{11} =R_1.
  \end{align}
 	\end{subequations} 
  The relay transmits 
  \begin{align}
  {\bf x}_{\rm R} = 	{\bf x}_{{\rm R1}} + {\bf x}_{{\rm R2}}
  \end{align}
  where $ {\mb x}_{{\rm R}i}$ is power-constrained by  ${1\over n}||{\mb x}_{{\rm R}i}||^2 \leq  p_{{\rm R}i}$, $i\in\{1,2\}$, and
 \begin{align}
 p_{\rm R1}+p_{\rm R2} =P_{\rm R}.
 \end{align}
 The encoding operation at the relay is shown in Fig. 4.

   \begin{figure*}
    	\centering
    	\includegraphics[width= 4.7  in]{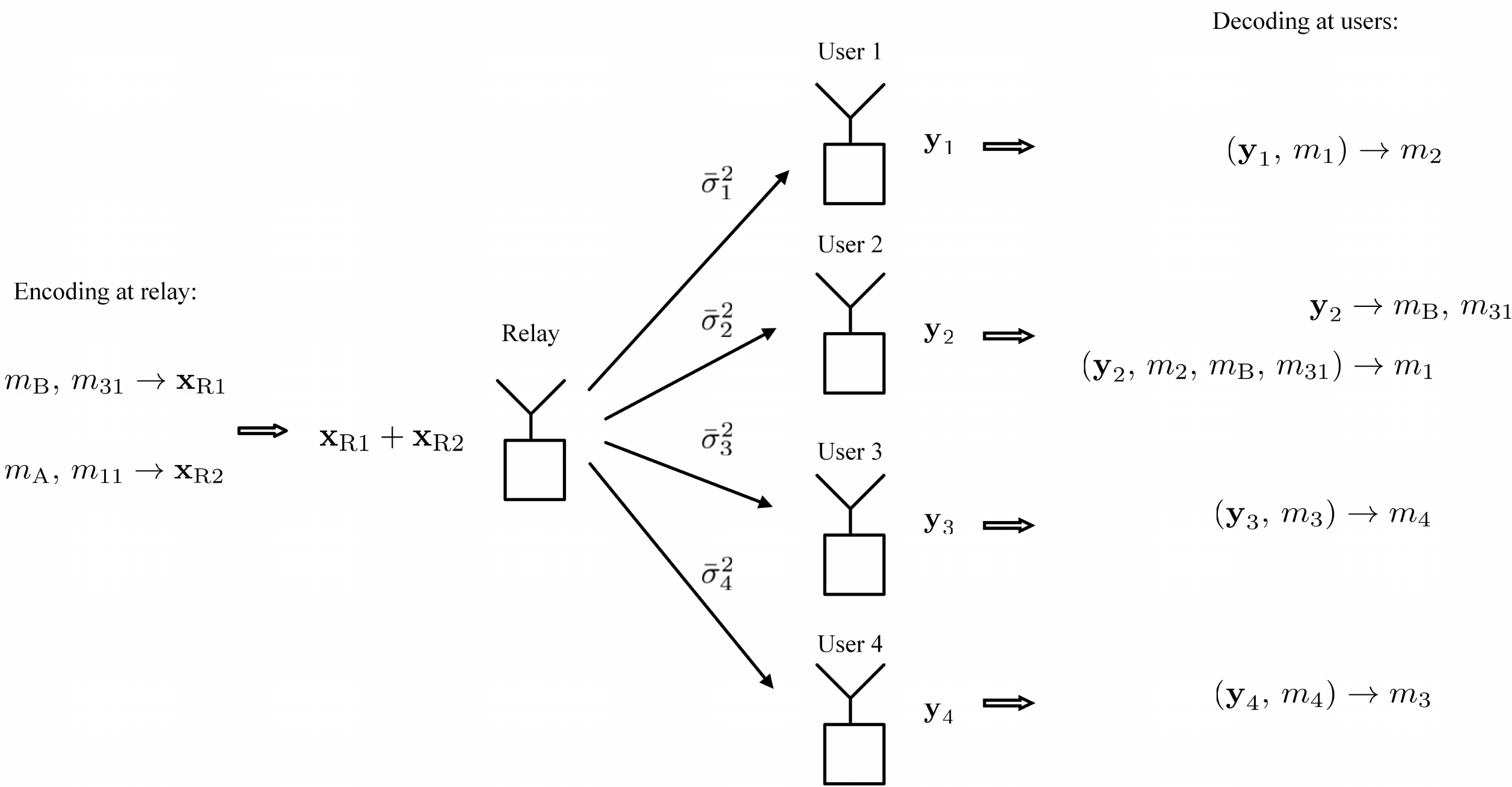}  \vspace{-2mm}
    	\caption{The message flow table of the second scheme for Case II: $\bar \sigma_3^2   \geq \bar \sigma_1^2    \geq  \bar \sigma_4^2 \geq \bar \sigma_2^2$. \vspace{-2mm}}\vspace{-2mm}
    \end{figure*}

 The received signal  at user 1 is given by
 \begin{align}
  \mb y_1 = g_1{\mb x}_{{\rm R}1} +  g_1{\mb x}_{{\rm R2}}  + \mb n_1.
 \end{align}
 By treating  ${\mb x}_{{\rm R1}}$ as noise,  user 1 decodes ${\mb x}_{{\rm R2}}$ (to acquire $m_2$) with the help of $m_1$.  The decoding operation at user 1 is shown in Fig. 4. 
  The decoding error probability goes to zero as $n\rightarrow \infty$, provided
  \begin{align} \label{Pro4.3_0_2}
    R_{2}  &  \leq    {1\over 2} \log \left(1+  {  p_{{\rm R2}}\over   p_{{\rm R1}} + \bar  \sigma_1^2}\right).
  \end{align}

  The received signal at user 2 is given by
  \begin{align}
  \mb y_2 = g_2{\mb x}_{{\rm R}1} +  g_2{\mb x}_{{\rm R2}}   + \mb n_2.
  \end{align}
  User	2 first decodes ${\mb x}_{{\rm R1}}$  to obtain $( m_{\rm B},\, m_{31})$ by treating $	{\bf x}_{{\rm R2}}$ as noise. The decoding error probability goes to zero as $n\rightarrow \infty$, provided
  \begin{align} \label{Pro4.3_0_R1}
  R_{\rm R1}    &\leq  {1\over 2} \log \left(1+ {p_{\rm R1}\over {p_{\rm R2}+\bar \sigma_2^2}}\right).
  \end{align}
  After removing ${\bf x}_{{\rm R1}}$ from the received signal, user 2 further decodes ${\bf x}_{{\rm R2}}$ with the help of its self-message.  
  The decoding operation at user 2 is shown in Fig. 4.
  From [2, Theorem 1], the decoding error probability goes to zero as $n\rightarrow \infty$,  provided  
  \begin{align} \label{Pro4.3_0_1}
   R_1 &  \leq {1\over 2} \log \left(1+ {p_{\rm R2}\over \bar \sigma_2^2}\right). 
  \end{align}

 The received signal  of user $i$ in pair B is given by
 \begin{align}
 \mb y_i = g_i \mb x_{\rm R1} + g_i\mb x_{\rm R2} + \mb n_i, \ i \in \mathcal{I}_{\rm B}.
 \end{align}
 By treating $ 	{\bf x}_{{\rm R2}}$ as noise,  users in pair B  decode $	{\bf x}_{{\rm R1}}$ with the help of their self-message  to obtain   the partner's message.  The decoding error probability goes to zero as $n\rightarrow \infty$, provided
 \begin{subequations} 
 	\begin{align}\label{Pro4.3_0_4} 
 	R_{4} &\leq {1\over 2} \log \left(1+ { p_{\rm R1}\over {p_{\rm R2}+\bar \sigma_3^2}}\right)\\  
 	\label{Pro4.3_0_3} 
 	 R_{3} &\leq {1\over 2} \log \left(1+ { p_{\rm R1}\over {p_{\rm R2}+\bar \sigma_4^2}}\right). 
 	\end{align}
 \end{subequations}

We are now ready to present the following proposition.

\emph {Proposition 4.3:} For Case II in (\ref{case2}), an achievable rate tuple for the relay-to-user link is given by
	 \begin{subequations} \label{Pro4.3}
	 	\begin{align} 
	 	\label{Pro4.3_1}
	 	R_1 &  \! = \! {1\over 2} \log \left(  1\!+\! {  p_{{\rm R2}}\over\bar  \sigma_2^2}\right)   \\ \label{Pro4.3_2}
	 	R_2 &   \! =  \!  {1\over 2} \log \left(1+  {  p_{{\rm R2}}\over   p_{{\rm R1}} + \bar  \sigma_1^2}\right)  \\ \label{Pro4.3_3}
	 	R_3 &   \! =  \!    {1\over 2} \log \left( 1+ {  p_{{\rm R1}}\over   p_{{\rm R2}} +\bar  \sigma_4^2}\right) \\ \label{Pro4.3_4}
	 	R_4 &  \! =  \!   {1\over 2} \log \left( 1+ {p_{{\rm R1}}\over  p_{{\rm R2}} + \bar  \sigma_3^2}\right).
	 	\end{align}
	 \end{subequations}

\emph {Proof:}
Note that (\ref{Pro4.3_1}) is from (\ref{Pro4.3_0_1}); (\ref{Pro4.3_2}) is from (\ref{Pro4.3_0_2}); (\ref{Pro4.3_4}) is from (\ref{Pro4.3_0_4}). For (\ref{Pro4.3_3}), we see that $R_3$ is subject to the constraints in (\ref{Pro4.3_0_R3}), (\ref{Pro4.3_0_R1}) and (\ref{Pro4.3_0_3}). Together with the fact that $\bar  \sigma_2^2 \le \bar  \sigma_4^2$, we obtain that $R_3$ in (\ref{Pro4.3_3})  is achievable. \hfill $\blacksquare$ 
\subsubsection{Case III}

\begin{figure*}
	\centering
	\includegraphics[width= 5.5 in]{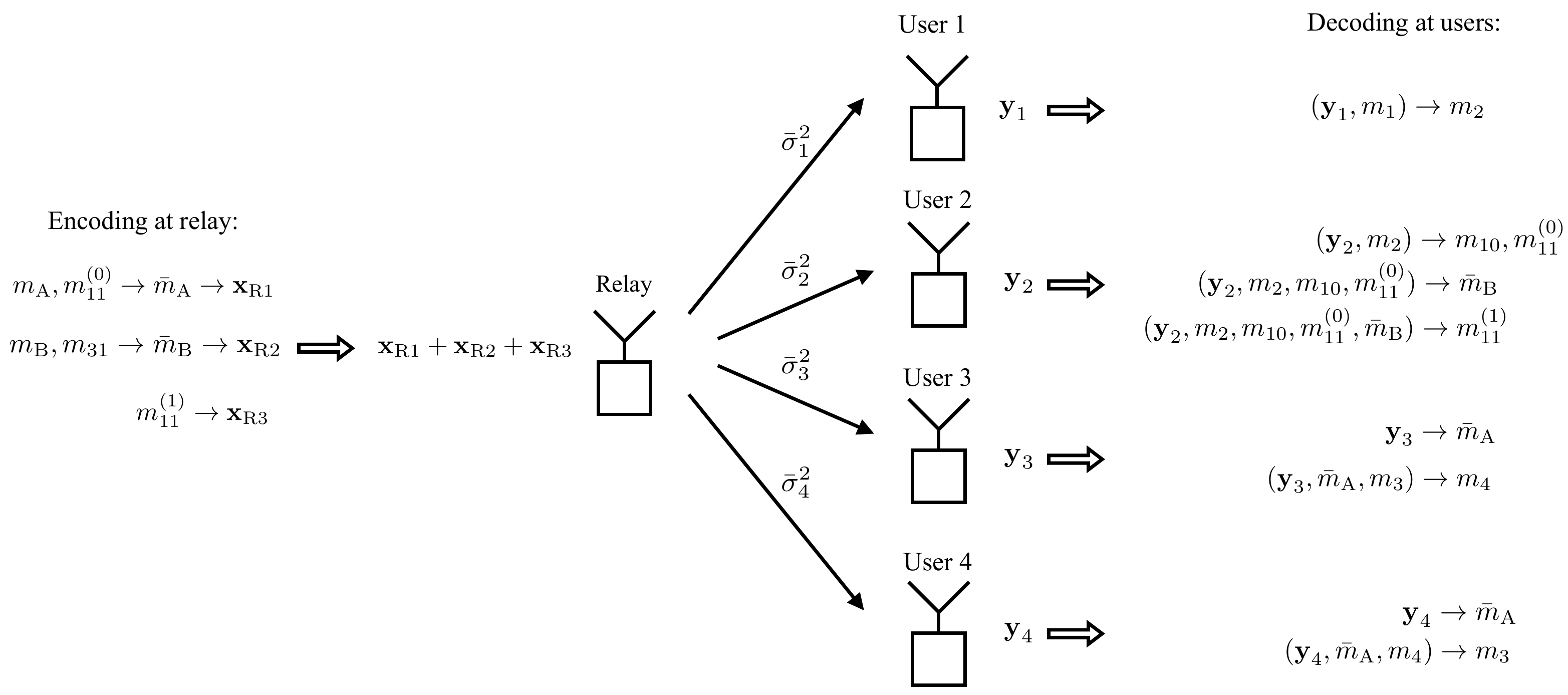}  \vspace{-2mm}
	\caption{ The message flow table  for Case III: $\bar \sigma_1^2   \geq \bar \sigma_3^2    \geq  \bar \sigma_4^2 \geq \bar \sigma_2^2$.   \vspace{-2mm}}\vspace{-2mm}
\end{figure*}

In this case, the relay splits the message $m_{11}$ into  $m_{11}^{(0)}$ and $m_{11}^{(1)}$, and then concatenates $m_{\rm A}$ and $m_{11}^{(0)}$ into  $\bar m_{\rm A}$. The relay concatenates $m_{\rm B}$ and $m_{31}$ into  $\bar m_{\rm B}$. Then, the relay maps $\bar m_{\rm A}$ to   $ {\mb x}_{\rm R1}$ using a Gaussian codebook of size $2^{n{R}_{\rm R1}}$,  maps  $\bar m_{\rm B}$  to ${\mb x}_{{\rm R2}}$ using a Gaussian codebook of size $2^{n{ R}_{{\rm R2}}}$, and maps $m_{11}^{(1)}$ to  ${\mb x}_{{\rm R}3}$ using a Gaussian codebook of size $2^{n{R}_{{\rm R}3}}$.
    The relay transmits a  superposition of these  three codewords:
    \begin{align}  
    {\bf x}_{\rm R} =  {\bf x}_{{\rm R}1}  +  {\bf x}_{{\rm R}2}   + {\bf x}_{{\rm R3}}
    \end{align} 
    where ${\mb x}_{{\rm R } i}$ is power-constrained by    $  {1\over n}|| {\mb x}_{{\rm R}i}||^2 \leq  p_{{\rm R}i}$, $i \in \{1,2,3\}$ and
    \begin{align}
    p_{\rm R1}+p_{\rm R2}+p_{\rm R3} =\!P_{\rm R}.
    \end{align}
 By construction, we have the following rate relations:
    \begin{subequations} \label{51}
    \begin{align}
    R_1  & = R_{\rm R1} + R_{\rm R3} \\
    R_2  & \leq  R_{\rm R1}  \\
    R_3  & = R_{\rm R2} \\
    R_4  & \leq R_{\rm R2}.
    \end{align}
    \end{subequations}

The encoding operation at the relay is shown in Fig. 5. The received signal  at user 1 is given by
\begin{align}
\mb y_1 = g_1{\mb x}_{{\rm R}1} +  g_1{\mb x}_{{\rm R2}} +   g_1{\mb x}_{{\rm R3}}  + \mb n_1.
\end{align}
User 1 decodes ${\mb x}_{{\rm R1}}$ (to acquire $m_2$) with the help of $m_1$. Note that given $m_1$, the rate of ${\mb x}_{{\rm R1}}$ is $R_2$, and ${\mb x}_{{\rm R3}}$ is known to user 1. The decoding operation at user 1 is shown in Fig. 5. 
The decoding error probability goes to zero as $n\rightarrow \infty$, provided
\begin{align} \label{49}
  R_{2}  &  \leq    {1\over 2} \log \left(1+  {  p_{{\rm R1}}\over   p_{{\rm R2}} + \bar  \sigma_1^2}\right).
\end{align}

The received signal at user 2 is given by
\begin{align}
\mb y_2 = g_2{\mb x}_{{\rm R}1} +  g_2{\mb x}_{{\rm R2}} +   g_2{\mb x}_{{\rm R3}}  + \mb n_2.
\end{align}	
User 2 sequentially decodes $  {\mb x}_{{\rm R1}}$, $  {\mb x}_{{\rm R2}}$,  and $  {\mb x}_{{\rm R3}}$  to obtain $\{m_{\rm A},m_{11}^{(0)}\}$, $\bar m_{\rm B}$ and $m_{11}^{(1)}$ of rates $R_{\rm R1}$, $R_{\rm R2}$, and $R_{\rm R3}$.  After decoding one codeword, user 2 removes it from the received signal, and then decodes the next one.   The decoding operation at user 2 is shown in Fig. 5. 
The decoding error probability goes to zero as $n\rightarrow \infty$, provided
\begin{subequations} \label{46}
\begin{align} \label{51a}
  R_{\rm R1}  &  \leq    {1\over 2} \log \left(1+  {  p_{{\rm R1}}\over   p_{{\rm R2}} +   p_{{\rm R3}} + \bar  \sigma_2^2}\right)   \\ \label{51b}
  R_{{\rm R2}} & \leq {1\over 2} \log \left( 1+ {  p_{{\rm R2}}\over   p_{{\rm R}3} +\bar  \sigma_2^2}\right) \\ \label{51c}
  R_{\rm R3} & \leq   {1\over 2} \log \left(  1+ {  p_{{\rm R3}}\over\bar  \sigma_2^2}\right).
\end{align}			
\end{subequations}

The received signal  of user $i$ in pair B is given by
\begin{align}
\mb y_i= g_i{\mb x}_{{\rm R}1} +  g_i{\mb x}_{{\rm R2}} +   g_i{\mb x}_{{\rm R3}} + \mb n_i, \ i\in\mathcal{I}_{\rm B}.
\end{align}			
Each user in pair B first decodes  $  {\mb x}_{{\rm R1}}$ to obtain $\bar m_{\rm A}$ of rate $R_{\rm R1}$ by treating $  {\bf x}_{{\rm R2}}$  and $  {\bf x}_{{\rm R3}}$ as noise. The decoding error probability goes to zero as $n\rightarrow \infty$, provided
\begin{align}  \label{47} 
  R_{{\rm R1}}  \leq    {1\over 2} \log \left(1+  {  p_{{\rm R1}}\over   p_{{\rm R2}}  +   p_{{\rm R3}} + \bar  \sigma_3^2}\right).
\end{align}		
After removing  $  {\mb x}_{{\rm R1}}$ from the received signal, each user in pair B further decodes $  {\bf x}_{{\rm R2}}$ with the help of its self-message. The decoding operation at pair B is shown in Fig. 5. The decoding error probability goes to zero as $n\rightarrow \infty$, provided
\begin{subequations} \label{48}
\begin{align}   \label{54a} 
  R_4 &  \leq  {1\over 2} \log \left(1+ {  p_{\rm R2}\over {  p_{{\rm R3}}+\bar \sigma_3^2}}\right) \\  \label{54b} 
  R_3 & \leq  {1\over 2} \log \left(1+ {  p_{\rm R2}\over {  p_{{\rm R3}}+\bar \sigma_4^2}}\right).
\end{align}		
\end{subequations}

We are now ready to present the following proposition.

\emph {Proposition 4.4:}  For Case III in (\ref{case3}), an achievable rate tuple for the relay-to-user link is given by
\begin{subequations} \label{Pro4.4}
\begin{align} \label{Pro4.4_1}
R_1 & = {1\over 2} \log \left( 1 \! + \! {  p_{{\rm R3}}\over \bar \sigma_2^2}\right)     \! +\!    {1\over 2} \log \left(1\!+\!  {  p_{{\rm R1}}\over   p_{{\rm R2}}  \!+\!   p_{{\rm R3}} \!+\! \bar  \sigma_3^2}\right)       \\ \label{Pro4.4_2}
R_2 & =  \min\left\{{1\over 2} \log \left(1+  {  p_{{\rm R1}}\over   p_{{\rm R2}} + \bar  \sigma_1^2}\right),\right. \nonumber\\
& \ \ \ \ \ \ \left. \frac{1}{2}\log\left(1+\frac{p_{\rm R1}}{p_{\rm R2}+p_{\rm R3}+\bar\sigma_3^2}\right)\right\}   \\ \label{Pro4.4_3}
R_3 &  = {1\over 2} \log \left(1+ {  p_{\rm R2}\over {  p_{{\rm R3}}+\bar \sigma_4^2}}\right)  \\  \label{Pro4.4_4}
R_4 &  =  {1\over 2} \log \left(1+ {  p_{\rm R2}\over {  p_{{\rm R3}}+\bar \sigma_3^2}}\right).
\end{align}
\end{subequations}

  \emph {Proof:}  To prove the above proposition, we need to show that (\ref{Pro4.4}) satisfies (\ref{51}),  (\ref{49}),  (\ref{46}), (\ref{47}) and  (\ref{48}). To this end, we first combine (\ref{46}) and (\ref{47}) by noting $\bar \sigma_1^2   \geq \bar \sigma_3^2    \geq  \bar \sigma_4^2 \geq \bar \sigma_2^2$ for Case III, yielding
\begin{subequations} \label{46x}
\begin{align} \label{51ax}
  R_{\rm R1}  &  \leq    {1\over 2} \log \left(1+  {  p_{{\rm R1}}\over   p_{{\rm R2}} +   p_{{\rm R3}} + \bar  \sigma_3^2}\right)   \\ \label{51bx}
  R_{{\rm R2}} & \leq {1\over 2} \log \left( 1+ {  p_{{\rm R2}}\over   p_{{\rm R}3} +\bar  \sigma_2^2}\right) \\ \label{51cx}
  R_{\rm R3} & \leq   {1\over 2} \log \left(  1+ {  p_{{\rm R3}}\over\bar  \sigma_2^2}\right).
\end{align}			
\end{subequations}	
Then, substituting (\ref{46x}) into (\ref{51}), together with (\ref{49}) and (\ref{48}), we obtain (\ref{Pro4.4}), which concludes the proof. \hfill $\blacksquare$

\section{Main Result}

We are now ready to present the main result of the paper. The proof is given in the next section.

 \emph {Theorem 1:}
\label{capacity}
For any rate tuple $(R_1, R_2, R_3, R_4)$ in the rate region of the outer bound (\ref{outer}), the rate tuple $(R_1- {1\over 2}, R_2- {1\over 2}, R_3- {1\over 2}, R_4- {1\over 2})$ is  achievable for the considered two-pair TWRC.

Prior to this work, the best known rate gap from the capacity of the two-pair TWRC is ${3\over 2}$ bits per user reported in \cite{Sezgin12}. Theorem 1 reduces the capacity gap to within $1\over 2$ bit per user. {A number of new techniques are employed to derive our capacity bounds. First, we derive a genie-aided outer bound for the Gaussian two-pair TWRC. This new bound is tighter than the cut-set outer bound used in \cite{Sezgin12}. Second, for the user-to-relay link, the relay appropriately scales its received signal for nested lattice decoding, so as to include the extra term $\frac{1}{2}$ in the logarithm of (\ref{network_decoding_r_10}), as compared to the scheme in \cite{Sezgin12}. Third, we present a message-reassembling strategy at the relay to decouple the coding design for the user-to-relay and relay-to-user transmissions. This provides more flexibility to the coding design for the relay-to-user link, so as to better accommodate the asymmetry of the channel conditions for the user-to-relay and relay-to-user links. Finally, more advanced power allocation and more intricate analysis techniques are involved in the proof, as seen in the subsequent section.}

\section{Proof of Theorem 1}
\subsection{Preliminaries}

The rate region specified by   (\ref{outer}) is a polytope, denoted by $\mathcal{R}$. Let $\mb R^{(j)} =  (R_1^{(j)}, R_2^{(j)}, R_3^{(j)}, R_4^{(j)})$,   $j\in\{1,2,...,K\}$ be the $j$-th vertex of $\mathcal{R}$, where $K$ is the total number of the vertices of $\mathcal{R}$. Then, a rate tuple in $\mathcal{R}$ can be generally represented as  
\begin{align} \label{56}
\mb R =\sum\limits_{j=1}^{K} \lambda_j\mb R^{(j)}  
\end{align}
where
\begin{align}
\sum\limits_{j=1}^{K} \lambda_j = 1 \   \mathrm{and} \ \lambda_j  \geq 0, \ j\in\{1,2,...,K\}.
\end{align}

We have the following proposition. The proof is straightforward using the technique of rate splitting \cite{Cover91}.

\emph {Proposition 5:}  If the vertices $\mb R^{(1)}$, $\mb R^{(2)}$, ..., $\mb R^{(K)}$ are achievable,  then  the convex combination  $\mb R$ in (\ref{56}) is also achievable.

We further have the following proposition.

\emph {Proposition 6:} If  all the vertices of $\mathcal{R}$ are achievable to within $\tau $ bits per dimension, then  any rate tuple in $\mathcal{R}$ is achievable to within  $\tau$  bits per dimension.

\emph {Proof}: For each  vertex  $\mb R^{(j)}= (R_1^{(j)}, R_2^{(j)}, R_3^{(j)}, R_4^{(j)})$,  $j\in\{1,...,K\} $,  there is an achievable rate tuple $\tilde {\mb R}^{(j)}= (\tilde R_1^{(j)}, \tilde R_2^{(j)}, \tilde R_3^{(j)}, \tilde R_4^{(j)})$, with
$  R_{i}^{(j)}\leq  \tilde R_{i}^{(j)}+ \tau$,   $\tau\geq 0$,  $i \in\{1,...,4\}$.  For any rate tuple $\mb R  \in \mathcal{R}$, we have 
\begin{align}
\mb R  & =\sum\limits_{j=1}^{K} \lambda_j\mb R^{(j)}   \cr
& \leq \sum\limits_{j=1}^{K} \lambda_j  ( \tilde {\mb R}^{(j)} + \tau\mb 1)  \cr
& = \sum\limits_{j=1}^{K} \lambda_j    \tilde {\mb R}^{(j)} + \tau\mb 1
\end{align}
where $\mb 1$ is an all-one vector with an appropriate size, and $``\leq"$ means ``entry-wise no greater than".

From Proposition 5,  the convex combination $\sum\limits_{j=1}^{K} \lambda_j    \tilde {\mb R}^{(j)}$ is achievable. Therefore, $\mb R$ is achievable to within $\tau$ bits per dimension. 
This completes the proof of   Proposition 6.  \hfill $\blacksquare$

With Proposition 6, to prove Theorem 1, it suffices to only consider the achievability of the vertices of $\mathcal{R}$. To further simplify the analysis, we  introduce the concept of \emph{maximal vertex}.

\emph {Definition 1:}  (\textit{Maximal vertex}):   A vertex $\mb R^{(j)}$ is a maximal vertex if there is no other vertex $\mb R^{(j')}$, $j'\neq j$, satisfying  $ \mb R^{(j')} \geq \mb R^{(j)}$,  where $``\geq"$ means ``entry-wise no less than".

	Clearly, if the {maximal vertices} are achievable to within $\tau$ bits per dimension, then all the vertices are achievable to within $\tau$ bits per dimension. Hence, we only need to consider  the achievability of the   {maximal vertices} of the outer bound.

\subsection {Achievability for User-to-Relay Link}

We start with the maximal vertices for the user-to-relay link. From Proposition 1, the rates  in the user-to-relay link  are outer bounded by
\begin{subequations} \label{60}
\begin{align} \label{62a}
R_{1}+R_{3}  &\leq   C_{13}  \\ \label{62b}
R_{1}+R_{4}  &\leq   C_{14}   \\ \label{62c}
R_{2}+R_{3} & \leq  C_{23} \\    \label{62d}
R_{2}+R_{4} & \leq  C_{24}   \\ \label{62e}
R_i &\leq C_i,  \ i \in \{1,...,4\} \\  \label{62f}
R_i &\geq 0,  \ i \in \{1,...,4\}.
\end{align}
\end{subequations}
Clearly, (\ref{60}) specifies a four-dimension polytope. The pivoting algorithm \cite{Bremner98}   can be used to determine the vertices of (\ref{60}). Among these vertices, six of them are maximal vertices with the rate tuples $(R'_1, R'_2, R'_3, R'_4 )$ given by
\begin{subequations} 
\begin{flalign}
{\rm (U.1)} \quad \quad \quad \quad\quad \quad \quad \quad \quad \quad  \quad  \quad 	R_1' &= C_{13} - C_3 &&\\
	R_2'&=   C_{23} - C_3  &&\\
	R_3' &= C_3 &&\\
	R_4'&= C_4, &&
 \end{flalign}
\end{subequations}
 \vspace{-6mm}
\begin{subequations}
\begin{flalign}
{\rm (U.2)} \quad \quad \quad \quad \quad  \quad \quad \quad \quad \quad \quad  \quad 	R_1' &= C_1  &&\\
R_2'&=  C_2  &&\\
R_3' &=  C_{13} - C_1 &&  \\
R_4'&= C_{14} - C_1,  &&
\end{flalign}
\end{subequations}
 \vspace{-6mm}
\begin{subequations}
\begin{flalign}
{\rm (U.3)}  \quad \quad \quad \quad \quad  \quad \quad \quad \quad \quad  \quad  \quad 	R_1' &=  C_{13}- C_{23}+C_{24}-C_4   &&\\
R_2'&=  C_{24}-C_4 &&\\
R_3' &= C_{23}- C_{24}+C_4 &&\\
R_4'&= C_4, &&
\end{flalign}
\end{subequations}
 \vspace{-6mm}
\begin{subequations}
\begin{flalign}
{\rm (U.4)} \quad \quad \quad \quad \quad  \quad \quad \quad \quad \quad \quad  \quad	R_1' &=  C_{14}-C_4  &&\\
R_2'&=   C_{24}-C_4 &&\\
R_3' &= C_{13}-C_{14}+C_4 &&\\
R_4'&= C_4, &&
\end{flalign}
\end{subequations}
 \vspace{-6mm}
\begin{subequations}
\begin{flalign}
{\rm (U.5)}  \quad \quad \quad \quad \quad  \quad \quad \quad \quad \quad \quad  \quad 	R_1' &=  C_{13}-C_{23}+C_2   &&\\
R_2'&=  C_2 && \\
R_3' &= C_{23}-C_2 &&\\
R_4'&= C_{24}-C_2, &&
\end{flalign}
\end{subequations}
 \vspace{-6mm}
\begin{subequations}
\begin{flalign}
{\rm (U.6)} \quad \quad \quad \quad \quad   \quad \quad \quad \quad \quad \quad  \quad 	R_1' &=  C_{14}-C_{24}+C_2   && \\
R_2'&=  C_2  &&\\
R_3' &= C_{13}-C_{14}+C_{24}-C_2  &&\\
R_4'&= C_{24}-C_2. &&
\end{flalign}
\end{subequations}

{\bf Remark 3}: Instead of using the pivoting algorithm  \cite{Bremner98}, we can determine the vertices of  (\ref{60}) by an exhaustive search.
 A vertex of (\ref{60}) has the following property: If  $(R_1, R_2, R_3, R_4)$ is a vertex, then exactly four inequalities of  (\ref{60}) become equality. Therefore, a brute forth way to enumerate vertices is to choose every four equations from (\ref{60}), find the solution, and then check whether the solution satisfies all the other inequalities in (\ref{60}). This is time-consuming since  there are $\binom{12}{4}$ combinations in total. To reduce the number of candidate vertices, we have the following observations: (i) For a maximal vertex, no equality can be selected from (\ref{62f}); (ii) for a vertex, at least one and at most two equalities are selected from (\ref{62e}), one for each user pair.
 Suppose that $R_1' = C_1$ holds. Then $R_3'$ is at most $C_{13}-C_1$, and $R_4'$ is at most $C_{14}-C_{1}$. As $h_1^2P_1\geq h_2^2P_2$, we obtain $C_{13}-C_{1}\leq C_{23}-C_2$ and $C_{14}-C_1\leq C_{24}-C_2$. Thus, for a vertex, we obtain $R_2'=C_2$. As a result, the only valid vertex is given by (U. 2). Now suppose that $R'_1<C_1$ and $R'_2=C_2$. It can be readily verified that (U.5) and (U.6) are the only possible vertices. The third case to consider is $\{R_1'<C_1, R_2'<C_2, R_3'=C_3\}$. Then, $R_1'$ is at most $C_{13}-C_{3}$, and $R_2'$ is at most $C_{23}-C_3$. As $h_3^2P_3\geq h_4^2P_4$, we obtain $C_{13}-C_3\leq C_{14}-C_4$ and  $C_{23}-C_3 \leq C_{24}-C_4$. Therefore, $R_4'=C_4$, which leads to vertex (U.1). What remains is the case of $\{R_1'<C_1, R_2'<C_2, R_3'<C_3, R_4'=C_4\}$.  In this case, (U.3) and (U.4) are the only possible choices.

We next show that the   maximal vertices  (U.1)-(U.6) are achievable to within $1\over 2$ bit per dimension. 
We start with (U.1). We  first rewrite   (U.1) as  
  \begin{subequations}  \label{67}
  	\begin{align}
  	R_1' & = R_{{\rm 1}0}' +R_{{\rm 1}1}'  \\ 
  	R_2'& = R_{{\rm 1}0}' \\ 
  	R_3' &=  R_{{\rm 3}0}' +R_{{\rm 3}1}'   \\ 
  	R_4' &=  R_{{\rm 3}0}', 
  	\end{align}
  \end{subequations}
  with 
 \begin{subequations} 
\begin{align} \label{67a}
R_{{\rm 1}0}'  &= C_{23} - C_{3} \geq 0    \\  \label{67b}
R_{{\rm 3}0}'&= C_4  \geq 0  \\ \label{67c}
R_{{\rm 1}1}'&=   C_{13} - C_{23 }   \geq 0 \\ \label{67d}
R_{{\rm 3}1}'&= C_{3} - C_4 \geq 0.
\end{align}
\end{subequations}
where the inequality in (\ref{67a}) follows from  (\ref{C_i}) and (\ref{C_ij}),  the inequality in (\ref{67c})  is due to $h_1^2P_1\geq h_2^2P_2$, and  the inequality in (\ref{67d}) is due to $h_3^2P_3\geq h_4^2 P_4$. 

The power parameters are set as
 \begin{subequations}  \label{66}
\begin{align}
p_{\rm 10 }  & = {1\over 2} h_2^2P_2 \\
p_{\rm 30 } & = {1\over 2} h_4^2P_4 \\
p_{\rm 11} &   =  h_1^2P_1 - h_2^2P_2 \\
p_{\rm 31} & = h_3^2P_3 - h_4^2P_4.
\end{align} 
\end{subequations}
With the decoding order of 
\begin{align}
\mb x_{11} \rightarrow  \{\mb x_{10}, \mb x_{2}\}  \rightarrow \mb x_{31} \rightarrow   \{\mb x_{30}, \mb x_{4}\}, 
\end{align}
we have the following achievable rates:
 \begin{subequations} 
 	\begin{align}
R_{11} & =  {1\over 2} \log \left(1 + {p_{11} \over 2p_{10} + 2p_{30} + p_{31} + \sigma_{\rm R}^2} \right) \cr
& =  {1\over 2} \log \left( { h_1^2P_{1} + h_3^2P_{3}  + \sigma_{\rm R}^2 \over h_2^2P_{2} + h_3^2P_{3}  + \sigma_{\rm R}^2} \right) \cr
& = C_{13}- C_{23} \\
R_{10} & =   {1\over 2}  \left[\log \left( {1\over 2} +  {p_{10}\over 2p_{30} + p_{31} + \sigma_{\rm R}^2 }\right)\right]^+ \cr
& \ge  {1\over 2}  \log \left(  {h_2^2P_{2}  + h_3^2P_{3}  + \sigma_{\rm R}^2 \over h_3^2P_{3}   + \sigma_{\rm R}^2 }\right) -{1\over 2} \cr
& = C_{23}- C_{3}-{1\over 2} \\
R_{31} & =  {1\over 2} \log \left(1 + {p_{31} \over  2p_{30}  + \sigma_{\rm R}^2} \right) \cr
& = C_{3}- C_{4} \\
R_{30} & =   {1\over 2} \left[\log \left( {1\over 2} +  {p_{30}\over   \sigma_{\rm R}^2 }\right)\right]                                                                              ^{+} \cr
& \ge C_{4}-{1\over 2}. 
\end{align} 
\end{subequations}
 Hence the rate tuple ($R_1,\,R_2,\,R_3,\,R_4$) with
  \begin{align} \label{70}
  R_{i}  & =  R_{i}'- {1\over 2} ,  \  i\in\{1,...,4\}
  \end{align}
is achievable.

 We now consider (U.2). Similarly to (U.1), we     rewrite (U.2) as 
   \begin{subequations}   
   	\begin{align}
   	R_1' & = R_{{\rm 1}0}' +R_{{\rm 1}1}'  \\ 
   	R_2'& = R_{{\rm 1}0}' \\ 
   	R_3' &=  R_{{\rm 3}0}' +R_{{\rm 3}1}'   \\ 
   	R_4' &=  R_{{\rm 3}0}', 
   	\end{align}
   \end{subequations}
   with  
\begin{subequations}
	\begin{align}
	R_{{\rm 1}0}'  &= C_2  \geq 0  \\
	R_{{\rm 3}0}'&=  C_{14} - C_1  \geq 0 \\
	R_{{\rm 1}1}'&=   C_1-C_2 \geq 0 \\
	R_{{\rm 3}1}'&= C_{13} - C_{14} \geq 0.  
	\end{align}
\end{subequations}
  The power parameters are still given by (\ref{66}).
With the decoding order of
 \begin{align}
 \mb x_{31} \rightarrow  \{\mb x_{30}, \mb x_{4}\}  \rightarrow \mb x_{11} \rightarrow   \{\mb x_{10}, \mb x_{2}\}, 
 \end{align}
we have the following achievable rates:
  \begin{subequations} 
  	\begin{align}
  	R_{31} & =  {1\over 2} \log \left(1 + {p_{31} \over 2p_{10} + 2p_{30} + p_{11} + \sigma_{\rm R}^2} \right) \cr
  	&  =  C_{13}- C_{14} \\
  	R_{30} & =   {1\over 2}  \log \left[\left( {1\over 2} +  {p_{30}\over 2p_{10} + p_{11} + \sigma_{\rm R}^2 }\right)\right]^{+} \cr
  	& \ge C_{14}- C_{1}-{1\over 2} \\
  	R_{11} & =  {1\over 2} \log \left(1 + {p_{11} \over  2p_{10}  + \sigma_{\rm R}^2} \right) \cr
  	& = C_{1}- C_{2} \\
  	R_{10} & =   {1\over 2}  \left[\log \left( {1\over 2} +  {p_{10}\over   \sigma_{\rm R}^2 }\right)\right]^{+} \cr
  	& \ge C_{2}-{1\over 2}. 
  	\end{align} 
  \end{subequations}
  Therefore, (\ref{70})  holds for (U.2).

For  (U.3), we     rewrite (U.3) as 
 \begin{subequations}   
 	\begin{align}
 	R_1' & = R_{{\rm 1}0}' +R_{{\rm 1}1}'  \\ 
 	R_2'& = R_{{\rm 1}0}' \\ 
 	R_3' &=  R_{{\rm 3}0}' +R_{{\rm 3}1}'   \\ 
 	R_4' &=  R_{{\rm 3}0}', 
 	\end{align}
 \end{subequations}
 with 
\begin{subequations}
	\begin{align}
	R_{{\rm 1}0}' & =  C_{24}-C_4 \geq 0 \\ 
	R_{{\rm 3}0}'&    = C_4 \geq 0 \\
	R_{{\rm 1}1}' & = C_{13} -C_{23} \geq 0 \\ 
	R_{{\rm 3}1}' & =   C_{23} -C_{24} \geq 0.  
	\end{align}				
\end{subequations}
The power parameters are   given by (\ref{66}).
With  the decoding order of
 \begin{align}
 \mb x_{11} \rightarrow \mb x_{31} \rightarrow   \{\mb x_{10}, \mb x_{2}\}  \rightarrow  \{\mb x_{30}, \mb x_{4}\},   
 \end{align}
 we have the following achievable rates
 \begin{subequations} 
 	\begin{align}
 	R_{11} & =  {1\over 2} \log \left(1 + {p_{11} \over 2p_{10} + 2p_{30} + p_{31} + \sigma_{\rm R}^2} \right) \cr
 	& = C_{13}- C_{23} \\
	 R_{31} & =  {1\over 2} \log \left(1 + {p_{31} \over  2p_{10} + 2p_{30}  + \sigma_{\rm R}^2} \right) \cr
 	 	& = C_{23}- C_{24} \\
 	R_{10} & =   {1\over 2} \left[ \log \left( {1\over 2} +  {p_{10}\over 2p_{30}  + \sigma_{\rm R}^2 }\right)\right]^{+} \cr
 	& \ge C_{24}- C_{4}-{1\over 2} \\
 	R_{30} & =   {1\over 2} \left[ \log \left( {1\over 2} +  {p_{30}\over   \sigma_{\rm R}^2 }\right)\right]^{+} \cr
 	& \ge C_{4}-{1\over 2}. 
 	\end{align} 
 \end{subequations}
  Therefore, (\ref{70}) holds for (U.3).

We now consider  (U.4), we     rewrite (U.4) as 
 \begin{subequations}   
 	\begin{align}
 	R_1' & = R_{{\rm 1}0}' +R_{{\rm 1}1}'  \\ 
 	R_2'& = R_{{\rm 1}0}' \\ 
 	R_3' &=  R_{{\rm 3}0}' +R_{{\rm 3}1}'   \\ 
 	R_4' &=  R_{{\rm 3}0}', 
 	\end{align}
 \end{subequations}
 with 
\begin{subequations}
	\begin{align}
	R_{{\rm 1}0}' & =  C_{24}-C_4 \\ 
	R_{{\rm 3}0}'&    = C_4 \\
	R_{{\rm 1}1}' & = C_{14} -C_{24} \\ 
	R_{{\rm 3}1}' & =  C_{13} -C_{14}.  
	\end{align}			\end{subequations}
  The power parameters are   given by (\ref{66}).
  With the decoding order of
 \begin{align}
 \mb x_{31} \rightarrow \mb x_{11} \rightarrow   \{\mb x_{10}, \mb x_{2}\}  \rightarrow  \{\mb x_{30}, \mb x_{4}\},   
 \end{align}
we have the following achievable rates:
 \begin{subequations} 
 	\begin{align}
 	R_{31} & =  {1\over 2} \log \left(1 + {p_{31} \over 2p_{10} + 2p_{30} + p_{11} + \sigma_{\rm R}^2} \right) \cr
 	& = C_{13}- C_{14} \\
 	R_{11} & =  {1\over 2} \log \left(1 + {p_{11} \over  2p_{10} + 2p_{30}  + \sigma_{\rm R}^2} \right) \cr
 	& = C_{14}- C_{24} \\
 	R_{10} & =   {1\over 2}  \left[\log \left( {1\over 2} +  {p_{10}\over 2p_{30}  + \sigma_{\rm R}^2 }\right)\right]^{+} \cr
 	& \ge C_{24}- C_{4}-{1\over 2} \\
 	R_{30} & =   {1\over 2}  \left[\log \left( {1\over 2} +  {p_{30}\over   \sigma_{\rm R}^2 }\right)\right]^{+} \cr
 	& \ge  C_{4}-{1\over 2}. 
 	\end{align} 
 \end{subequations}
  Therefore, (\ref{70})  holds for (U.4).

We now consider  (U.5), we     rewrite (U.5) as 
 \begin{subequations}   
 	\begin{align}
 	R_1' & = R_{{\rm 1}0}' +R_{{\rm 1}1}'  \\ 
 	R_2'& = R_{{\rm 1}0}' \\ 
 	R_3' &=  R_{{\rm 3}0}' +R_{{\rm 3}1}'   \\ 
 	R_4' &=  R_{{\rm 3}0}', 
 	\end{align}
 \end{subequations}
 where 
\begin{subequations}
	\begin{align}
	R_{{\rm 1}0}' & =  C_{2}\\ 
	R_{{\rm 3}0}'&    = C_{24} -C_2 \\
	R_{{\rm 1}1}' & = C_{13} -C_{23} \\ 
	R_{{\rm 3}1}' & =   C_{23} -C_{24}.  
	\end{align}				\end{subequations}
  The power parameters are   given by (\ref{66}).
  With the decoding order of
  \begin{align}
  \mb x_{11} \rightarrow \mb x_{31} \rightarrow  \{\mb x_{30}, \mb x_{4}\}    \rightarrow   \{\mb x_{10}, \mb x_{2}\},  
  \end{align}
 we have the following achievable rates:
   \begin{subequations} 
   	\begin{align}
   	R_{11} & =  {1\over 2} \log \left(1 + {p_{11} \over 2p_{10} + 2p_{30} + p_{31} + \sigma_{\rm R}^2} \right) \cr
   	& = C_{13}- C_{23} \\
   	R_{31} & =  {1\over 2} \log \left(1 + {p_{31} \over  2p_{10} + 2p_{30}  + \sigma_{\rm R}^2} \right) \cr
   	& = C_{23}- C_{24} \\
   	R_{30} & =   {1\over 2}  \left[\log \left( {1\over 2} +  {p_{30}\over 2p_{10}  + \sigma_{\rm R}^2 }\right)\right]^{+} \cr
   	& \ge C_{24}- C_{2}-{1\over 2} \\
   	R_{10} & =   {1\over 2}  \log \left[\left( {1\over 2} +  {p_{10}\over   \sigma_{\rm R}^2 }\right)\right]^{+} \cr
   	& \ge  C_{2}-{1\over 2}. 
   	\end{align} 
   \end{subequations}
     Therefore, (\ref{70}) holds for (U.5).

We now consider (U.6), we     rewrite (U.6) as 
 \begin{subequations}   
 	\begin{align}
 	R_1' & = R_{{\rm 1}0}' +R_{{\rm 1}1}'  \\ 
 	R_2'& = R_{{\rm 1}0}' \\ 
 	R_3' &=  R_{{\rm 3}0}' +R_{{\rm 3}1}'   \\ 
 	R_4' &=  R_{{\rm 3}0}', 
 	\end{align}
 \end{subequations}
 where 
\begin{subequations}
	\begin{align}
	R_{{\rm 1}0}' & =  C_{2} \geq 0  \\ 
	R_{{\rm 3}0}'&    = C_{24} -C_2 \geq 0\\
	R_{{\rm 1}1}' & = C_{14} -C_{24} \geq 0 \\ 
	R_{{\rm 3}1}' & =  C_{13} -C_{14} \geq 0.  
	\end{align}		
		\end{subequations}
  The power parameters are   given by (\ref{66}).
  With the decoding order of
  \begin{align}
  \mb x_{31} \rightarrow \mb x_{11} \rightarrow  \{\mb x_{30}, \mb x_{4}\}    \rightarrow   \{\mb x_{10}, \mb x_{2}\},  
  \end{align}
 we have the following achievable rates:
    \begin{subequations} 
    	\begin{align}
    	R_{31} & =  {1\over 2} \log \left(1 + {p_{31} \over 2p_{10} + 2p_{30} + p_{11} + \sigma_{\rm R}^2} \right) \cr
    	& = C_{13}- C_{14} \\
    	R_{11} & =  {1\over 2} \log \left(1 + {p_{11} \over  2p_{10} + 2p_{30}  + \sigma_{\rm R}^2} \right) \cr
    	& = C_{14}- C_{24} \\
    	R_{30} & =   {1\over 2} \left[ \log \left( {1\over 2} +  {p_{30}\over 2p_{10}  + \sigma_{\rm R}^2 }\right)\right]^{+} \cr
    	& \geq C_{24}- C_{2}-{1\over 2} \\
    	R_{10} & =   {1\over 2}  \left[\log \left( {1\over 2} +  {p_{10}\over   \sigma_{\rm R}^2 }\right)\right]^{+} \cr
    	& \ge C_{2}-{1\over 2}. 
    	\end{align} 
    \end{subequations}
      Therefore, (\ref{70})  holds for (U.6). This concludes the proof for the user-to-relay link.

 Considering all the above six cases,  we conclude that 
the rate gap between the inner   and   outer bounds is at most $1\over 2$ bit per user in the user-to-relay link.

\subsection{Achievability for Relay-to-User Link}

For relay-to-user link, we analyze the rate gap for each of the  three cases in (\ref{case}).

\subsubsection{Case I}

With (\ref{case1}), we   simplify the outer bound in (\ref{outer}) for the relay-to-user link as:

\begin{subequations} \label{c1}
\begin{align} \label{c1a}
	R_1 +R_3 & \leq    D_2 \\ \label{c1b}
	R_1+R_4 &  \leq  D_2\\ \label{c1c}
	R_2 +R_3& \leq  D_1 \\ \label{c1d}
	R_2 +R_4& \leq   D_1 \\ \label{c1e}
	R_3&  \leq  D_4\\ \label{c1f}
	R_4&  \leq   D_3 \\ \label{c1g}
	R_i & \geq 0, \ i\in\mathcal{I}.
\end{align}
\end{subequations}
Note that $R_1 \leq D_2$ and $R_2 \leq D_1$ are implied by (\ref{c1a}) and (\ref{c1c}), and so are not included in (\ref{c1}).
For  the polytope defined by (\ref{c1}), we have  three maximal vertices with the rate tuples ($R_1', R_2', R_3', R_4'$) given by
\begin{subequations} \label{74}
\begin{flalign} 
\label{D1.1_R1'}
{\rm (D1.1)}  \quad \quad \quad \quad \quad \quad\quad \quad \quad \quad \quad  \quad  \quad 	R_1' &= D_2 &&\\
\label{D1.1_R2'}
R_2'&= D_1 &&\\
\label{D1.1_R3'}
R_3' &= 0 &&\\
\label{D1.1_R4'}
R_4'&= 0 &&
\end{flalign}\end{subequations}
 \vspace{-6mm}
\begin{subequations} \label{75}
\begin{flalign}
\label{D1.2_R1'}
{\rm (D1.2)}   \quad\quad \quad \quad \quad \quad\quad \quad \quad \quad \quad  \quad  \quad 	R_1' &= D_2 -D_4 &&\\
\label{D1.2_R2'}
R_2' &= D_1- D_4 &&\\
\label{D1.2_R3'}
R_3'& = D_4&&\\
\label{D1.2_R4'}
R_4'& = D_3 &&
\end{flalign}
\end{subequations}
 \vspace{-6mm}
\begin{subequations}\label{76}
\begin{flalign}
\label{D1.3_R1'}
{\rm (D1.3)}   \quad\quad \quad \quad \quad \quad\quad \quad \quad \quad \quad  \quad  \quad 	R_1' &= D_2 -D_3 &&\\
\label{D1.3_R2'}
R_2' &= D_1 - D_3 &&\\
\label{D1.3_R3'}
R_3' &= D_3 &&\\
\label{D1.3_R4'}
R_4' &= D_3. &&
\end{flalign}
\end{subequations}

{\bf Remark 4}: The maximal vertices of  (\ref {c1}) can  be determined as follows. Suppose that both the equalities in (\ref{c1a}) and (\ref{c1b}) hold for a maximal vertex. Then, from (\ref{c1a}) and (\ref{c1b}), we obtain $R_3'=R_4'$. This implies that the equalities in (\ref{c1c}) and (\ref{c1d}) either both hold or both fail. If the former is true, we obtain the maximal vertices (D1.1) and (D1.3); for latter, there are no other two equalities in (\ref{c1e}) to (\ref{c1g}), together with (\ref{c1a}) and (\ref{c1b}), to yield a maximal vertex. 

Therefore, except for (D1.1) and (D1.3), at most one of equalities in (\ref{c1a}) and (\ref{c1b}) holds. By noting $R_3'\geq R_4'$, we further see that only the equality in (\ref{c1a}) can hold. From similar arguments, for (\ref{c1c}) and (\ref{c1d}), only the equality in (\ref{c1c}) can hold. Together with the equalities in (\ref{c1e}) and (\ref{c1f}), we obtain the last maximal vertex (D1.2).

We need to prove that these three maximal vertices (D1.1)-(D1.3) are achievable to within   $1\over 2$ bit. We use the achievable rates in Proposition 4.1 for the proof.

For (D1.1), we set $p_{\rm R1} = 0$ and $p_{\rm R2} = P_{\rm R}$ in (\ref{Pro4.1}). Then 
\begin{subequations}
\begin{align}
R_1 + {1\over 2} & = {1\over 2}\log \left(1+  {P_{\rm R}\over \bar \sigma_2^2}\right) + {1\over 2}  > R_1' \\
R_2 + {1\over 2} & = {1\over 2}\log \left(1+  {P_{\rm R}\over \bar \sigma_1^2}\right) + {1\over 2}  > R_2'  \\
R_3 + {1\over 2} & = {1\over 2} > 0 =  R_3' \\
R_4 + {1\over 2} &={1\over 2}  > 0 = R_4'. 
\end{align}
\end{subequations}
Therefore, (D1.1) is achievable  to within $1\over 2$ bit.

We now consider (D1.2).  We   set $p_{\rm R2}  = \min(P_{\rm R}, \bar \sigma_4^2)$ and $p_{\rm R1}  =  [P_{\rm R}-\bar \sigma_4^2]^+$ in (\ref{Pro4.1}).
For  $P_{\rm R}\geq \bar \sigma_4^2$,  we obtain $p_{\rm R2}  = \bar \sigma_4^2$. Then
\begin{subequations}
\begin{flalign}\label{D1.2_CASE1_1}
   \quad\quad\quad \quad\quad\quad\quad\quad\quad\quad	R_1  +{1\over 2}& = {1\over 2} \log \left(2{  \bar \sigma_2^2+ \bar \sigma_4^2\over \bar \sigma_2^2}\right)  & \\
   \label{D1.2_CASE1_1_1}
    &	 \geq  { 1\over 2} \log \left(  { {P_{\rm R} + \bar \sigma_2^2}\over \bar \sigma_2^2 }  \cdot  {  \bar \sigma_4^2 \over {P_{\rm R}+ \bar \sigma_4^2}  } \right)   &\\  
    \label{D1.2_CASE1_1_2}
    &  =R'_1 &
\end{flalign}
\end{subequations}
\vspace{-6mm}
\begin{subequations}
\begin{flalign}\label{D1.2_CASE1_2}
   \quad\quad\quad\quad\quad\quad\quad\quad\quad\quad	R_2 +{1\over 2}& = {1\over 2} \log \left(2{  \bar \sigma_1^2+ \bar \sigma_4^2\over \bar \sigma_1^2}\right)  & \\
   \label{D1.2_CASE1_2_1}
   &	\geq  { 1\over 2} \log \left(  { {P_{\rm R} + \bar \sigma_1^2}\over \bar \sigma_1^2 }  \cdot  {  \bar \sigma_4^2 \over {P_{\rm R}+ \bar \sigma_4^2}  } \right)   & \\ 
   \label{D1.2_CASE1_2_2}
    & =R'_2 &
\end{flalign}
\end{subequations}
\vspace{-6mm}
\begin{subequations}
\begin{flalign}
    \label{D1.2_CASE1_3}
    \quad\quad\quad\quad\quad\quad\quad\quad\quad\quad R_3+{1\over 2} & =   {1\over 2} \log \left({P_{\rm R} +  \bar \sigma_4^2  \over {\bar \sigma_4^2}}\right)  &\\
    \label{D1.2_CASE1_3_1} 
   & = R_3'&
\end{flalign}
\end{subequations}
\vspace{-6mm}
\begin{subequations}
\begin{flalign}
    \label{D1.2_CASE1_4}
    \quad\quad\quad\quad\quad\quad\quad\quad\quad\quad R_4+{1\over 2} & =  {1\over 2} \log \left(2{P_{\rm R} +  \bar \sigma_3^2  \over {\bar \sigma_3^2+    \bar \sigma_4^2}}\right)   & \\
    \label{D1.2_CASE1_4_1} 
   &  \geq    {1\over 2} \log \left( { {P_{\rm R} + \bar \sigma_3^2}\over \bar \sigma_3^2 }\right)  & \\ 
   \label{D1.2_CASE1_4_2}
    & = R_4'.&
\end{flalign}
\end{subequations}
In the above, step (\ref{D1.2_CASE1_1}) follows from (\ref{Pro4.1_1}); step (\ref{D1.2_CASE1_1_1}) from $\bar \sigma_4^2 \geq \bar \sigma_2^2$; and step (\ref{D1.2_CASE1_1_2}) from (\ref{D1.2_R1'}). Step (\ref{D1.2_CASE1_2}) follows from (\ref{Pro4.1_2}); step (\ref{D1.2_CASE1_2_1}) from $\bar \sigma_4^2 \geq \bar \sigma_1^2$; and step (\ref{D1.2_CASE1_2_2}) from (\ref{D1.2_R2'}). Step (\ref{D1.2_CASE1_3}) follows from (\ref{Pro4.1_3}); and  step (\ref{D1.2_CASE1_3_1}) from (\ref{D1.2_R3'}). Step (\ref{D1.2_CASE1_4}) follows from (\ref{Pro4.1_4}); step (\ref{D1.2_CASE1_4_1}) from $\bar \sigma_3^2 \geq \bar \sigma_4^2$; and step (\ref{D1.2_CASE1_4_2}) from (\ref{D1.2_R4'}).

For $P_{\rm R}< \bar \sigma_4^2$, we have $p_{\rm R2}  = P_{\rm R}$. Then
\begin{subequations}
\begin{align}
R_1 + {1\over 2} & = {1\over 2}\log \left(1+  {P_{\rm R}\over \bar \sigma_2^2}\right) + {1\over 2}  > R_1' \\
R_2 + {1\over 2} & = {1\over 2}\log \left(1+  {P_{\rm R}\over \bar \sigma_1^2}\right) + {1\over 2}  > R_2'  \\
R_3 + {1\over 2} & = {1\over 2} > {1\over 2}\log \left(1+  {P_{\rm R}\over \bar \sigma_4^2}\right) =  R_3' \\
R_4 + {1\over 2} &={1\over 2}  > {1\over 2}\log \left(1+  {P_{\rm R}\over \bar \sigma_3^2}\right) = R_4'. 
\end{align}
\end{subequations}
Therefore, (D1.2) is achievable to within ${1\over 2}$ bit.

We next consider (D1.3). We    set $p_{\rm R2}  = \min (\bar \sigma_3^2, P_{\rm R})$ and $p_{\rm R1}  =  [P_{\rm R}-\bar \sigma_3^2]^+$ in (\ref{Pro4.1}). 

For $P_{\rm R}\geq \bar \sigma_3^2$, we obtain $p_{\rm R2}  = \bar \sigma_3^2$. Then
\begin{subequations}
\begin{flalign}\label{D1.3_CASE1_1}
    \quad\quad\quad\quad\quad\quad\quad\quad\quad\quad	R_1  +{1\over 2}& = {1\over 2} \log \left(2{  \bar \sigma_2^2+ \bar \sigma_3^2\over \bar \sigma_2^2}\right)  & \\
    \label{D1.3_CASE1_1_1}
	&	  \geq  { 1\over 2} \log \left(  { {P_{\rm R} + \bar \sigma_2^2}\over \bar \sigma_2^2 }  \cdot  {  \bar \sigma_3^2 \over {P_{\rm R}+ \bar \sigma_3^2}  } \right) & \\ 
	\label{D1.3_CASE1_1_2}
	&  =R'_1 &
\end{flalign}
\end{subequations}
\vspace{-6mm}
\begin{subequations}
\begin{flalign}\label{D1.3_CASE1_2}
   \quad\quad\quad\quad\quad\quad\quad\quad\quad\quad	R_2 +{1\over 2}& = {1\over 2} \log \left(2{  \bar \sigma_1^2+ \bar \sigma_3^2\over \bar \sigma_1^2}\right)  & \\
   \label{D1.3_CASE1_2_1}
	&  	\geq  { 1\over 2} \log \left(  { {P_{\rm R} + \bar \sigma_1^2}\over \bar \sigma_1^2 }  \cdot  {  \bar \sigma_3^2 \over {P_{\rm R}+ \bar \sigma_3^2}  } \right)  &  \\ 
	\label{D1.3_CASE1_2_2} 
	& =R'_2 &
\end{flalign}
\end{subequations}
\vspace{-6mm}
\begin{subequations}
\begin{flalign}\label{D1.3_CASE1_3}	
  \quad\quad\quad\quad\quad\quad\quad\quad\quad\quad R_3+{1\over 2} & =   {1\over 2} \log \left(2{P_{\rm R} +  \bar \sigma_4^2  \over {\bar \sigma_3^2 + \bar \sigma_4^2}}\right)& \\
  \label{D1.3_CASE1_3_1} 
  &   \geq {1\over 2} \log \left( { {P_{\rm R} + \bar \sigma_3^2}\over \bar \sigma_3^2 }\right)&   \\ 
  \label{D1.3_CASE1_3_2}
   & = R_3'&
\end{flalign}
\end{subequations}
\vspace{-6mm}
\begin{subequations}
\begin{flalign}\label{D1.3_CASE1_4}  
  \quad\quad\quad\quad\quad\quad\quad\quad\quad\quad R_4+{1\over 2} & =  {1\over 2} \log \left({P_{\rm R} +  \bar \sigma_3^2  \over {\bar \sigma_3^2}}\right)  &\\
  \label{D1.3_CASE1_4_1}  
   & = R_4'.&
	\end{flalign}
\end{subequations}
In the above, step (\ref{D1.3_CASE1_1}) follows from (\ref{Pro4.1_1}); step (\ref{D1.3_CASE1_1_1}) from $\bar \sigma_3^2 \geq \bar \sigma_2^2$; and step (\ref{D1.3_CASE1_1_2}) from (\ref{D1.3_R1'}). Step (\ref{D1.3_CASE1_2}) follows from (\ref{Pro4.1_2}); step (\ref{D1.3_CASE1_2_1}) from $\bar \sigma_3^2 \geq \bar \sigma_1^2$; and step (\ref{D1.3_CASE1_2_2}) from (\ref{D1.3_R2'}). Step (\ref{D1.3_CASE1_3}) follows from (\ref{Pro4.1_3}); step (\ref{D1.3_CASE1_3_1}) from $\bar \sigma_3^2 \geq \bar \sigma_4^2$; and  step (\ref{D1.3_CASE1_3_2}) from (\ref{D1.3_R3'}). Step (\ref{D1.3_CASE1_4}) follows from (\ref{Pro4.1_4});  and step (\ref{D1.3_CASE1_4_1}) from (\ref{D1.3_R4'}).

For $P_{\rm R}< \bar \sigma_3^2$, we have $p_{\rm R2}  = P_{\rm R}$. Then
\begin{subequations}
\begin{align}
R_1 + {1\over 2} & = {1\over 2}\log \left(1+  {P_{\rm R}\over \bar \sigma_2^2}\right) + {1\over 2}  > R_1' \\
R_2 + {1\over 2} & = {1\over 2}\log \left(1+  {P_{\rm R}\over \bar \sigma_1^2}\right) + {1\over 2}  > R_2'  \\
R_3 + {1\over 2} & = {1\over 2} > {1\over 2}\log \left(1+  {P_{\rm R}\over \bar \sigma_3^2}\right) =  R_3' \\
R_4 + {1\over 2} &={1\over 2}  > {1\over 2}\log \left(1+  {P_{\rm R}\over \bar \sigma_3^2}\right) = R_4'. 
\end{align}
\end{subequations}
Therefore, (D1.3) is achievable to within ${1\over 2}$ bit.
 This concludes the proof for Case I of the relay-to-user link.



\subsubsection{Case  II}

With (\ref{case2}),  the outer bound of the relay-to-user link in (\ref{outer}) can be written as 
\begin{subequations} \label{c2}
\begin{align} \label{c2a}
R_1 +R_3 &\leq  D_2,  \\ \label{c2b}
R_1 +R_4 &\leq  D_2, \\ \label{c2c}
R_2 +R_3 &\leq  D_4, \\ \label{c2d}
R_2 +R_4 &\leq  D_1, \\ \label{c2e}
R_4 &  \leq D_3,   \\ \label{c2f}
R_i & \geq 0, \ i\in\mathcal{I}.
\end{align}
\end{subequations}
Note that $R_1 \leq D_2$, $R_2\leq D1$, and $R_3\leq D_4$ are implied by (\ref{c2a}), (\ref{c2c}), and (\ref{c2d}), and so are not included here. For  the polytope defined by (\ref{c2}), we have the following five  maximal vertices with the rate tuples $(R_{1}',R_2',R_3',R_4')$ given by
\begin{subequations}   \label{81}
\begin{flalign}
\label{D2.1_R1'}
{\rm ({D2}.1)}  \quad \quad \quad \quad \quad  \quad \quad \quad \quad \quad  \quad  \quad 	R_1'  & = D_2 &&\\
\label{D2.1_R2'}
R_2' & = D_1  &&\\
\label{D2.1_R3'}
R_3' & = 0 &&\\
\label{D2.1_R4'}
R_4' & = 0, &&
\end{flalign}\end{subequations}
 \vspace{-6mm}
\begin{subequations}   \label{82}
\begin{flalign}
\label{D2.2_R1'}
{\rm (D2.2)}  \quad \quad \quad \quad \quad \quad \quad \quad \quad \quad   \quad  \quad 	R_1'& = D_2 -D_4 &&\\
\label{D2.2_R2'}
R_2' & = 0 &&\\
\label{D2.2_R3'}
R_3' & = D_4 &&\\
\label{D2.2_R4'}
R_4' & = D_3, &&
\end{flalign}\end{subequations}
 \vspace{-6mm}
\begin{subequations}   \label{83}
\begin{flalign}
\label{D2.3_R1'}
{\rm (D2.3)}   \quad \quad \quad \quad \quad \quad \quad \quad \quad \quad  \quad  \quad 	R_1' & = D_2 - D_4 +D_1  &&\\
\label{D2.3_R2'}
R_2' & = D_1 &&\\
\label{D2.3_R3'}
R_3' & = D_4 - D_1  && \\
\label{D2.3_R4'}
R_4' & = 0, &&
\end{flalign}\end{subequations}
 \vspace{-6mm}
\begin{subequations}   \label{84}
\begin{flalign}
\label{D2.4_R1'}
	{\rm (D2.4)}   \quad \quad \quad \quad \quad \quad \quad \quad \quad \quad  \quad  \quad 	R_1' & = D_2 - D_3 &&\\
	\label{D2.4_R2'}
	R_2' & = D_1 -D_3 &&\\
	\label{D2.4_R3'}
	R_3' & = D_3 &&\\
	\label{D2.4_R4'}
	R_4' & = D_3, &&
\end{flalign}\end{subequations} \vspace{-6mm}
\begin{subequations}    \label{85}
\begin{flalign}
\label{D2.5_R1'}
{\rm (D2.5)}   \quad \quad \quad \quad \quad \quad \quad \quad \quad \quad  \quad  \quad 	 R_1' & =D_2\!-\!D_4 \!+\!D_1 -D_3 &&\\
\label{D2.5_R2'}
R_2' & =D_1 -D_3 && \\
\label{D2.5_R3'}
R_3' & = D_4 - D_1 + D_3 &&\\
\label{D2.5_R4'}
R_4' & = D_3. &&
\end{flalign}\end{subequations}

We need to prove that the five maximal vertices (D2.1)-(D2.5) are achievable to within   $1\over 2$ bit.	Note that the proofs of  (D2.1), (D2.3), and (D2.4) only involve the achievable rates in Proposition 4.2, (D2.2) only involve the achievable rates in Proposition 4.3, while the proof of (D2.5) involves both Propositions 4.2 and 4.3.

For ({D2}.1), we use the achievable rates in Proposition 4.2, we  set $  p_{{\rm R}4} = \min(\bar \sigma_1^2, P_{\rm R})$, $  p_{{\rm R}2}= [P_{\rm R} - \bar \sigma_1^2]^+$, and   $p_{{\rm R}1}=   p_{{\rm R}3}=0$ in (\ref{Pro4.2}).
For   $P_{\rm R} \geq \bar \sigma_1^2$,  we obtain $p_{{\rm R}4} = \bar \sigma_1^2$. Then
\begin{subequations}
\begin{flalign}\label{D2.1_CASE1_1}
     \quad\quad\quad\quad\quad\quad\quad\quad\quad	R_1  +{1\over 2}& = {1\over 2} \log \left(2{  \bar \sigma_2^2+ \bar \sigma_1^2\over \bar \sigma_2^2} \cdot    {  P_{\rm R}+ \bar \sigma_4^2  \over \bar  \sigma_4^2 + \sigma_1^2 } \right)   &\\
     \label{D2.1_CASE1_1_1}
     & \geq {1\over 2} \log \left({  \bar \sigma_2^2+ \bar \sigma_1^2\over \bar \sigma_1^2} \cdot    {  P_{\rm R}+ \bar \sigma_4^2  \over \bar   \sigma_2^2 } \right)&\\
     \label{D2.1_CASE1_1_2}
	 &  \geq    {1\over 2} \log \left(  { {P_{\rm R} + \bar \sigma_2^2}\over \bar \sigma_2^2 }  \right)
	 &\\ 
	 \label{D2.1_CASE1_1_3}
	 & = R_1' &
\end{flalign}
\end{subequations}
\vspace{-6mm}
\begin{subequations}
\begin{flalign}\label{D2.1_CASE1_2} 	 
	\quad\quad\quad\quad\quad\quad\quad\quad\quad R_2  +{1\over 2}   &  =    {1\over 2} \log \left({  {  P_{\rm R}+ \bar \sigma_1^2  \over    \bar \sigma_1^2 } }\right)   &\\
	\label{D2.1_CASE1_2_1} 
	& = R_2'.&
\end{flalign}
\end{subequations}
In the above, step (\ref{D2.1_CASE1_1}) follows from (\ref{Pro4.2_1}); step (\ref{D2.1_CASE1_1_1}) from $\bar \sigma_4^2 \leq \bar \sigma_1^2$; step (\ref{D2.1_CASE1_1_2}) from $\bar \sigma_4^2 \geq \bar \sigma_2^2$; and step (\ref{D2.1_CASE1_1_3}) from (\ref{D2.1_R1'}). Step (\ref{D2.1_CASE1_2}) follows from (\ref{Pro4.2_2}); and step (\ref{D2.1_CASE1_2_1}) from  (\ref{D2.1_R2'}).

For  $P_{\rm R}  <  \sigma_1^2$,  we have $p_{{\rm R}4} = P_{\rm R}$. Then
 \begin{subequations}
 	\begin{align}
 		R_1 + {1\over 2} & > R_1' \\
		R_2 +{1\over 2} & = {1\over 2} >R_2'\\
 		R_3  +{1\over 2}  &  \ge 0 = R_3'\\
 		R_4  +{1\over 2}   & \ge 0 = R_4'.
 	\end{align}
 \end{subequations}
 Therefore, (D2.1) is achievable to within ${1\over 2}$ bit.

We now consider ({D2}.2). In this case, we use the achievable rates in Proposition 4.3. For  $P_{\rm R}\geq \bar \sigma_4^2$,  we   set $p_{{\rm R}1}=P_{\rm R}-\bar \sigma_4^2$, $  p_{{\rm R}2}= \bar \sigma_4^2$ in (\ref{Pro4.3}). 
Then 
\begin{subequations}
\begin{flalign}\label{D2.2_CASE1_1}
    \quad\quad\quad\quad\quad\quad\quad\quad\quad R_1 \!+\! {1\over 2} & =  {1\over 2} \log \left(   2 {  \bar \sigma_4^2 +\bar  \sigma_2^2  \over\bar  \sigma_2^2}\right)  &\\
    \label{D2.2_CASE1_1_1}
    &      \geq      {1\over 2} \log \left(   { {P_{\rm R} + \bar \sigma_2^2}\over \bar \sigma_2^2 } \cdot   {{  \bar \sigma_4^2}\over P_{\rm R}+  \bar \sigma_4^2 }   \right) &\\ 
    \label{D2.2_CASE1_1_2}
     &  = R'_1 &
\end{flalign}
\end{subequations}
\vspace{-6mm}
\begin{flalign}
\quad\quad\quad\quad\quad\quad\quad\quad\quad R_2 \!+\! {1\over 2}  &  >0  = R_2' &
\end{flalign}
\vspace{-6mm}
\begin{subequations}
\begin{flalign}\label{D2.2_CASE1_3}
\quad\quad\quad\quad\quad\quad\quad\quad\quad R_3 \!+\! {1\over 2} & =      {1\over 2} \log \left(    {  P_{\rm R} +\bar  \sigma_4^2  \over   \bar  \sigma_4^2}\right)      & \\
\label{D2.2_CASE1_3_1}
&  = R_3' &  
\end{flalign}
\end{subequations}
\vspace{-6mm}
\begin{subequations}
\begin{flalign}\label{D2.2_CASE1_4}
\quad\quad\quad\quad\quad\quad\quad\quad\quad R_4\!+\! {1\over 2} &= \frac{1}{2}\log \left( 2\frac{P_{\rm R}  + \bar{\sigma}_3^2}{ \bar{\sigma}_3^2 + \bar  \sigma_4^2 }\right) &\\
\label{D2.2_CASE1_4_1} 
&    \geq     {1\over 2} \log \left(    {{P_{\rm R} + \bar \sigma_3^2}\over \bar \sigma_3^2 }    \right)    &\\
\label{D2.2_CASE1_4_2}
 & = R_4'.&
\end{flalign}
\end{subequations}
In the above, step (\ref{D2.2_CASE1_1}) follows from (\ref{Pro4.3_1}); step (\ref{D2.2_CASE1_1_1}) from $\bar \sigma_4^2 \geq \bar \sigma_2^2$;  and step (\ref{D2.2_CASE1_1_2}) from (\ref{D2.2_R1'}). 
Step (\ref{D2.2_CASE1_3}) follows from (\ref{Pro4.3_3}); and step (\ref{D2.2_CASE1_3_1}) from  (\ref{D2.2_R3'}).
 Step (\ref{D2.2_CASE1_4}) follows from (\ref{Pro4.3_4}); step (\ref{D2.2_CASE1_4_1}) from $\bar \sigma_3^2 \geq \bar \sigma_4^2$; and step (\ref{D2.2_CASE1_4_2}) from  (\ref{D2.2_R4'}).

For $P_{\rm R}< \bar \sigma_4^2$,  we   set $p_{{\rm R}1}=0$, $  p_{{\rm R}2}= P_{\rm R}$ in (\ref{Pro4.3}). 
Then
\begin{subequations}
	\begin{align}
R_1 + {1\over 2}  &  > R_1' \\
R_2  + {1\over 2} & > {1\over 2}  > R_2' \\
R_3 +{1\over 2} & ={1\over 2} > R_3' \\
R_4 +{1\over 2} & ={1\over 2} > R_4'. 
\end{align}
\end{subequations}
Therefore, (D2.2) is achievable to within ${1\over 2}$ bit.

We next consider ({D2}.3). In this case, we use the achievable rates in Proposition 4.2. The proof is divided into three subcases:
	\begin{subequations} \label{D2.3_CASE_i}
	 \begin{align} 
 \rm {(i)} \; 	 &	P_{\rm R}\ge  \bar{\sigma}_1^2 \\
 \rm {(ii)} \; & \bar{\sigma}_4^2\le P_{\rm R}< \bar{\sigma}_1^2  \\ 
 \rm {(iii)} \;  	&	P_{\rm R}< \bar{\sigma}_4^2.
		\end{align}
	\end{subequations}
We now consider  the three subcases in (\ref{D2.3_CASE_i}) one by one.

	$\rm {(i)}\; P_{\rm R}\ge  \bar{\sigma}_1^2$:  
 We   set $p_{{\rm R}1} = 0$, $p_{{\rm R}2} =P_{\rm R} - \bar \sigma_1^2$ and  $p_{{\rm R}3} + p_{{\rm R}4} =\bar \sigma_1^2$ in (\ref{Pro4.2}).  Then
\begin{subequations}
\begin{flalign}\label{D2.3_CASE1_2}
\quad\quad\quad\quad\quad\quad\quad\quad\quad\quad\quad\quad R_2 \!+\! {1\over 2} &  \!=\!  {1\over 2}\log \left(      {P_{\rm R}  + \bar \sigma_1^2 \over  \bar \sigma_1^2}   \right)  &\\ 
 \label{D2.3_CASE1_2_1}
& \!=\!R_2'&    
\end{flalign}
\end{subequations}
 \vspace{-6mm}
\begin{flalign} 
\quad\quad\quad\quad\quad\quad\quad\quad\quad\quad\quad\quad R_4\!+\! {1\over 2} &  \!=\! {1\over 2} >  R_4', &    
\end{flalign}
where 
step (\ref{D2.3_CASE1_2}) follows from (\ref{Pro4.2_2}), and step (\ref{D2.3_CASE1_2_1}) from  (\ref{D2.3_R2'}).

 Then we need to show that there exists $p_{\rm{R}3}$ and $p_{\rm{R}4}$ satisfying the following  two inequalities:
 \begin{subequations}
 \label{D2.3_CASE1_1_0}
 \begin{eqnarray}\label{D2.3_CASE1_1}
 \frac{1}{2}\log \left(2\frac{p_{\rm{R}4} + \bar{\sigma}_2^2}{\bar{\sigma}_2^2}\cdot
 \frac{P_{\rm R} + \bar{\sigma}_4^2}{p_{\rm{R}3} + p_{\rm{R}4} + \bar{\sigma}_4^2} \right) &  \geq &   	{1\over 2}\log \left(   { {P_{\rm R} \!+\! \bar \sigma_2^2}\over \bar \sigma_2^2 }\cdot   {{  \bar \sigma_4^2}\over P_{\rm R}\!+\!  \bar \sigma_4^2 }\cdot  {{P_{\rm R} \!+\! \bar \sigma_1^2}\over \bar \sigma_1^2 }   \right) \\
 \frac{1}{2}\log \left( 2 \frac{p_{\rm{R}3} + p_{\rm{R}4} +  \bar{\sigma}_4^2}{ p_{\rm{R}4} +  \bar{\sigma}_4^2}\right)
 \label{D2.3_CASE1_3}
 &  \geq &    {1\over 2}\log \left(    {{P_{\rm R} + \bar \sigma_4^2}\over \bar \sigma_4^2 }  {{  \bar \sigma_1^2}\over P_{\rm R}+ \bar \sigma_1^2 }  \right),   
 \end{eqnarray}
 \end{subequations}
 where the left hand side (LHS) of (\ref{D2.3_CASE1_1}) is equal to $R_1 + {1\over 2}$ with $R_1$ given by (\ref{Pro4.2_1}), the right hand side (RHS) of (\ref{D2.3_CASE1_1}) is equal to $R_1^{'}$ given by  (\ref{D2.3_R1'}), the LHS of (\ref{D2.3_CASE1_3}) is equal to $R_3 + {1\over 2}$ with $R_3$ given by (\ref{Pro4.2_3}), and the RHS of (\ref{D2.3_CASE1_3}) is equal to $R_3^{'}$ given by  (\ref{D2.3_R3'}).
 Note that (\ref{D2.3_CASE1_1_0}) can be rewritten as   
 \begin{subequations}
 \label{D2.3_CASE1_13}
 \begin{align}
   \frac{2(p_{\rm{R}4} + \bar{\sigma}_2^2)(P_{\rm R} + \bar{\sigma}_4^2)}{\bar{\sigma}_2^2(p_{\rm{R}3} + p_{\rm{R}4} + \bar{\sigma}_4^2)} \ge& \frac{(P_{\rm R} + \bar{\sigma}_2^2)\bar{\sigma}_4^2(P_{\rm R} + \bar{\sigma}_1^2)}{\bar{\sigma}_2^2(P_{\rm R} + \bar{\sigma}_4^2)\bar{\sigma}_1^2}\\
  2\frac{p_{\rm{R}3} + p_{\rm{R}4} + \bar{\sigma}_4^2}{p_{\rm{R}4} + \bar{\sigma}_4^2} \ge & \frac{(P_{\rm R} + \bar{\sigma}_4^2)\bar{\sigma}_1^2}{\bar{\sigma}_4^2(P_{\rm R} + \bar{\sigma}_1^2)}.
  \end{align}
  \end{subequations}
Tother with $p_{{\rm R}3} + p_{{\rm R}4} =\bar \sigma_1^2$, we can further write (\ref{D2.3_CASE1_13}) as
  \begin{subequations}
  \label{D2.3_CASE1_maxmin}
  \begin{flalign}
     \quad\quad\quad\quad\quad\quad\quad\quad  p_{\rm{R}4} \ge {}& \frac{\bar{\sigma}_4^2(P_{\rm R} + \bar{\sigma}_1^2)}{(P_{\rm R} + \bar{\sigma}_4^2)\bar{\sigma}_1^2}(\bar{\sigma}_1^2 + \bar{\sigma}_4^2)\!\cdot\! \frac{P_{\rm R} + \bar{\sigma}_2^2}{2(P_{\rm R} + \bar{\sigma}_4^2)} 
      \! -\! \bar{\sigma}_2^2\!\!& \\
      \stackrel{\vartriangle }{=}{}& p_{\rm{R}4,\textrm{min}} &
  \end{flalign}
 \end{subequations}
   \vspace{-6mm} 
  \begin{subequations}
  	\label{D2.3_CASE1_maxmin_1}
  	\begin{flalign}
     \quad\quad\quad\quad\quad\quad\quad\quad   p_{\rm{R}4}\le {}& \frac{\bar{\sigma}_4^2(P_{\rm R} + \bar{\sigma}_1^2)}{(P_{\rm R} + \bar{\sigma}_4^2)\bar{\sigma}_1^2}(\bar{\sigma}_1^2 + \bar{\sigma}_4^2)\cdot 2 - \bar{\sigma}_4^2&\\
      \stackrel{\vartriangle }{=}{}& p_{\rm{R}4,\textrm{max}}.&
  \end{flalign}
  \end{subequations}
  To prove that there exists $p_{\rm{R}4}$ satisfying (\ref{D2.3_CASE1_maxmin}) and (\ref{D2.3_CASE1_maxmin_1}), we need to show that the following inequalities hold: 
  \begin{subequations}
  \label{D2.3_CASE1_max&min}
  \begin{align}
  \label{D2.3_CASE1_max&min_1}
  {}& p_{\rm{R}4,\textrm{max}}- p_{\rm{R}4,\textrm{min}}\ge 0\\
  \label{D2.3_CASE1_max&min_2}
  {}& p_{\rm{R}4,\textrm{max}}\ge  0\\
  \label{D2.3_CASE1_max&min_3}
  {}& p_{\rm{R}4,\textrm{min}}\le p_{\rm{R}3} + p_{\rm{R}4}.
  \end{align}
  \end{subequations}
  We have the following results: 
  \begin{subequations}
  \begin{flalign}
       \quad \quad \quad p_{\rm{R}4,\textrm{max}}&- p_{\rm{R}4,\textrm{min}}=  \frac{\bar{\sigma}_4^2(P_{\rm R} + \bar{\sigma}_1^2)}{(P_{\rm R} + \bar{\sigma}_4^2)\bar{\sigma}_1^2}(\bar{\sigma}_1^2 + \bar{\sigma}_4^2) \cdot \left(2-\frac{P_{\rm R} + \bar{\sigma}_2^2}{2(P_{\rm R} + \bar{\sigma}_4^2)}\right)+ \bar{\sigma}_2^2 -  \bar{\sigma}_4^2&\\
            \label{D2.3_CASE1_max-min_1} 
     &  \ge  \frac{\bar{\sigma}_4^2}{\bar{\sigma}_1^2}(\bar{\sigma}_1^2 + \bar{\sigma}_4^2)\!\cdot\!\frac{3}{2}  \!+ \!\bar{\sigma}_2^2 \!- \! \bar{\sigma}_4^2&\\
       \label{D2.3_CASE1_max-min_2}
      & \ge  0 &
  \end{flalign}
  \end{subequations}
  \vspace{-6mm} 
  \begin{subequations}
  \begin{flalign}
     \quad \quad \quad p_{\rm{R}4,\textrm{max}}&=  \frac{\bar{\sigma}_4^2(P_{\rm R} + \bar{\sigma}_1^2)}{(P_{\rm R} + \bar{\sigma}_4^2)\bar{\sigma}_1^2}(\bar{\sigma}_1^2 + \bar{\sigma}_4^2)\cdot 2 - \bar{\sigma}_4^2&\\
       \label{D2.3_CASE1_max_1} 
     & \ge  2\frac{\bar{\sigma}_4^2}{\bar{\sigma}_1^2}(\bar{\sigma}_1^2 + \bar{\sigma}_4^2)   \!- \! \bar{\sigma}_4^2&\\
      \label{D2.3_CASE1_max_2}   
     &  \ge  0&
  \end{flalign}
  \end{subequations}
  \vspace{-6mm} 
  \begin{subequations}
  \begin{flalign}
    \quad \quad \quad p_{\rm{R}4,\textrm{min}}& =  \frac{\bar{\sigma}_4^2(P_{\rm R} \! +\!  \bar{\sigma}_1^2)}{(P_{\rm R}\!  +\!  \bar{\sigma}_4^2)\bar{\sigma}_1^2}(\bar{\sigma}_1^2 \! +\!  \bar{\sigma}_4^2)\!\cdot\! \frac{P_{\rm R} \! +\!  \bar{\sigma}_2^2}{2(P_{\rm R} \! +\!  \bar{\sigma}_4^2)} 
           \! -\! \bar{\sigma}_2^2& \\
           \label{D2.3_CASE1_min_1}
       &\le \frac{1}{2}(\bar{\sigma}_1^2 + \bar{\sigma}_4^2) -  \bar{\sigma}_2^2&\\
      \label{D2.3_CASE1_min_2}
      &\le \bar{\sigma}_1^2 &\\
      & = p_{\rm{R}3} + p_{\rm{R}4}. &
  \end{flalign}
  \end{subequations}
  In the above, step (\ref{D2.3_CASE1_max-min_1}) follows from  $\bar{\sigma}_4^2 \ge \bar{\sigma}_2^2 $ and $\bar{\sigma}_1^2 \ge \bar{\sigma}_4^2 $.
 Step (\ref{D2.3_CASE1_max_1}) follows from  $\bar{\sigma}_1^2 \ge \bar{\sigma}_4^2$.
 Step (\ref{D2.3_CASE1_min_1}) from  $\bar{\sigma}_4^2 \le \bar{\sigma}_1^2$ and $\bar{\sigma}_2^2 \le \bar{\sigma}_4^2$, and step (\ref{D2.3_CASE1_min_2}) from  $\bar{\sigma}_4^2 \le \bar{\sigma}_1^2$. This proves the existence of $p_{\rm{R}3}$ and $p_{\rm{R}4}$ satisfying (\ref{D2.3_CASE1_1_0}).

	$\rm {(ii)}\;\bar \sigma_4^2 \le P_{\rm R} < \bar \sigma_1^2$:  We set $p_{{\rm R}1} = p_{{\rm R}2} = 0$, $p_{{\rm R}3} =P_{\rm R} - \bar \sigma_4^2$ and  $p_{{\rm R}4} =\bar \sigma_4^2$ in (\ref{Pro4.2}). Then
\begin{subequations}
\begin{flalign}
\label{D2.3_CASE2_1}
  \quad\quad\quad \quad\quad \quad\quad \quad \quad  R_1 \!+\! {1\over 2}  & \!=\! {1\over 2}\log \left(2{\bar \sigma_4^2+ \bar \sigma_2^2 \over \bar \sigma_2^2} \right) &\\
  \label{D2.3_CASE2_1_1}
  &    \!\geq    \!	{1\over 2}\log \!\left( \!  { {P_{\rm R} \!+\! \bar \sigma_2^2}\over \bar \sigma_2^2 }\cdot   {{  \bar \sigma_4^2}\over P_{\rm R}\!+\!  \bar \sigma_4^2 }\cdot  {{P_{\rm R} \!+\! \bar \sigma_1^2}\over \bar \sigma_1^2 }   \!\right) \!  & \\ 
  \label{D2.3_CASE2_1_2}
  & =  R_1'&  
\end{flalign}
\end{subequations}
\vspace{-6mm}
\begin{flalign}
  \quad \quad\quad \quad\quad \quad\quad \quad \quad R_2\!+ \!{1\over 2}  &\! =\! {1\over 2} >   R_2'  & 
\end{flalign}
\vspace{-6mm}
\begin{subequations}
\begin{flalign} 
\label{D2.3_CASE2_3}  
  \quad\quad\quad \quad\quad \quad\quad \quad \quad R_3\!+ \!{1\over 2}  &  \!=\! 	{1\over 2}\log \left(   {P_{\rm R}  + \bar \sigma_4^2 \over    \bar \sigma_4^2}    \right)  &\\
  \label{D2.3_CASE2_3_1}
  &    \geq     {1\over 2}\log \left(    {{P_{\rm R} + \bar \sigma_4^2}\over \bar \sigma_4^2 }  {{  \bar \sigma_1^2}\over P_{\rm R}+ \bar \sigma_1^2 }  \right)   &\\ 
  \label{D2.3_CASE2_3_2}
  & = R_3'&
\end{flalign}
\end{subequations}
\vspace{-6mm}
\begin{flalign} 
  \quad\quad \quad\quad\quad \quad\quad \quad \quad R_4\!+ \!{1\over 2}  &\! =\! {1\over 2} >  R_4'. &    
 \end{flalign}
In the above, step (\ref{D2.3_CASE2_1}) follows from (\ref{Pro4.2_1}); step (\ref{D2.3_CASE2_1_1}) from ${ {P_{\rm R} + \bar \sigma_2^2}\over {P_{\rm R} + \bar \sigma_4^2} } \le 1$ and ${ {P_{\rm R} + \bar \sigma_1^2}\over \bar \sigma_1^2 } \le 2$;  and step (\ref{D2.3_CASE2_1_2}) from (\ref{D2.3_R1'}). 
Step (\ref{D2.3_CASE2_3}) follows from (\ref{Pro4.2_3}); and step (\ref{D2.3_CASE2_3_2}) from  (\ref{D2.3_R3'}).

	$\rm {(iii)}\; P_{\rm R} < \bar \sigma_4^2$:  We set $p_{{\rm R}1} = p_{{\rm R}2} = p_{{\rm R}3} = 0$ and  $p_{{\rm R}4} =P_{\rm R}$ in (\ref{Pro4.2}). Then
\begin{subequations}
 	 	\begin{align}
 	 	\label{D2.3_CASE3_1}
 	 	R_1 + {1\over 2}  & >R_1'  \\
 	 	R_2+ {1\over 2}  & = {1\over 2} >   R_2'   \\
 	 	R_3+ {1\over 2}  &  = {1\over 2}>  R_3'\\
 	 	R_4+ {1\over 2}  &  ={1\over 2}> R_4'.     
 	 	\end{align}
\end{subequations}
Therefore, (D2.3) is achievable to within ${1\over 2}$ bit.

We next consider ({D2}.4). In this case, we use the achievable rates in Proposition 4.2.  










The proof is divided into five subcases:
	\begin{subequations} \label{case_i}
	 \begin{align} \label{casei_1}
 \rm {(i)} \; 	 &	P_{\rm R}< \bar{\sigma}_4^2 \\
 \rm {(ii)} \; & \bar{\sigma}_4^2\le P_{\rm R}< \bar{\sigma}_3^2 \; \textrm{and} \; \bar{\sigma}_4^2 \ge 2\bar{\sigma}_2^2  \\ \label{casei_2}
 \rm {(iii)} \; &   \bar{\sigma}_4^2\le P_{\rm R}< \bar{\sigma}_3^2\; \textrm{and} \; \bar{\sigma}_4^2 < 2\bar{\sigma}_2^2 \\ \label{casei_3}
 \rm {(iv)} \;  	&	P_{\rm R}\ge \bar{\sigma}_3^2 \;\textrm{and}\;  \bar{\sigma}_3^2 \ge  2\bar{\sigma}_1^2\\
 \label{casei_4}
  \rm {(v)} \; & P_{\rm R}\ge \bar{\sigma}_3^2\; \textrm{and} \; \bar{\sigma}_3^2 <  2\bar{\sigma}_1^2.
		\end{align}
	\end{subequations}
We now consider  the five subcases in (\ref{case_i}) one by one.

	$\rm {(i)}\; P_{\rm R}< \bar{\sigma}_4^2$:   
	 We set $p_{{\rm R1}} =  p_{\rm{R}2} = p_{\rm{R}3} = 0$ and $p_{R_4
	} = P_{\rm R}$ in (\ref{Pro4.2}). Then
	\begin{subequations}
	\begin{flalign}\label{D2.4_CASE0_1}
	       \quad\quad \quad\quad \quad\quad \quad\quad\quad  R_1 \!+\! \frac{1}{2} =& \frac{1}{2}\log \left(2\frac{P_{\rm R} + \bar{\sigma}_2^2}{\bar{\sigma}_2^2}
	         \right)&\\ 
	         \label{D2.4_CASE0_1_1}
	        \ge&  \frac{1}{2}\log \left( \frac{P_{\rm R} + \bar{\sigma}_2^2}{ \bar{\sigma}_2^2} \cdot\frac{\bar{\sigma}_3^2}{P_{\rm R} + \bar{\sigma}_3^2}\right)&\\
	        \label{D2.4_CASE0_1_2}
	        =& R_1^{'} 
	\end{flalign}
	\end{subequations}
	\vspace{-6mm}
	\begin{subequations}
	\begin{flalign}\label{D2.4_CASE0_2}
	        \quad \quad \quad\quad \quad\quad \quad\quad\quad R_2 \!+\! \frac{1}{2} =& \frac{1}{2}&\\ 
	            \label{D2.4_CASE0_2_1}           
	            \ge & \frac{1}{2}\log \left( \frac{P_{\rm R} + \bar{\sigma}_1^2}{ \bar{\sigma}_1^2}\cdot\frac{\bar{\sigma}_3^2}{P_{\rm R} + \bar{\sigma}_3^2}\right)&\\
	            \label{D2.4_CASE0_2_2}
	           = & R_2^{'}
	\end{flalign}
	\end{subequations}
	\vspace{-6mm}
	\begin{flalign}\label{D2.4_CASE0_3}
	\quad \quad\quad \quad\quad \quad\quad\quad R_3 \!+\! \frac{1}{2} =& \frac{1}{2}
	 > \frac{1}{2}\log \left(1 + \frac{P_{\rm R}}{\bar{\sigma}_3^2}\right) 
	 = R_3^{'}&
	\end{flalign}
	\vspace{-6mm}
	\begin{subequations}
	\begin{flalign}\label{D2.4_CASE0_4}
	   \quad\quad\quad \quad\quad \quad\quad \quad\quad\quad R_4 \!+\! \frac{1}{2} 
	    = &\frac{1}{2} & \\
	    \label{D2.4_CASE0_4_1}
	   > &\frac{1}{2}\log \left(1 + \frac{P_{\rm R}}{\bar{\sigma}_3^2}\right)&\\ 
	   \label{D2.4_CASE0_4_2}
	   =& R_4^{'}.&
	\end{flalign}
	\end{subequations}
	In the above, step (\ref{D2.4_CASE0_1}) follows from (\ref{Pro4.2_1});  and step (\ref{D2.4_CASE0_1_2}) from (\ref{D2.4_R1'}). Step (\ref{D2.4_CASE0_2}) follows from (\ref{Pro4.2_2}); step (\ref{D2.4_CASE0_2_1}) from
	\ $P_{\rm R} < 
	 \bar{\sigma}_4^2$ and $\bar{\sigma}_4^2 \le 
	 	 \bar{\sigma}_1^2$; and step (\ref{D2.4_CASE0_2_2}) from (\ref{D2.4_R2'}).  Step (\ref{D2.4_CASE0_4}) follows from (\ref{Pro4.2_4}); step (\ref{D2.4_CASE0_4_1}) from $P_{\rm R} < \bar \sigma_3^2$; and step (\ref{D2.4_CASE0_4_2}) from (\ref{D2.4_R4'}).

$\rm {(ii)}\; \bar{\sigma}_4^2\le P_{\rm R}< \bar{\sigma}_3^2$ and  $\bar{\sigma}_4^2 \ge 2\bar{\sigma}_2^2$: 
 We set $p_{{\rm R1}} = p_{\rm{R}3} = 0$ and $p_{R_4
} = \bar{\sigma}_4^2 - 2\bar{\sigma}_2^2$ in (\ref{Pro4.2}). Then
\begin{subequations}
\begin{flalign}\label{D2.4_CASE1_1}
       \quad\quad\quad \quad\quad \quad\quad\quad  R_1 \!+\! \frac{1}{2} =& \frac{1}{2}\log \left(2\frac{p_{\rm{R}4} + \bar{\sigma}_2^2}{\bar{\sigma}_2^2}\cdot
        \frac{P_{\rm R} + \bar{\sigma}_4^2}{p_{\rm{R}4} + \bar{\sigma}_4^2} \right)&\\ 
        \label{D2.4_CASE1_1_1}           
         = & \frac{1}{2}\log \left( \frac{P_{\rm R} + \bar{\sigma}_4^2}{ \bar{\sigma}_2^2}\right)&\\
         \label{D2.4_CASE1_1_2}
        \ge&  \frac{1}{2}\log \left( \frac{P_{\rm R} + \bar{\sigma}_2^2}{ \bar{\sigma}_2^2}\right)&\\
        \label{D2.4_CASE1_1_3}
        \ge& R_1^{'} 
\end{flalign}
\end{subequations}
\vspace{-6mm}
\begin{subequations}
\begin{flalign}\label{D2.4_CASE1_2}
         \quad\quad\quad \quad\quad \quad\quad\quad R_2 \!+\! \frac{1}{2} =& \frac{1}{2}\log \left(2\frac{P_{\rm R} + \bar{\sigma}_1^2}{p_{\rm{R}4} + \bar{\sigma}_1^2}  \right)&\\ 
            \label{D2.4_CASE1_2_1}           
            \ge & \frac{1}{2}\log \left( \frac{P_{\rm R} + \bar{\sigma}_1^2}{ \bar{\sigma}_1^2}\right)&\\
            \label{D2.4_CASE1_2_2}
           \ge & R_2^{'}
\end{flalign}
\end{subequations}
\vspace{-6mm}
\begin{flalign}\label{D2.4_CASE1_3}
\quad\quad\quad \quad\quad \quad\quad\quad R_3 \!+\! \frac{1}{2} =& \frac{1}{2}
 > \frac{1}{2}\log \left(1 + \frac{P_{\rm R}}{\bar{\sigma}_3^2}\right) 
 = R_3^{'}&
\end{flalign}
\vspace{-6mm}
\begin{subequations}
\begin{flalign}\label{D2.4_CASE1_4}
   \quad\quad\quad \quad\quad \quad\quad\quad R_4 \!+\! \frac{1}{2} \ge& \frac{1}{2}\log \!\left(\!1 \!+\!\frac{p_{{\rm R1}}}{p_{\rm{R}2} + p_{\rm{R}3} + p_{\rm{R}4} + \bar{\sigma}_3^2}\!\right)\! \!+\! \frac{1}{2} &\\
   \label{D2.4_CASE1_4_1}
    = &\frac{1}{2} & \\
    \label{D2.4_CASE1_4_2}
   > &\frac{1}{2}\log \left(1 + \frac{P_{\rm R}}{\bar{\sigma}_3^2}\right)&\\ 
   \label{D2.4_CASE1_4_3}
   =& R_4^{'}.&
\end{flalign}
\end{subequations}
In the above, step (\ref{D2.4_CASE1_1}) follows from (\ref{Pro4.2_1}); step (\ref{D2.4_CASE1_1_2}) from $\bar \sigma_4^2 \geq \bar \sigma_2^2$; and step (\ref{D2.4_CASE1_1_3}) from (\ref{D2.4_R1'}). Step (\ref{D2.4_CASE1_2}) follows from (\ref{Pro4.2_2}); step (\ref{D2.4_CASE1_2_1}) from $ p_{\rm{R}4} =  \bar{\sigma}_4^2 - 2\bar{\sigma}_2^2 \le \bar{\sigma}_1^2$; and step (\ref{D2.4_CASE1_2_2}) from (\ref{D2.4_R2'}).  Step (\ref{D2.4_CASE1_4}) follows from (\ref{Pro4.2_4}) and $\bar \sigma_3^2 \geq \bar \sigma_1^2$; step (\ref{D2.4_CASE1_4_2}) from $P_{\rm R} < \bar \sigma_3^2$; and step (\ref{D2.4_CASE1_4_3}) from (\ref{D2.4_R4'}).

$\rm {(iii)}\; \bar{\sigma}_4^2\le P_{\rm R}< \bar{\sigma}_3^2$ and  $\bar{\sigma}_4^2 < 2\bar{\sigma}_2^2$: 
 We set $p_{{\rm R1}} = p_{\rm{R}3} = p_{R_4
 }= 0$ and $p_{R_2
} = P_{\rm R}$ in (\ref{Pro4.2}). Then
\begin{subequations}
\begin{flalign}\label{D2.4_CASE2_1}
       \quad\quad\quad \quad\quad \quad\quad\quad  R_1 + \frac{1}{2}           
               & = \frac{1}{2}\log \left( 2\frac{P_{\rm R} + \bar{\sigma}_4^2}{ \bar{\sigma}_4^2}\right)&\\
               \label{D2.4_CASE2_1_1}
               & \ge \frac{1}{2}\log \left( \frac{P_{\rm R} + \bar{\sigma}_4^2}{ \bar{\sigma}_2^2}\right)&\\
               \label{D2.4_CASE2_1_2}
               & \ge \frac{1}{2}\log \left( \frac{P_{\rm R} + \bar{\sigma}_2^2}{ \bar{\sigma}_2^2}\right)\\
               \label{D2.4_CASE2_1_3}
               & \ge R_1^{'}&
\end{flalign}
\end{subequations}
\vspace{-6mm}
\begin{subequations}
\begin{flalign}\label{D2.4_CASE2_2}
         \quad\quad\quad \quad\quad \quad\quad\quad R_2 + \frac{1}{2} &= \frac{1}{2}\log \left(\frac{P_{\rm R} + \bar{\sigma}_1^2}{\bar{\sigma}_1^2}  \right) + \frac{1}{2}&\\
         \label{D2.4_CASE2_2_1}            
                     & \ge R_2^{'}&
\end{flalign}
\end{subequations}
\vspace{-6mm}
\begin{flalign}\label{D2.4_CASE2_3}
\quad\quad\quad\quad \quad\quad \quad\quad R_3 + \frac{1}{2} =& \frac{1}{2}
 > \frac{1}{2}\log \left(1 + \frac{P_{\rm R}}{\bar{\sigma}_3^2}\right) 
 = R_3^{'}&
\end{flalign}
\vspace{-6mm}
\begin{subequations}
\begin{flalign}\label{D2.4_CASE2_4}
  \quad\quad\quad \quad\quad \quad\quad \quad R_4 + \frac{1}{2} \ge& \frac{1}{2}\log \!\left(\!1 \!+\!\frac{p_{{\rm R1}}}{p_{\rm{R}2} \!+\! p_{\rm{R}3} \!+\! p_{\rm{R}4} \!+\! \bar{\sigma}_3^2}\!\right)\! \!+\! \frac{1}{2} &\\
   \label{D2.4_CASE2_4_1}
    = &\frac{1}{2} & \\
    \label{D2.4_CASE2_4_2}
   > &\frac{1}{2}\log \left(1 + \frac{P_{\rm R}}{\bar{\sigma}_3^2}\right)&\\ 
   \label{D2.4_CASE2_4_3}
   =& R_4^{'}.&
\end{flalign}
\end{subequations}
In the above, step (\ref{D2.4_CASE2_1}) follows from (\ref{Pro4.2_1}); step (\ref{D2.4_CASE2_1_1}) from $\bar \sigma_4^2 \leq 2\bar \sigma_2^2$; step (\ref{D2.4_CASE2_1_2}) from $\bar \sigma_4^2 \geq \bar \sigma_2^2$; and step (\ref{D2.4_CASE2_1_3}) from (\ref{D2.4_R1'}). Step (\ref{D2.4_CASE2_2}) follows from (\ref{Pro4.2_2});  and step (\ref{D2.4_CASE2_2_1}) from (\ref{D2.4_R2'}).  Step (\ref{D2.4_CASE2_4}) follows from (\ref{Pro4.2_4}) and $\bar \sigma_3^2 \geq \bar \sigma_1^2$; step (\ref{D2.4_CASE2_4_2}) from $P_{\rm R} < \bar \sigma_3^2$; and step (\ref{D2.4_CASE2_4_3}) from (\ref{D2.4_R4'}).

$\rm {(iv)}\; P_{\rm R}\ge  \bar{\sigma}_3^2$ and  $\bar{\sigma}_3^2 \ge 2\bar{\sigma}_1^2$: 
 We set $p_{{\rm R1}} = P_{\rm R} - \bar{\sigma}_3^2$, $p_{\rm{R}4} = \frac{\bar{\sigma}_3^2 + 2\bar{\sigma}_1^2 - \bar{\sigma}_4^2}{\bar{\sigma}_3^2 + \bar{\sigma}_1^2}\bar{\sigma}_1^2$, $p_{\rm{R}3} + p_{\rm{R}4} = \frac{\bar{\sigma}_3^2 + 2\bar{\sigma}_1^2}{\bar{\sigma}_3^2}\bar{\sigma}_1^2$ and $ p_{R_2
 } + p_{\rm{R}3} + p_{\rm{R}4}= \bar{\sigma}_3^2$ in (\ref{Pro4.2}).     Clearly, $p_{\rm{R}3} + p_{\rm{R}4} = \frac{\bar{\sigma}_3^2 + 2\bar{\sigma}_1^2}{\bar{\sigma}_3^2}\bar{\sigma}_1^2 \le 2 \bar{\sigma}_1^2\le \bar{\sigma}_3^2$. Thus, $p_{\rm{R}2} = \bar{\sigma}_3^2 - p_{\rm{R}3} - p_{\rm{R}4} \ge 0$.  Then
\begin{subequations}
\begin{flalign}\label{D2.4_CASE3_1}
        \quad \quad \quad \quad R_1 \!\!+\!\! \frac{1}{2}           
       \!=&\frac{1}{2}\!\log \left( 2\frac{p_{\rm{R}4} + \bar{\sigma}_2^2}{ \bar{\sigma}_2^2}\cdot \frac{\bar{\sigma}_3^2 + \bar{\sigma}_4^2}{p_{\rm{R}3} + p_{\rm{R}4} +  \bar{\sigma}_4^2}\right)&\\
       \label{D2.4_CASE3_1_1}
       \!=&\frac{1}{2}\!\log \!\left(\! \!2\!\frac{\frac{\bar{\sigma}_3^2 + 2\bar{\sigma}_1^2 - \bar{\sigma}_4^2}{\bar{\sigma}_3^2 \!+\! \bar{\sigma}_1^2}\bar{\sigma}_1^2 \!+ \! \bar{\sigma}_2^2}{\bar{\sigma}_2^2}\!\!\cdot\!\! \frac{\bar{\sigma}_3^2 + \bar{\sigma}_4^2}{\frac{\bar{\sigma}_3^2 + 2\bar{\sigma}_1^2}{\bar{\sigma}_3^2}\bar{\sigma}_1^2 \!+\!  \bar{\sigma}_4^2} \!\right)\!\!\!&\\
       \label{D2.4_CASE3_1_2}
       \!\ge&\frac{1}{2}\!\log \left(2\frac{\bar{\sigma}_3^2 \!+\! \bar{\sigma}_4^2}{\bar{\sigma}_3^2\! +\! \bar{\sigma}_1^2}\!\cdot\! \frac{\bar{\sigma}_3^2\bar{\sigma}_1^2 \!+\! 2\bar{\sigma}_1^4 \!-\! \bar{\sigma}_4^2\bar{\sigma}_1^2}{\bar{\sigma}_3^2\bar{\sigma}_1^2 \!+\! 2\bar{\sigma}_1^4 \!+\! \bar{\sigma}_3^2\bar{\sigma}_4^2}\!\cdot\! \frac{\bar{\sigma}_3^2}{\bar{\sigma}_2^2} \right)\!\!  &\\
       \label{D2.4_CASE3_1_3}
       \!=&\frac{1}{2}\log \left( \frac{\bar{\sigma}_3^2\bar{\sigma}_1^2 + 2\bar{\sigma}_1^4 + \bar{\sigma}_3^2\bar{\sigma}_1^2 + 2\bar{\sigma}_1^2(\bar{\sigma}_1^2 - \bar{\sigma}_4^2) }{\bar{\sigma}_3^2\bar{\sigma}_1^2 + 2\bar{\sigma}_1^4 + \bar{\sigma}_3^2\bar{\sigma}_4^2} \cdot\frac{\bar{\sigma}_3^2 + \bar{\sigma}_4^2}{\bar{\sigma}_3^2 + \bar{\sigma}_1^2} \cdot \frac{\bar{\sigma}_3^2}{\bar{\sigma}_2^2}\right) &\\
       \label{D2.4_CASE3_1_4} 
       \!\ge& \frac{1}{2}\log \left( \frac{\bar{\sigma}_3^2}{\bar{\sigma}_2^2}\right)&\\
       \label{D2.4_CASE3_1_5} 
       \!\ge&\frac{1}{2}\log \left( \frac{P_{\rm R} + \bar{\sigma}_2^2}{\bar{\sigma}_2^2}\cdot\frac{\bar{\sigma}_3^2}{P_{\rm R} + \bar{\sigma}_3^2}\right)&\\
       \label{D2.4_CASE3_1_6}
        \!=& R_1^{'}&
\end{flalign}
\end{subequations}
\vspace{-6mm}
\begin{subequations}
\begin{flalign}\label{D2.4_CASE3_2}
       \quad \quad \quad \quad R_2 \!+\! \frac{1}{2}  \!=&\frac{1}{2}\log \left(\frac{\bar{\sigma}_3^2 + \bar{\sigma}_1^2}{p_{\rm{R}3} + p_{\rm{R}4} +  \bar{\sigma}_1^2}\right) + \frac{1}{2}&\\
         \label{D2.4_CASE3_2_1}
         \!=&\frac{1}{2}\log \left(2\frac{\bar{\sigma}_3^2 + \bar{\sigma}_1^2}{\frac{\bar{\sigma}_3^2 + 2\bar{\sigma}_1^2}{\bar{\sigma}_3^2}\bar{\sigma}_1^2 +  \bar{\sigma}_1^2}\right) &\\
         \label{D2.4_CASE3_2_2}
        \!= &\frac{1}{2}\log \left( \frac{\bar{\sigma}_3^2}{\bar{\sigma}_1^2}\right)&\\
        \label{D2.4_CASE3_2_3} 
         \!\ge& \frac{1}{2}\log \left( \frac{P_{\rm R} + \bar{\sigma}_1^2}{\bar{\sigma}_1^2}\cdot\frac{\bar{\sigma}_3^2}{P_{\rm R} + \bar{\sigma}_3^2}\right)&\\
         \label{D2.4_CASE3_2_4}
         \!=& R_2^{'}&
\end{flalign}
\end{subequations}
\vspace{-6mm}
\begin{subequations}
\begin{flalign}\label{D2.4_CASE3_3}
       \quad \quad \quad \quad R_3 \!+\! \frac{1}{2}           
       \!=&\frac{1}{2}\log \left( 2\frac{P_{\rm R} + \bar{\sigma}_1^2}{ \bar{\sigma}_3^2 + \bar{\sigma}_1^2}\cdot\frac{p_{\rm{R}3} + p_{\rm{R}4} +  \bar{\sigma}_4^2}{ p_{\rm{R}4} +  \bar{\sigma}_4^2}\right) &\\
       \label{D2.4_CASE3_3_1} 
       \!=&\frac{1}{2}\log \left( 2\frac{P_{\rm R} + \bar{\sigma}_1^2}{ \bar{\sigma}_3^2 + \bar{\sigma}_1^2}\cdot
       \frac{\frac{\bar{\sigma}_3^2 + 2\bar{\sigma}_1^2}{\bar{\sigma}_3^2}\bar{\sigma}_1^2 +  \bar{\sigma}_4^2}{ \frac{\bar{\sigma}_3^2 + 2\bar{\sigma}_1^2 - \bar{\sigma}_4^2}{\bar{\sigma}_3^2 + \bar{\sigma}_1^2}\bar{\sigma}_1^2 +  \bar{\sigma}_4^2} \right)\!\! &\\
       \label{D2.4_CASE3_3_2} 
       \!= &\frac{1}{2}\log \left( 2\frac{P_{\rm R}  + \bar{\sigma}_1^2}{\bar{\sigma}_3^2}\right)&\\
       \label{D2.4_CASE3_3_3} 
       \!\ge&  \frac{1}{2}\log \left( \frac{P_{\rm R}  + \bar{\sigma}_3^2}{\bar{\sigma}_3^2}\right)&\\
       \label{D2.4_CASE3_3_4} 
       \!=& R_3^{'}&
\end{flalign}
\end{subequations}
\vspace{-6mm}
\begin{subequations}
\begin{flalign}\label{D2.4_CASE3_4}
   \quad \quad \quad \quad R_4 \!+\! \frac{1}{2} \!\ge & \frac{1}{2}\log \left(1 + \frac{p_{{\rm R1}}}{p_{\rm{R}2} + p_{\rm{R}3} + p_{\rm{R}4} + \bar{\sigma}_3^2}\right) \!+\! \frac{1}{2}& \\
   \label{D2.4_CASE3_4_1}
   Ê \!=& \frac{1}{2}\log \left(1 + \frac{P_{\rm R} - \bar{\sigma}_3^2} {2\bar{\sigma}_3^2}\right) +\frac{1}{2}Ê  &\\
   Ê \label{D2.4_CASE3_4_2}
     \!=& \frac{1}{2}\log \left(1 + \frac{P_{\rm R}}{\bar{\sigma}_3^2}\right) &\\
     \label{D2.4_CASE3_4_3}
     \!=& R_4^{'}.&
\end{flalign}
\end{subequations}
In the above, step (\ref{D2.4_CASE3_1}) follows from (\ref{Pro4.2_1}); step (\ref{D2.4_CASE3_1_4}) from
\begin{align}
\frac{\bar{\sigma}_3^2\bar{\sigma}_1^2 + 2\bar{\sigma}_1^4 + \bar{\sigma}_3^2\bar{\sigma}_1^2 + 2\bar{\sigma}_1^2(\bar{\sigma}_1^2 - \bar{\sigma}_4^2) }{\bar{\sigma}_3^2\bar{\sigma}_1^2 + 2\bar{\sigma}_1^4 + \bar{\sigma}_3^2\bar{\sigma}_4^2} \!\cdot\! \frac{\bar{\sigma}_3^2 + \bar{\sigma}_4^2}{\bar{\sigma}_3^2 + \bar{\sigma}_1^2}   \ge 1
\end{align}
by noting that 
 $\bar \sigma_1^2 \geq \bar \sigma_4^2$, $\bar \sigma_3^2 \geq \bar \sigma_1^2$, and $\bar \sigma_3^2 \geq 2\bar \sigma_1^2$; step (\ref{D2.4_CASE3_1_5}) from $\bar \sigma_3^2 \geq \bar \sigma_2^2$; and step (\ref{D2.4_CASE3_1_6}) from (\ref{D2.4_R1'}). Step (\ref{D2.4_CASE3_2}) follows from (\ref{Pro4.2_2}); step (\ref{D2.4_CASE3_2_3}) from $\bar \sigma_3^2 \geq \bar \sigma_1^2$; and step (\ref{D2.4_CASE3_2_4}) from (\ref{D2.4_R2'}). Step (\ref{D2.4_CASE3_3}) follows from (\ref{Pro4.2_3}); step (\ref{D2.4_CASE3_3_3}) from $P_{\rm R} \ge  \bar \sigma_3^2$; and step (\ref{D2.4_CASE3_3_4}) from (\ref{D2.4_R3'}). Step (\ref{D2.4_CASE3_4}) follows from (\ref{Pro4.2_4}) and $\bar \sigma_3^2 \geq \bar \sigma_1^2$;  and step (\ref{D2.4_CASE3_4_3}) from (\ref{D2.4_R4'}).

$\rm {(v)}\; P_{\rm R}\ge  \bar{\sigma}_3^2$ and  $\bar{\sigma}_3^2 < 2\bar{\sigma}_1^2$: 
 We set $p_{{\rm R1}} = P_{\rm R} - \bar{\sigma}_3^2$, $p_{\rm{R}2} = 0$, and $p_{\rm{R}3} = p_{\rm{R}4} = \frac{\bar{\sigma}_3^2}{2}$ in (\ref{Pro4.2}).   Then
\begin{subequations}
\begin{flalign}\label{D2.4_CASE4_1}
       \quad \quad \quad \quad \quad \quad\quad\quad R_1 \!+\! \frac{1}{2}           
       = & \frac{1}{2}\log \left( 2\frac{p_{\rm{R}4}+ \bar{\sigma}_2^2}{ \bar{\sigma}_2^2}\cdot \frac{\bar{\sigma}_3^2 + \bar{\sigma}_4^2}{p_{\rm{R}3} + p_{\rm{R}4} +  \bar{\sigma}_4^2}\right) &\\
        \label{D2.4_CASE4_1_1}
       =& \frac{1}{2}\log \left(\frac{ \bar{\sigma}_3^2 + 2\bar{\sigma}_2^2}{ \bar{\sigma}_2^2}\right)&\\
       \label{D2.4_CASE4_1_2}
       \ge & \frac{1}{2}\log \left( \frac{\bar{\sigma}_3^2}{\bar{\sigma}_2^2}\right)&\\
       \label{D2.4_CASE4_1_3}
       \ge&\frac{1}{2}\log \left( \frac{P_{\rm R} + \bar{\sigma}_2^2}{\bar{\sigma}_2^2}\cdot\frac{\bar{\sigma}_3^2}{P_{\rm R} + \bar{\sigma}_3^2}\right)&\\
       \label{D2.4_CASE4_1_4}
       =& R_1^{'}&
\end{flalign}
\end{subequations}
\vspace{-6mm}
\begin{subequations}
\begin{flalign}\label{D2.4_CASE4_2}
     \quad \quad \quad \quad \quad \quad\quad\quad R_2 + \frac{1}{2} =& \frac{1}{2}\log \left(\frac{\bar{\sigma}_3^2 + \bar{\sigma}_1^2}{p_{\rm{R}3} + p_{\rm{R}4} +  \bar{\sigma}_1^2}\right) + \frac{1}{2}&\\
     \label{D2.4_CASE4_2_1}
     =& \frac{1}{2}&\\
     \label{D2.4_CASE4_2_2}
     \ge &\frac{1}{2}\log \left( \frac{P_{\rm R} + \bar{\sigma}_1^2}{\bar{\sigma}_1^2}\cdot\frac{\bar{\sigma}_3^2}{P_{\rm R} + \bar{\sigma}_3^2}\right)&\\
     \label{D2.4_CASE4_2_3}
     = & R_2^{'}&
\end{flalign}
\end{subequations}
\vspace{-6mm}
\begin{subequations}
\begin{flalign}\label{D2.4_CASE4_3}
       \quad \quad \quad \quad \quad \quad\quad\quad R_3 \!+\! \frac{1}{2}           
       = &\frac{1}{2}\log \left( 2\frac{P_{\rm R} + \bar{\sigma}_1^2}{ \bar{\sigma}_3^2 + \bar{\sigma}_1^2}\cdot \frac{p_{\rm{R}3} + p_{\rm{R}4} +  \bar{\sigma}_4^2}{ p_{\rm{R}4} +  \bar{\sigma}_4^2}\right) &\\
       \label{D2.4_CASE4_3_1}
       =&\frac{1}{2}\log \left( 2\frac{P_{\rm R} + \bar{\sigma}_1^2}{ \bar{\sigma}_3^2 + \bar{\sigma}_1^2}\cdot
       \frac{\bar{\sigma}_3^2 + \bar{\sigma}_4^2}{\frac{\bar{\sigma}_3^2}{2} + \bar{\sigma}_4^2} \right) &\\
       \label{D2.4_CASE4_3_2}
       \ge&  \frac{1}{2}\log \left( \frac{P_{\rm R}  + \bar{\sigma}_3^2}{\bar{\sigma}_3^2}\right)&\\
       \label{D2.4_CASE4_3_3}
       =& R_3^{'}&
\end{flalign}
\end{subequations}
\vspace{-6mm}
\begin{subequations}
\begin{flalign}\label{D2.4_CASE4_4}
   \quad \quad \quad \quad \quad \quad\quad\quad R_4 \!+\! \frac{1}{2} \ge & \frac{1}{2}\log \left(1 \!+\! \frac{p_{{\rm R1}}}{p_{\rm{R}2} \!+\! p_{\rm{R}3} \!+\! p_{\rm{R}4} \!+\! \bar{\sigma}_3^2}\right)\! + \!\frac{1}{2}& \\
   \label{D2.4_CASE4_4_1}
   Ê =& \frac{1}{2}\log \left(1 + \frac{P_{\rm R} - \bar{\sigma}_3^2} {2\bar{\sigma}_3^2}\right) +\frac{1}{2}Ê  &\\
   Ê \label{D2.4_CASE4_4_2}
     =& \frac{1}{2}\log \left(1 + \frac{P_{\rm R}}{\bar{\sigma}_3^2}\right) &\\
     \label{D2.4_CASE4_4_3}
     =& R_4^{'}.&
\end{flalign}
\end{subequations}
In the above, step (\ref{D2.4_CASE4_1}) follows from (\ref{Pro4.2_1}); step (\ref{D2.4_CASE4_1_3}) from
 $\bar \sigma_3^2 \geq \bar \sigma_2^2$;  and step (\ref{D2.4_CASE4_1_4}) from (\ref{D2.4_R1'}). Step (\ref{D2.4_CASE4_2}) follows from (\ref{Pro4.2_2}); step (\ref{D2.4_CASE4_2_2}) from $2\bar \sigma_1^2 \geq \bar \sigma_3^2$ and $\bar \sigma_3^2 \geq \bar \sigma_1^2$; and step (\ref{D2.4_CASE4_2_3}) from (\ref{D2.4_R2'}). Step (\ref{D2.4_CASE4_3}) follows from (\ref{Pro4.2_3}); step (\ref{D2.4_CASE4_3_2}) from 
 \begin{align}
 2\frac{\bar{\sigma}_3^2}{P_{\rm R}  + \bar{\sigma}_3^2}\cdot \frac{P_{\rm R} + \bar{\sigma}_1^2}{ \bar{\sigma}_3^2 + \bar{\sigma}_1^2}\cdot
 \frac{\bar{\sigma}_3^2 + \bar{\sigma}_4^2}{\frac{\bar{\sigma}_3^2}{2} + \bar{\sigma}_4^2} \ge \frac{\bar{\sigma}_3^2 + \bar{\sigma}_4^2}{\frac{\bar{\sigma}_3^2}{2} + \bar{\sigma}_4^2} \ge 1,
 \end{align}
  in the above, the inequality $2\frac{\bar{\sigma}_3^2}{P_{\rm R}  + \bar{\sigma}_3^2}\cdot \frac{P_{\rm R} + \bar{\sigma}_1^2}{ \bar{\sigma}_3^2 + \bar{\sigma}_1^2} \ge 1$ follows from $P_{\rm R} \ge \bar{\sigma}_3^2$, together with the fact that for $\bar{\sigma}_3^2\ge \bar{\sigma}_1^2$, $\frac{x + \bar{\sigma}_1^2}{ x + \bar{\sigma}_3^2}$ is monotonically increasing in $x$; and step (\ref{D2.4_CASE4_3_3}) from (\ref{D2.4_R3'}). Step (\ref{D2.4_CASE4_4}) follows from (\ref{Pro4.2_4}) and $\bar \sigma_3^2 \geq \bar \sigma_1^2$;  and step (\ref{D2.4_CASE4_4_3}) from (\ref{D2.4_R4'}). 
Therefore, (D2.4) is achievable to within ${1\over 2}$ bit.


We next consider ({D2}.5). The proof is divided into eight subcases:
	\begin{subequations} \label{D2.5_case_i}
	 \begin{align} \label{D2.5_casei_1}
 \rm {(i)} \; &  \bar{\sigma}_3^2 \ge 3\bar{\sigma}_1^2 \; \textrm{and} \; P_{\rm R}\ge  \bar{\sigma}_3^2   \\ \label{D2.5_casei_2}
 \rm {(ii)} \; &     \bar{\sigma}_3^2 \ge 3\bar{\sigma}_1^2 \; \textrm{and} \; P_{\rm R} \in \left(\frac{\bar{\sigma}_1^2\bar{\sigma}_3^2}{\bar{\sigma}_3^2 - 2 \bar{\sigma}_1^2}, \bar{\sigma}_3^2\right)   \\ \label{D2.5_casei_3}
 \rm {(iii)} \;  	&	\bar{\sigma}_3^2 \ge 3\bar{\sigma}_1^2 \; \textrm{and} \; P_{\rm R}\le  \frac{\bar{\sigma}_1^2\bar{\sigma}_3^2}{\bar{\sigma}_3^2 - 2 \bar{\sigma}_1^2} \\
 \label{D2.5_casei_4}
  \rm {(iv)} \; & 2\bar{\sigma}_1^2 \le \bar{\sigma}_3^2 < 3\bar{\sigma}_1^2 \; \textrm{and} \; P_{\rm R}\ge  \frac{\bar{\sigma}_1^2\bar{\sigma}_3^2}{\bar{\sigma}_3^2 - 2 \bar{\sigma}_1^2}\\
  \rm {(v)} \; & 2\bar{\sigma}_1^2 \le \bar{\sigma}_3^2 < 3\bar{\sigma}_1^2 \; \textrm{and} \; P_{\rm R} \in \!\left(\!\bar{\sigma}_3^2, \frac{\bar{\sigma}_1^2\bar{\sigma}_3^2}{\bar{\sigma}_3^2 - 2 \bar{\sigma}_1^2}\!\right)\!\\
  \rm {(vi)} \; & 2\bar{\sigma}_1^2 \le \bar{\sigma}_3^2 < 3\bar{\sigma}_1^2 \; \textrm{and} \; P_{\rm R}\le  \bar{\sigma}_3^2 \\
  \rm {(vii)} \; &  \bar{\sigma}_3^2 < 2\bar{\sigma}_1^2 \; \textrm{and}\; P_{\rm R} \ge \bar{\sigma}_4^2\\
  \rm {(viii)} \;  & \bar{\sigma}_3^2 < 2\bar{\sigma}_1^2\; \textrm{and}\; P_{\rm R} <\bar{\sigma}_4^2.
		\end{align}
	\end{subequations}
We now consider the eight subcases in (\ref{D2.5_case_i}) one by one.

$\rm {(i)} \;   \bar{\sigma}_3^2 \ge 3\bar{\sigma}_1^2 \; \textrm{and} \; P_{\rm R}\ge  \bar{\sigma}_3^2$: Using the achievability rates in Proposition 4.2,  we set $p_{{\rm R1}} = P_{\rm R} - \bar{\sigma}_3^2,\, p_{\rm{R}2
} + p_{\rm{R}3} + p_{\rm{R}4}= \bar{\sigma}_3^2$ and $p_{\rm{R}3} + p_{\rm{R}4} = \frac{\bar{\sigma}_1^2(P_{\rm R} + \bar{\sigma}_3 ^2) }{(P_{\rm R} + \bar{\sigma}_1 ^2)\bar{\sigma}_3^2}\cdot 2(\bar{\sigma}_3^2 + \bar{\sigma}_1^2) - \bar{\sigma}_1^2$ in (\ref{Pro4.2}). Note that 
\begin{subequations}
\label{D2.5_CASE1_0}
\begin{align}
 p_{\rm{R}3} + p_{\rm{R}4} &= \frac{\bar{\sigma}_1^2(P_{\rm R} + \bar{\sigma}_3 ^2) }{(P_{\rm R} + \bar{\sigma}_1 ^2)\bar{\sigma}_3^2}\cdot 2(\bar{\sigma}_3^2 + \bar{\sigma}_1^2) - \bar{\sigma}_1^2 \\
 \label{D2.5_CASE1_0_1}
 &\le 3\bar{\sigma}_1^2 \\
 &\le \bar{\sigma}_3^2,
 \end{align}
 \end{subequations}
where (\ref{D2.5_CASE1_0_1}) follows from  $\frac{(P_{\rm R} + \bar{\sigma}_3 ^2)(\bar{\sigma}_3^2 + \bar{\sigma}_1^2) }{(P_{\rm R} + \bar{\sigma}_1 ^2)\bar{\sigma}_3^2} \le 2$ for $P_{\rm R} \ge \bar{\sigma}_3^2$. Thus, $p_{\rm{R}2} = \bar{\sigma}_3^2 - p_{\rm{R}3} - p_{\rm{R}4}\ge 0 $. 	
 Then 
 \begin{subequations}
 \begin{flalign}\label{D2.5_CASE1_2}
      \quad \quad \quad \quad \quad \quad\quad\quad R_2 + \frac{1}{2} &= \frac{1}{2}\log \left(\frac{ \bar{\sigma}_3^2 + \bar{\sigma}_1^2} {p_{\rm{R}3} +p_{\rm{R}4} + \bar{\sigma}_1^2} \right) +\frac{1}{2}Ê  &\\
      \label{D2.5_CASE1_2_1}
      &= \frac{1}{2}\log \!\left(\!\frac{ \bar{\sigma}_3^2 + \bar{\sigma}_1^2} {\frac{\bar{\sigma}_1^2(P_{\rm R} + \bar{\sigma}_3 ^2) }{(P_{\rm R} + \bar{\sigma}_1 ^2)\bar{\sigma}_3^2}\cdot 2(\bar{\sigma}_3^2 + \bar{\sigma}_1^2)} \!\right)\!\! +\! \frac{1}{2} &\\
      \label{D2.5_CASE1_2_2}
      & = \frac{1}{2}\log \left(\frac{P_{\rm R} + \bar{\sigma}_1^2 }{\bar{\sigma}_1^2} \cdot \frac{\bar{\sigma}_3^2}{P_{\rm R} + \bar{\sigma}_3^2} \right)&\\
      \label{D2.5_CASE1_2_3}
      &= R_2^{'}&
 \end{flalign}
 \end{subequations}
\vspace{-6mm}
 \begin{subequations}
 \begin{flalign}
      \quad \quad \quad \quad \quad \quad\quad\quad \label{D2.5_CASE1_4}
       R_4 \!+\! \frac{1}{2} 
       &\ge \frac{1}{2}\log \!\left(\!1 \!+\! \frac{p_{{\rm R1}}}{p_{\rm{R}2} + p_{\rm{R}3} + p_{\rm{R}4} + \bar{\sigma}_3^2}\!\right)\! \!+\! \frac{1}{2} &\\
       \label{D2.5_CASE1_4_1}
       &= \frac{1}{2}\log \left(1 + \frac{P_{\rm R} - \bar{\sigma}_3^2} {2\bar{\sigma}_3^2}\right) +\frac{1}{2}Ê  &\\
       \label{D2.5_CASE1_4_2}
       &= \frac{1}{2}\log \left(1 + \frac{P_{\rm R}}{\bar{\sigma}_3^2}\right) &\\
       \label{D2.5_CASE1_4_3}
       &= R_4^{'}.&
 \end{flalign}
 \end{subequations}
In the above, step (\ref{D2.5_CASE1_2}) follows from (\ref{Pro4.2_2}); and step (\ref{D2.5_CASE1_2_3}) from
(\ref{D2.5_R2'}). Step (\ref{D2.5_CASE1_4}) follows from (\ref{Pro4.2_4}) and $\bar \sigma_3^2 \geq \bar \sigma_1^2$;  and step (\ref{D2.5_CASE1_4_3}) from (\ref{D2.5_R4'}).

What remains is to show that there exists $p_{\rm{R}3}$ and $p_{\rm{R}4}$ satisfying the following  two inequalities:
\begin{subequations}
\label{D2.5_CASE1_13_0}
\begin{align}
\label{D2.5_CASE1_13_1}
 	   & \frac{1}{2}\log \left( 2\frac{p_{\rm{R}4} + \bar{\sigma}_2^2}{ \bar{\sigma}_2^2}\cdot\frac{\bar{\sigma}_3^2 + \bar{\sigma}_4^2}{p_{\rm{R}3} + p_{\rm{R}4} +  \bar{\sigma}_4^2}\right)   \ge \frac{1}{2}\log \!\left(\!\frac{P_{\rm R}\! +\! \bar{\sigma}_2^2}{\bar{\sigma}_2^2}\!\cdot\! \frac{\bar{\sigma}_4^2}{P_{\rm R} \!+\! \bar{\sigma}_4^2}\!\cdot\! \frac{P_{\rm R} \!+\! \bar{\sigma}_1^2}{\bar{\sigma}_1^2}\!\cdot\! \frac{\bar{\sigma}_3^2}{P_{\rm R} \!+\! \bar{\sigma}_3^2}
 	    \!\right)\\
 	    \label{D2.5_CASE1_13_3}
 	    & \frac{1}{2}\log \left( 2\frac{P_{\rm R} + \bar{\sigma}_1^2}{ \bar{\sigma}_3^2 + \bar{\sigma}_1^2}\cdot\frac{p_{\rm{R}3} + p_{\rm{R}4} +  \bar{\sigma}_4^2}{ p_{\rm{R}4} +  \bar{\sigma}_4^2}\right) \ge   \frac{1}{2}\log \left(\frac{P_{\rm R} + \bar{\sigma}_4^2}{\bar{\sigma}_4^2}\!\cdot\! \frac{\bar{\sigma}_1^2}{P_{\rm R} + \bar{\sigma}_1^2}\!\cdot\! \frac{P_{\rm R} + \bar{\sigma}_3^2}{\bar{\sigma}_3^2}
 	     	 \right),
\end{align}
\end{subequations}
where the left hand side (LHS) of (\ref{D2.5_CASE1_13_1}) is equal to $R_1 + {1\over 2}$ with $R_1$ given by (\ref{Pro4.2_1}), the right hand side (RHS) of (\ref{D2.5_CASE1_13_1}) is equal to $R_1^{'}$ given by  (\ref{D2.5_R1'}), the LHS of (\ref{D2.5_CASE1_13_3}) is equal to $R_3 + {1\over 2}$ with $R_3$ given by (\ref{Pro4.2_3}), and the RHS of (\ref{D2.5_CASE1_13_3}) is equal to $R_3^{'}$ given by  (\ref{D2.5_R3'}). Note that (\ref{D2.5_CASE1_13_0}) can be rewritten as 
\begin{subequations}
\label{D2.5_CASE1_13_0_1}
\begin{align} 	  
 	& 2\frac{(p_{\rm{R}4} \!+\! \bar{\sigma}_2^2)(\bar{\sigma}_3^2 \!+\! \bar{\sigma}_4^2)}{\bar{\sigma}_2^2(p_{\rm{R}3} \!+\! p_{\rm{R}4} \!+\!  \bar{\sigma}_4^2)} \ge   \frac{(P_{\rm R} \!+\! \bar{\sigma}_2^2)\bar{\sigma}_4^2(P_{\rm R}\! +\! \bar{\sigma}_1^2)\bar{\sigma}_3^2}{\bar{\sigma}_2^2(P_{\rm R} \!+\! \bar{\sigma}_4^2)\bar{\sigma}_1^2(P_{\rm R}\! +\! \bar{\sigma}_3^2)}  \\	  
 	& 2\frac{\!(\!P_{\rm R} \!+\! \bar{\sigma}_1^2)(p_{\rm{R}3} \!+\! p_{\rm{R}4} \!+\!  \bar{\sigma}_4^2)}{ (\bar{\sigma}_3^2 \!+\! \bar{\sigma}_1^2)(p_{\rm{R}4} \!+\!  \bar{\sigma}_4^2)} \!\ge\! 
 	 \frac{(P_{\rm R} \!+\! \bar{\sigma}_4^2)\bar{\sigma}_1^2(P_{\rm R} \!+\! \bar{\sigma}_3^2)}{\bar{\sigma}_4^2(P_{\rm R} \!+\! \bar{\sigma}_1^2)\bar{\sigma}_3^2}. 	
\end{align}
 \end{subequations}

We can further write (\ref{D2.5_CASE1_13_0_1}) as
\begin{subequations}
\label{D2.5_CASE1_maxmin}
\begin{flalign}
 	     p_{\rm{R}4} \ge {}& \frac{(P_{\rm R} + \bar{\sigma}_1^2)\bar{\sigma}_4^2\bar{\sigma}_3^2}{\bar{\sigma}_1^2(P_{\rm R} +  \bar{\sigma}_4^2)(P_{\rm R} + \bar{\sigma}_3^2)}(p_{\rm{R}3} + p_{\rm{R}4} + \bar{\sigma}_4^2)   \cdot \frac{P_{\rm R} + \bar{\sigma}_2^2}{2(\bar{\sigma}_3^2 + \bar{\sigma}_4^2)} - \bar{\sigma}_2^2\\
	    \stackrel{\vartriangle }{=}{}& p_{\rm{R}4,\textrm{min}}
\end{flalign}
\end{subequations}
\vspace{-6mm}
\begin{subequations}
\label{D2.5_CASE1_maxmin_1}
\begin{flalign}
 	     p_{\rm{R}4}\le {}& \frac{(P_{\rm R} + \bar{\sigma}_1^2)\bar{\sigma}_4^2\bar{\sigma}_3^2}{\bar{\sigma}_1^2(P_{\rm R} +  \bar{\sigma}_4^2)(P_{\rm R} + \bar{\sigma}_3^2)}(p_{\rm{R}3} + p_{\rm{R}4} + \bar{\sigma}_4^2)   \cdot \frac{2(P_{\rm R} + \bar{\sigma}_1^2)}{\bar{\sigma}_3^2 + \bar{\sigma}_1^2}  - \bar{\sigma}_4^2\\
 	    \stackrel{\vartriangle }{=}{}& p_{\rm{R}4,\textrm{max}}.
\end{flalign}
\end{subequations}
To prove the existence of  $p_{\rm{R}4}$ satisfying (\ref{D2.5_CASE1_maxmin}) and (\ref{D2.5_CASE1_maxmin_1}), we need to show that the following inequalities hold:  
\begin{subequations}
\label{D2.5_CASE1_max&min}
\begin{align}
\label{D2.5_CASE1_max&min_1}
{}& p_{\rm{R}4,\textrm{max}}- p_{\rm{R}4,\textrm{min}}\ge 0\\
\label{D2.5_CASE1_max&min_2}
{}& p_{\rm{R}4,\textrm{max}}\ge  0\\
\label{D2.5_CASE1_max&min_3}
{}& p_{\rm{R}4,\textrm{min}}\le p_{\rm{R}3} + p_{\rm{R}4}.
\end{align}
\end{subequations}
 We have the following results:   
\begin{subequations}
\begin{flalign}
   p_{\rm{R}4,\textrm{max}}&-  p_{\rm{R}4,\textrm{min}}
  = \left(\frac{2(P_{\rm R} + \bar{\sigma}_1^2)}{\bar{\sigma}_3^2 + \bar{\sigma}_1^2} - \frac{P_{\rm R} + \bar{\sigma}_2^2}{2(\bar{\sigma}_3^2 + \bar{\sigma}_4^2)}\right)  \cdot \frac{(P_{\rm R} + \bar{\sigma}_1^2)\bar{\sigma}_4^2\bar{\sigma}_3^2}{\bar{\sigma}_1^2(P_{\rm R} +  \bar{\sigma}_4^2)(P_{\rm R} + \bar{\sigma}_3^2)} (p_{\rm{R}3} + p_{\rm{R}4} + \bar{\sigma}_4^2)\nonumber&\\
  &\ \ \ \ \ \   + \bar{\sigma}_2^2 -  \bar{\sigma}_4^2&\\ 
  &= \frac{P_{\rm R}(3\bar{\sigma}_3^2 \!+\! 4\bar{\sigma}_4^2 \!-\! \bar{\sigma}_1^2) \!+\! 4\bar{\sigma}_1^2(\bar{\sigma}_3^2\! +\! \bar{\sigma}_4^2)\! -\! \bar{\sigma}_2^2(\bar{\sigma}_3^2 \!+\! \bar{\sigma}_1^2)} {2(\bar{\sigma}_3^2 \!+\! \bar{\sigma}_1^2)(\bar{\sigma}_3^2 \!+\! \bar{\sigma}_4^2)} \cdot\frac{(P_{\rm R} + \bar{\sigma}_1^2)\bar{\sigma}_4^2\bar{\sigma}_3^2}{\bar{\sigma}_1^2(P_{\rm R} +  \bar{\sigma}_4^2)(P_{\rm R} + \bar{\sigma}_3^2)} \nonumber&\\
  & \ \ \ \ \ \ \cdot \left(\frac{\bar{\sigma}_1^2(P_{\rm R} + \bar{\sigma}_3 ^2) }{(P_{\rm R} + \bar{\sigma}_1 ^2)\bar{\sigma}_3^2}\cdot 2(\bar{\sigma}_3^2 + \bar{\sigma}_1^2) - \bar{\sigma}_1^2 + \bar{\sigma}_4^2\right) + \bar{\sigma}_2^2 -  \bar{\sigma}_4^2&\\
  \label{D2.5_CASE1_max-min_1}
  &\ge \frac{(P_{\rm R} + \bar{\sigma}_1^2)\bar{\sigma}_4^2}{(P_{\rm R} +  \bar{\sigma}_4^2)(P_{\rm R} + \bar{\sigma}_3^2)} \cdot \left(\frac{P_{\rm R} + \bar{\sigma}_3 ^2 }{P_{\rm R} + \bar{\sigma}_1 ^2}\cdot 2(\bar{\sigma}_3^2 + \bar{\sigma}_1^2) - \bar{\sigma}_3^2 + \frac{\bar{\sigma}_4^2\bar{\sigma}_3^2}{\bar{\sigma}_1^2}\right)\nonumber&\\
  &\ \ \ \ \ \ \!\cdot\! \frac{2P_{\rm R}(\bar{\sigma}_3^2 \!+\! \bar{\sigma}_4^2 ) \!+\! 2\bar{\sigma}_1^2(\bar{\sigma}_3^2 \!+\! \bar{\sigma}_4^2) } {2(\bar{\sigma}_3^2 \!+\! \bar{\sigma}_1^2)(\bar{\sigma}_3^2 \!+\! \bar{\sigma}_4^2)}\!+\! \bar{\sigma}_2^2 \!- \! \bar{\sigma}_4^2\!\!\!&\\
  \label{D2.5_CASE1_max-min_2}
    &\ge \frac{(P_{\rm R} + \bar{\sigma}_1^2)\bar{\sigma}_4^2}{(P_{\rm R} +  \bar{\sigma}_4^2)(P_{\rm R} + \bar{\sigma}_3^2)}  \cdot \left(\frac{P_{\rm R} + \bar{\sigma}_3 ^2 }{P_{\rm R} + \bar{\sigma}_1 ^2}\cdot (\bar{\sigma}_3^2 + \bar{\sigma}_1^2) + \frac{\bar{\sigma}_4^2\bar{\sigma}_3^2}{\bar{\sigma}_1^2}\right)\cdot\! \frac{2P_{\rm R}(\bar{\sigma}_3^2 \!+\! \bar{\sigma}_4^2 ) \!+\! 2\bar{\sigma}_1^2(\bar{\sigma}_3^2 \!+\! \bar{\sigma}_4^2) } {2(\bar{\sigma}_3^2 \!+\! \bar{\sigma}_1^2)(\bar{\sigma}_3^2 \!+\! \bar{\sigma}_4^2)}\nonumber\\&\ \ \ \ \ \  - \! \bar{\sigma}_4^2\!\!\!&\\
  &  \ge \!\left(\!\frac{\bar{\sigma}_1^2 + \bar{\sigma}_3^2}{P_{\rm R} + \bar{\sigma}_4^2}\!\cdot\!\frac{2P_{\rm R}(\bar{\sigma}_3^2 \!+\! \bar{\sigma}_4^2 ) \!+\! 2\bar{\sigma}_1^2(\bar{\sigma}_3^2 \!+\! \bar{\sigma}_4^2) } {2(\bar{\sigma}_3^2 \!+\! \bar{\sigma}_1^2)(\bar{\sigma}_3^2 \!+\! \bar{\sigma}_4^2)} \!-\! 1\!\right)\! \bar{\sigma}_4^2 &\\
  & = \left(\frac{P_{\rm R} + \bar{\sigma}_1^2}{P_{\rm R} + \bar{\sigma}_4^2} - 1\right)\bar{\sigma}_4^2&\\
    \label{D2.5_CASE1_max-min_3}
  &\ge  0 &
 \end{flalign}
 \end{subequations}
 \vspace{-6mm}
 \begin{subequations}
 \begin{flalign}
  p_{\rm{R}4,\textrm{max}}&=  \frac{(P_{\rm R} + \bar{\sigma}_1^2)\bar{\sigma}_4^2\bar{\sigma}_3^2}{\bar{\sigma}_1^2(P_{\rm R} +  \bar{\sigma}_4^2)(P_{\rm R} + \bar{\sigma}_3^2)}(p_{\rm{R}3} + p_{\rm{R}4} + \bar{\sigma}_4^2)  \cdot \frac{2(P_{\rm R} + \bar{\sigma}_1^2)}{\bar{\sigma}_3^2 + \bar{\sigma}_1^2}  - \bar{\sigma}_4^2&\\
 &\ge  \frac{(P_{\rm R} + \bar{\sigma}_1^2)\bar{\sigma}_4^2\bar{\sigma}_3^2}{\bar{\sigma}_1^2(P_{\rm R} +  \bar{\sigma}_4^2)(P_{\rm R} + \bar{\sigma}_3^2)}(p_{\rm{R}3} + p_{\rm{R}4})  \cdot \frac{2(P_{\rm R} + \bar{\sigma}_1^2)}{\bar{\sigma}_3^2 + \bar{\sigma}_1^2}  - \bar{\sigma}_4^2&\\
 &= \frac{2(P_{\rm R} + \bar{\sigma}_1^2)\bar{\sigma}_3^2}{(\bar{\sigma}_3^2 + \bar{\sigma}_1^2)\bar{\sigma}_1^2}\cdot \frac{(P_{\rm R} + \bar{\sigma}_1^2)\bar{\sigma}_4^2}{(P_{\rm R} +  \bar{\sigma}_4^2)(P_{\rm R} + \bar{\sigma}_3^2)}\cdot \!\left(\!\frac{\bar{\sigma}_1^2(P_{\rm R} \!+\! \bar{\sigma}_3 ^2) }{(P_{\rm R} \!+\! \bar{\sigma}_1 ^2)\bar{\sigma}_3^2}\!\cdot\! 2(\bar{\sigma}_3^2 \!+\! \bar{\sigma}_1^2) \!-\! \bar{\sigma}_1^2 \!\right) \! -\! \bar{\sigma}_4^2&\\
 &=  2\left(2(P_{\rm R} + \bar{\sigma}_3^2) - \frac{P_{\rm R} + \bar{\sigma}_1^2}{\bar{\sigma}_1^2 + \bar{\sigma}_3^2}\bar{\sigma}_3^2 \right) \cdot \frac{(P_{\rm R} + \bar{\sigma}_1^2)\bar{\sigma}_4^2}{(P_{\rm R} +  \bar{\sigma}_4^2)(P_{\rm R} + \bar{\sigma}_3^2)} - \bar{\sigma}_4^2 &\\
 & =\!\left(\!\frac{2\!\left(\!\frac{2\bar{\sigma}_1^2 + \bar{\sigma}_3^2}{\bar{\sigma}_1^2 + \bar{\sigma}_3^2}P_{\rm R} + \frac{\bar{\sigma}_1^2 + 2\bar{\sigma}_3^2}{\bar{\sigma}_1^2 + \bar{\sigma}_3^2}\bar{\sigma}_3^2\!\right)\!(P_{\rm R} + \bar{\sigma}_1^2)}{(P_{\rm R} +  \bar{\sigma}_4^2)(P_{\rm R} + \bar{\sigma}_3^2)} \!-\! 1\!\right)\!\bar{\sigma}_4^2&\\
 \label{D2.5_CASE1_max_1}
 &\ge  0&
 \end{flalign}
 \end{subequations}
 \vspace{-6mm} 	
 \begin{subequations}
 \begin{flalign}
  p_{\rm{R}3} &+   p_{\rm{R}4} -  p_{\rm{R}4,\textrm{min}} \nonumber&\\
  &\quad\;  = p_{\rm{R}3} +   p_{\rm{R}4} -\frac{(P_{\rm R} + \bar{\sigma}_1^2)\bar{\sigma}_4^2\bar{\sigma}_3^2}{\bar{\sigma}_1^2(P_{\rm R} +  \bar{\sigma}_4^2)(P_{\rm R} + \bar{\sigma}_3^2)} \cdot(p_{\rm{R}3} + p_{\rm{R}4} + \bar{\sigma}_4^2)  \frac{P_{\rm R} + \bar{\sigma}_2^2}{2(\bar{\sigma}_3^2 + \bar{\sigma}_4^2)} + \bar{\sigma}_2^2 &\\
   \label{D2.5_CASE1_min_1}
 &\quad\;\ge   p_{\rm{R}3} + p_{\rm{R}4} - \frac{p_{\rm{R}3} + p_{\rm{R}4} + \bar{\sigma}_4^2}{2} + \bar{\sigma}_2^2&\\
 &\quad\;= \frac{p_{\rm{R}3} + p_{\rm{R}4} - \bar{\sigma}_4^2 + 2\bar{\sigma}_2^2}{2}&\\
 &\quad\;= \frac{1}{2} \!\!\left(\!\frac{\bar{\sigma}_1^2(P_{\rm R} \!+\! \bar{\sigma}_3 ^2) }{(P_{\rm R} \!+\! \bar{\sigma}_1 ^2)\bar{\sigma}_3^2}\!\cdot\!\! 2(\bar{\sigma}_3^2 \!+\! \bar{\sigma}_1^2) \!- \!\!\bar{\sigma}_1^2 \!-\!\! \bar{\sigma}_4^2 \!+ \!\!2\bar{\sigma}_2^2\!\right)\!\!\!\!&\\
 &\quad\;=\frac{1}{2}\!\left(\!\frac{2(P_{\rm R}\! +\! \bar{\sigma}_3^2)}{P_{\rm R} \!+\! \bar{\sigma}_1^2}\!\left(\!\!\bar{\sigma}_1^2\! +\! \frac{\bar{\sigma}_1^4}{\bar{\sigma}_3^2}\!\right) \!-\! \bar{\sigma}_1^2 \!-\!\bar{\sigma}_4^2 \!+\! 2\bar{\sigma}_2^2\!\right)\!\!\!&\\
 \label{D2.5_CASE1_min_2}
 &\quad\;\ge\frac{1}{2}\!\left(\!2\left(\bar{\sigma}_1^2\! +\! \frac{\bar{\sigma}_1^4}{\bar{\sigma}_3^2}\right) \!-\! \bar{\sigma}_1^2 \!-\!\bar{\sigma}_4^2 \!+\! 2\bar{\sigma}_2^2\!\right)\!&\\
  &\quad\;\ge\frac{1}{2}\!\left(\! \bar{\sigma}_1^2 \!-\!\bar{\sigma}_4^2 \!+\! 2\bar{\sigma}_2^2\!\right)\!&\\
    \label{D2.5_CASE1_min_3}
 &\quad\;\ge 0.&
 \end{flalign}
 \end{subequations} 		
 In the above, step (\ref{D2.5_CASE1_max-min_1}) follows from 
 \begin{align}
 3\bar{\sigma}_3^2 + 4\bar{\sigma}_4^2 - \bar{\sigma}_1^2 \ge 2\bar{\sigma}_3^2 + 2\bar{\sigma}_4^2
 \end{align}
  for $\bar{\sigma}_3^2\ge \bar{\sigma}_1^2$ and
  \begin{subequations}
  \begin{align}
  &2\bar{\sigma}_1^2(\bar{\sigma}_3^2 + \bar{\sigma}_4^2) - \bar{\sigma}_2^2(\bar{\sigma}_1^2 + \bar{\sigma}_3^2) \\&\ge \bar{\sigma}_3^2 (\bar{\sigma}_1^2 - \bar{\sigma}_2^2) + \bar{\sigma}_1^2(\bar{\sigma}_4^2 - \bar{\sigma}_2^2)\\ &\ge 0
  \end{align} 
  \end{subequations} for $\bar{\sigma}_1^2\ge \bar{\sigma}_2^2$ and $\bar{\sigma}_4^2\ge \bar{\sigma}_2^2$; step (\ref{D2.5_CASE1_max-min_2}) from
  \begin{align}
  \frac{P_{\rm R} + \bar{\sigma}_3^2}{P_{\rm R} + \bar{\sigma}_1^2}(\bar{\sigma}_3^2 + \bar{\sigma}_1^2) - \bar{\sigma}_3^2 \ge 0
  \end{align}   for $\bar{\sigma}_3^2 \ge \bar{\sigma}_1^2$; and step (\ref{D2.5_CASE1_max-min_3}) from $\bar{\sigma}_1^2 \ge \bar{\sigma}_4^2$.
  Step (\ref{D2.5_CASE1_max_1})  follows from $\frac{2\bar{\sigma}_1^2 + \bar{\sigma}_3^2}{\bar{\sigma}_1^2 + \bar{\sigma}_3^2} \ge 1$ and $\frac{\bar{\sigma}_1^2 + 2\bar{\sigma}_3^2}{\bar{\sigma}_1^2 + \bar{\sigma}_3^2} \ge 1 $ and $\bar{\sigma}_1^2 \ge \bar{\sigma}_4^2$.
  Step (\ref{D2.5_CASE1_min_1}) follows from
  \begin{align}
  \frac{(P_{\rm R} + \bar{\sigma}_2^2)(P_{\rm R} + \bar{\sigma}_1^2)\bar{\sigma}_3^2\bar{\sigma}_4^2}{(P_{\rm R} + \bar{\sigma}_4^2)(P_{\rm R} + \bar{\sigma}_3^2)(\bar{\sigma}_3^2+ \bar{\sigma}_4^2)\bar{\sigma}_1^2} \le 1
  \end{align} 
  for $ \bar{\sigma}_2^2 \le \bar{\sigma}_4^2,\,\bar{\sigma}_1^2 \le \bar{\sigma}_3^2$ and $\bar{\sigma}_4^2 \le \bar{\sigma}_1^2$; step (\ref{D2.5_CASE1_min_2}) from $\bar{\sigma}_3^2 \ge \bar{\sigma}_1^2$; and step (\ref{D2.5_CASE1_min_3}) from $\bar{\sigma}_1^2 \ge \bar{\sigma}_4^2$. This proves the existence of $p_{\rm{R}3}$ and $p_{\rm{R}4}$ satisfying (\ref{D2.5_CASE1_13_0}).

  (ii)\;$\bar{\sigma}_3^2 \ge 3\bar{\sigma}_1^2 \; \textrm{and} \; P_{\rm R} \in \left(\frac{\bar{\sigma}_1^2\bar{\sigma}_3^2}{\bar{\sigma}_3^2 - 2 \bar{\sigma}_1^2}, \bar{\sigma}_3^2\right)$: Using the achievable rates in  Proposition 4.2, we set $p_{{\rm R1}} = 0 $ and $ p_{\rm{R}3} + p_{\rm{R}4} = \frac{2P_{\rm R} + \bar{\sigma}_3^2}{\bar{\sigma}_3^2}\bar{\sigma}_1^2$ in (\ref{Pro4.2}). Note that $p_{\rm{R}3} + p_{\rm{R}4} - P_{\rm R} = \frac{P_{\rm R}(2\bar{\sigma}_1^2- \bar{\sigma}_3^2) + \bar{\sigma}_1^2\bar{\sigma}_3^2}{\bar{\sigma}_3^2} < 0 $ for $P_{\rm R} > \frac{\bar{\sigma}_1^2\bar{\sigma}_3^2}{\bar{\sigma}_3^2 - 2 \bar{\sigma}_1^2}$. Thus, $p_{\rm{R}2} = P_{\rm R}- p_{\rm{R}3}- p_{\rm{R}4} >0$. Then 
   \begin{subequations}
   \begin{flalign}\label{D2.5_CASE2_2}
       \quad\quad\quad\quad\quad\quad\quad\quad R_2 \!+\! \frac{1}{2} =& \frac{1}{2}\log \left(\frac{P_{\rm R} + \bar{\sigma}_1^2}{p_{\rm{R}3} + p_{\rm{R}4} + \bar{\sigma}_1^2}   \right) + \frac{1}{2}&\\ 
        \label{D2.5_CASE2_2_1}
        = &\frac{1}{2}\log \left(\frac{P_{\rm R} + \bar{\sigma}_1^2}{\frac{P_{\rm R} + \bar{\sigma}_3^2}{\bar{\sigma}_3^2}2\bar{\sigma}_1^2}  \right) + \frac{1}{2}&\\
        \label{D2.5_CASE2_2_2}
        = &\frac{1}{2}\log \left(\frac{P_{\rm R} + \bar{\sigma}_1^2}{\bar{\sigma}_1^2}\cdot\frac{\bar{\sigma}_3^2}{P_{\rm R} + \bar{\sigma}_3^2} \right)&\\
        \label{D2.5_CASE2_2_3}
        =& R_2 ^{'}&
   \end{flalign}
   \end{subequations}
   \vspace{-6mm}
   \begin{subequations}
   \begin{flalign}
         \label{D2.5_CASE2_4}
          \quad\quad\quad\quad\quad\quad\quad\quad R_4 \!+\! \frac{1}{2} &\ge \frac{1}{2}\log \!\left(\!1 \!+\! \frac{p_{{\rm R1}}}{p_{\rm{R}2} + p_{\rm{R}3} + p_{\rm{R}4} + \bar{\sigma}_3^2}\!\right)\! \!+\! \frac{1}{2} &\\
         \label{D2.5_CASE2_4_1}
         &= \frac{1}{2}&\\
         \label{D2.5_CASE2_4_2}
         &\ge  \frac{1}{2}\log \left(1 + \frac{P_{\rm R}}{\bar{\sigma}_3^2}\right)&\\
         \label{D2.5_CASE2_4_3}
         &= R_4 ^{'}.&
   \end{flalign}
   \end{subequations}
  In the above, step (\ref{D2.5_CASE2_2}) follows from (\ref{Pro4.2_2}); and step (\ref{D2.5_CASE2_2_3}) from
  (\ref{D2.5_R2'}). Step (\ref{D2.5_CASE2_4}) follows from (\ref{Pro4.2_4}) and $\bar \sigma_3^2 \geq \bar \sigma_1^2$;  and step (\ref{D2.5_CASE2_4_3}) from (\ref{D2.5_R4'}).

  Then we need to show that there exists $p_{\rm{R}3}$ and $p_{\rm{R}4}$  satisfying the following  two inequalities:
\begin{subequations}
\label{D2.5_CASE2_13_0}
\begin{flalign}
 \label{D2.5_CASE2_13_1}
 	    \frac{1}{2}\log \left( 2\frac{p_{\rm{R}4} + \bar{\sigma}_2^2}{ \bar{\sigma}_2^2}\cdot\frac{P_{\rm R} + \bar{\sigma}_4^2}{p_{\rm{R}3} + p_{\rm{R}4} +  \bar{\sigma}_4^2}\right) &\ge \frac{1}{2}\log \!\left(\!\frac{P_{\rm R}\! +\! \bar{\sigma}_2^2}{\bar{\sigma}_2^2}\!\cdot\! \frac{\bar{\sigma}_4^2}{P_{\rm R} \!+\! \bar{\sigma}_4^2}\!\cdot\! \frac{P_{\rm R} \!+\! \bar{\sigma}_1^2}{\bar{\sigma}_1^2}\!\cdot\! \frac{\bar{\sigma}_3^2}{P_{\rm R} \!+\! \bar{\sigma}_3^2}
 	    \!\right)\\
 	    \label{D2.5_CASE2_13_3}
 	     \frac{1}{2}\log \left( 2\frac{p_{\rm{R}3} + p_{\rm{R}4} +  \bar{\sigma}_4^2}{ p_{\rm{R}4} +  \bar{\sigma}_4^2}\right) &\ge   \frac{1}{2}\log \left(\frac{P_{\rm R} + \bar{\sigma}_4^2}{\bar{\sigma}_4^2}\!\cdot\! \frac{\bar{\sigma}_1^2}{P_{\rm R} + \bar{\sigma}_1^2}\!\cdot\! \frac{P_{\rm R} + \bar{\sigma}_3^2}{\bar{\sigma}_3^2}
 	     	 \right),
\end{flalign}
\end{subequations}
where the left hand side (LHS) of (\ref{D2.5_CASE2_13_1}) is equal to $R_1 + {1\over 2}$ with $R_1$ given by (\ref{Pro4.2_1}), the right hand side (RHS) of (\ref{D2.5_CASE2_13_1}) is equal to $R_1^{'}$ given by  (\ref{D2.5_R1'}), the LHS of (\ref{D2.5_CASE2_13_3}) is equal to $R_3 + {1\over 2}$ with $R_3$ given by (\ref{Pro4.2_3}), and the RHS of (\ref{D2.5_CASE2_13_3}) is equal to $R_3^{'}$ given by  (\ref{D2.5_R3'}). Note that (\ref{D2.5_CASE2_13_0}) can be rewritten as 
  \begin{subequations}
  	 \label{D2.5_CASE2_13_0_1}
  	 \begin{align}
  \frac{2(p_{\rm{R}4} + \bar{\sigma}_2^2)(P_{\rm R} + \bar{\sigma}_4^2)}{\bar{\sigma}_2^2(p_{\rm{R}3} + p_{\rm{R}4} + \bar{\sigma}_4^2)} \!\ge& \frac{(P_{\rm R} + \bar{\sigma}_2^2)\bar{\sigma}_4^2(P_{\rm R} + \bar{\sigma}_1^2)\bar{\sigma}_3^2}{\bar{\sigma}_2^2(P_{\rm R} + \bar{\sigma}_4^2)\bar{\sigma}_1^2(P_{\rm R} + \bar{\sigma}_3^2)}\!\!\\
    2\frac{p_{\rm{R}3} + p_{\rm{R}4} + \bar{\sigma}_4^2}{p_{\rm{R}4} + \bar{\sigma}_4^2} \ge & \frac{(P_{\rm R} + \bar{\sigma}_4^2)\bar{\sigma}_1^2(P_{\rm R} + \bar{\sigma}_3^2)}{\bar{\sigma}_4^2(P_{\rm R} + \bar{\sigma}_1^2)\bar{\sigma}_3^2}.
  \end{align}
	  \end{subequations}
Tother with $p_{{\rm R}3} + p_{{\rm R}4} =\frac{2P_{\rm R} + \bar{\sigma}_3^2}{\bar{\sigma}_3^2}\bar{\sigma}_1^2$, we can further write (\ref{D2.5_CASE2_13_0_1}) as	  
\begin{subequations}
\label{D2.5_CASE2_maxmin}
\begin{flalign}
     \quad \quad\quad \quad p_{\rm{R}4} \!\ge {}& \!\left(\!\frac{2P_{\rm R}\bar{\sigma}_1^2}{\bar{\sigma}_3^2} + \bar{\sigma}_1^2 + \bar{\sigma}_4^2\!\right)\!\!\frac{\bar{\sigma}_4^2(P_{\rm R} + \bar{\sigma}_1^2)\bar{\sigma}_3^2}{(P_{\rm R} +  \bar{\sigma}_4^2)\bar{\sigma}_1^2(P_{\rm R} + \bar{\sigma}_3^2)}   \cdot \frac{P_{\rm R} + \bar{\sigma}_2^2}{2(P_{\rm R} + \bar{\sigma}_4^2)} - \bar{\sigma}_2^2&\\
      \stackrel{\vartriangle }{=}{}& p_{R_4,\textrm{min}}&
\end{flalign}
\end{subequations}
\vspace{-6mm}
\begin{subequations}
\label{D2.5_CASE2_maxmin_1}
\begin{flalign}
       \quad \quad\quad \quad p_{\rm{R}4}\!\le{} & \!\left(\!\frac{2P_{\rm R}\bar{\sigma}_1^2}{\bar{\sigma}_3^2} + \bar{\sigma}_1^2 + \bar{\sigma}_4^2\!\right)\!\!\frac{\bar{\sigma}_4^2(P_{\rm R} + \bar{\sigma}_1^2)\bar{\sigma}_3^2}{(P_{\rm R} +  \bar{\sigma}_4^2)\bar{\sigma}_1^2(P_{\rm R} + \bar{\sigma}_3^2)}  \cdot 2  - \bar{\sigma}_4^2&\\
      \stackrel{\vartriangle }{=}{}& p_{\rm{R}4,\textrm{max}}.&
\end{flalign}
\end{subequations}
 To prove that there exists $p_{\rm{R}4}$ satisfying (\ref{D2.5_CASE2_maxmin}) and (\ref{D2.5_CASE2_maxmin_1}), we need to show that the following inequalities hold: 
  \begin{subequations}
  \label{D2.5_CASE2_max&min}
  \begin{align}
  \label{D2.5_CASE2_max&min_1}
  {}& p_{\rm{R}4,\textrm{max}}- p_{\rm{R}4,\textrm{min}}\ge 0\\
  \label{D2.5_CASE2_max&min_2}
  {}& p_{\rm{R}4,\textrm{max}}\ge  0\\
  \label{D2.5_CASE2_max&min_3}
  {}& p_{\rm{R}4,\textrm{min}}\le p_{\rm{R}3} + p_{\rm{R}4}.
\end{align}
\end{subequations}
 We have the following results:  
\begin{subequations}
\begin{flalign}
      \quad\; p_{\rm{R}4,\textrm{max}}&- p_{\rm{R}4,\textrm{min}}=  \left(\frac{2P_{\rm R}\bar{\sigma}_1^2}{\bar{\sigma}_3^2} + \bar{\sigma}_1^2 + \bar{\sigma}_4^2\right)\frac{\bar{\sigma}_4^2(P_{\rm R} + \bar{\sigma}_1^2)}{(P_{\rm R} +  \bar{\sigma}_4^2)\bar{\sigma}_1^2}  \!\cdot\! \frac{\bar{\sigma}_3^2}{P_{\rm R} \!+\! \bar{\sigma}_3^2}\!\cdot\! \!\left(\!2\!-\!\frac{P_{\rm R} + \bar{\sigma}_2^2}{2(P_{\rm R} + \bar{\sigma}_4^2)}\!\right)\nonumber\\&\ \ \ \ \ \ + 
      \bar{\sigma}_2^2 \!- \! \bar{\sigma}_4^2\!\!&\\ 
      &=  \left(\frac{2P_{\rm R}\bar{\sigma}_1^2}{\bar{\sigma}_3^2} + \bar{\sigma}_1^2 + \bar{\sigma}_4^2\right)\frac{\bar{\sigma}_4^2(P_{\rm R} + \bar{\sigma}_1^2)}{(P_{\rm R} +  \bar{\sigma}_4^2)\bar{\sigma}_1^2}\cdot \frac{\bar{\sigma}_3^2}{P_{\rm R} + \bar{\sigma}_3^2} 
      \cdot \left(\frac{3P_{\rm R} + 4\bar{\sigma}_4^2- \bar{\sigma}_2^2}{2(P_{\rm R} + \bar{\sigma}_4^2)}\right)\nonumber\\&\ \ \ \ \ \  + \bar{\sigma}_2^2 -  \bar{\sigma}_4^2&\\
       \label{D2.5_CASE2_max-min_1}
      &\ge \!\left(\!\frac{2P_{\rm R}\bar{\sigma}_1^2}{\bar{\sigma}_3^2}\! +\! \bar{\sigma}_1^2\!\right)\!\frac{\bar{\sigma}_4^2  \bar{\sigma}_3^2}{\bar{\sigma}_1^2(P_{\rm R} \!+ \! \bar{\sigma}_3^2)}\!\cdot \!\frac{3}{2} \!+\! \bar{\sigma}_2^2 \!- \! \bar{\sigma}_4^2&\\ 
      &=  \frac{3}{2}\frac{(2P_{\rm R} + \bar{\sigma}_3^2)\bar{\sigma}_4^2}{P_{\rm R} + \bar{\sigma}_3^2} + \bar{\sigma}_2^2 -  \bar{\sigma}_4^2&\\ 
      &\ge \frac{3}{2}\bar{\sigma}_4^2 + \bar{\sigma}_2^2 -\bar{\sigma}_4^2 &\\
      &\ge 0 &
\end{flalign}
\end{subequations}
\vspace{-6mm}
\begin{subequations}
\begin{flalign}
      \quad\; p_{\rm{R}4,\textrm{max}}&= 2\left(2P_{\rm R} + \bar{\sigma}_3^2 + \frac{\bar{\sigma}_4^2\bar{\sigma}_3^2}{\bar{\sigma}_1^2}\right)\frac{\bar{\sigma}_4^2(P_{\rm R} + \bar{\sigma}_1^2)}{(P_{\rm R} +  \bar{\sigma}_4^2)(P_{\rm R} + \bar{\sigma}_3^2)}  - \bar{\sigma}_4^2&\\
     &\ge \left(\frac{2(2P_{\rm R} + \bar{\sigma}_3^2)(P_{\rm R} + \bar{\sigma}_1^2)}{(P_{\rm R} +  \bar{\sigma}_4^2)(P_{\rm R} + \bar{\sigma}_3^2)} - 1\right)\bar{\sigma}_4^2&\\
         \label{D2.5_CASE2_max_1}
      &\ge 0&
\end{flalign}
\end{subequations}
\vspace{-6mm}  
\begin{subequations}
\begin{flalign}
      \quad\; p_{\rm{R}4,\textrm{min}}&=  \left(\frac{2P_{\rm R}\bar{\sigma}_1^2}{\bar{\sigma}_3^2} + \bar{\sigma}_1^2 + \bar{\sigma}_4^2\right)\frac{\bar{\sigma}_4^2(P_{\rm R} + \bar{\sigma}_1^2)\bar{\sigma}_3^2}{(P_{\rm R} +  \bar{\sigma}_4^2)\bar{\sigma}_1^2(P_{\rm R} + \bar{\sigma}_3^2)} \cdot \frac{P_{\rm R} + \bar{\sigma}_2^2}{2(P_{\rm R} + \bar{\sigma}_4^2)} - \bar{\sigma}_2^2&\\
      \label{D2.5_CASE2_min_1}
      &\le \frac{1}{2} \left(\frac{2P_{\rm R}\bar{\sigma}_1^2}{\bar{\sigma}_3^2} + \bar{\sigma}_1^2 + \bar{\sigma}_4^2\right) -  \bar{\sigma}_2^2&\\
      \label{D2.5_CASE2_min_2}
      &\le  \frac{2P_{\rm R}\bar{\sigma}_1^2}{\bar{\sigma}_3^2} + \bar{\sigma}_1^2 &\\
      &=  p_{\rm{R}3} + p_{\rm{R}4}.&
\end{flalign}
\end{subequations}
In the above, step (\ref{D2.5_CASE2_max-min_1}) follows from $\bar{\sigma}_1^2 \ge \bar{\sigma}_4^2$ and $\bar{\sigma}_4^2 \ge \bar{\sigma}_2^2$.
Step (\ref{D2.5_CASE2_max_1}) follows from $\bar{\sigma}_1^2 \ge \bar{\sigma}_4^2$.
Step (\ref{D2.5_CASE2_min_1}) follows from 
\begin{align}
\frac{(P_{\rm R} + \bar{\sigma}_2^2)\bar{\sigma}_4^2(P_{\rm R} + \bar{\sigma}_1^2)\bar{\sigma}_3^2}{(P_{\rm R} + \bar{\sigma}_4^2)(P_{\rm R} +  \bar{\sigma}_4^2)\bar{\sigma}_1^2(P_{\rm R} + \bar{\sigma}_3^2)} \le 1
\end{align} for $\bar{\sigma}_2^2 \le \bar{\sigma}_4^2$ and $\bar{\sigma}_4^2 \le \bar{\sigma}_1^2$; and step (\ref{D2.5_CASE2_min_2}) from $\bar{\sigma}_4^2 \le \bar{\sigma}_1^2$. This proves the existence of $p_{\rm{R}3}$ and $p_{\rm{R}4}$ satisfying (\ref{D2.5_CASE2_13_0}).

$ \rm {(iii):} \quad 	\bar{\sigma}_3^2 \ge 3\bar{\sigma}_1^2 \; \textrm{and} \; P_{\rm R}\le  \frac{\bar{\sigma}_1^2\bar{\sigma}_3^2}{\bar{\sigma}_3^2 - 2 \bar{\sigma}_1^2} $: Using achievable rates in Proposition 4.2, we set $p_{\rm{R}1} =p_{\rm{R}2} = 0$ and $ p_{\rm{R}3} + p_{\rm{R}4} = P_{\rm R}$. Then
 \begin{subequations}
 	\begin{flalign}
 	\label{D2.5_CASE3_2}   
 	\quad\quad\quad\quad\quad\quad\quad\quad R_2 \! +\!  \frac{1}{2} =& \frac{1}{2}&\\ 
 	\label{D2.5_CASE3_2_1}
 	\ge &\frac{1}{2}\log \left(\frac{P_{\rm R} + \bar{\sigma}_1^2}{\bar{\sigma}_1^2}\cdot\frac{\bar{\sigma}_3^2}{P_{\rm R} + \bar{\sigma}_3^2} \right)&\\
 	\label{D2.5_CASE3_2_2} 
 	=& R_2 ^{'} &
 	\end{flalign}
 \end{subequations}
 \vspace{-6mm}  
 \begin{subequations}
 \begin{flalign}
 \label{D2.5_CASE3_4}
     \quad\quad\quad\quad\quad\quad\quad\quad R_4 \!+\! \frac{1}{2} &\ge \frac{1}{2}\log \! \left(\! 1 \! +\!  \frac{p_{{\rm R1}}}{p_{\rm{R}2} + p_{\rm{R}3} + p_{\rm{R}4} + \bar{\sigma}_3^2}\! \right)\!  \! + \! \frac{1}{2} &\\
     \label{D2.5_CASE3_4_1}
     &=  \frac{1}{2}&\\
     \label{D2.5_CASE3_4_2}
     &\ge  \frac{1}{2}\log \left(1 + \frac{P_{\rm R}}{\bar{\sigma}_3^2}\right)&\\
     \label{D2.5_CASE3_4_3}
     &= R_4 ^{'}.&
 \end{flalign}
 \end{subequations}
 In the above, step (\ref{D2.5_CASE3_2}) follows from (\ref{Pro4.2_2}); step (\ref{D2.5_CASE3_2_1}) from 
 \begin{align}
 \label{D2.5_CASE3_2_0_1}
 \frac{P_{\rm R} + \bar{\sigma}_1^2}{\bar{\sigma}_1^2}\cdot\frac{\bar{\sigma}_3^2}{P_{\rm R} + \bar{\sigma}_3^2} \le 2
 \end{align}
 for $P_{\rm R} \le \frac{\bar{\sigma}_1^2\bar{\sigma}_3^2}{\bar{\sigma}_3^2 - 2 \bar{\sigma}_1^2}$ and $\bar{\sigma}_3^2\ge \bar{\sigma}_1^2$; 
 and step (\ref{D2.5_CASE3_2_2}) from
 (\ref{D2.5_R2'}). Step (\ref{D2.5_CASE3_4}) follows from (\ref{Pro4.2_4}) and $\bar \sigma_3^2 \geq \bar \sigma_1^2$;  and step (\ref{D2.5_CASE3_4_3}) from (\ref{D2.5_R4'}).

 Then we need to show the existence of $p_{\rm{R}3}$ and $p_{\rm{R}4}$ satisfying the following two inequalities:
 \begin{subequations}
  \label{D2.5_CASE3_13_0}
  \begin{flalign}
   \label{D2.5_CASE3_13_1}
   	    \frac{1}{2}\log \left( 2\frac{p_{\rm{R}4} + \bar{\sigma}_2^2}{ \bar{\sigma}_2^2}\right)  & \ge \frac{1}{2}\log \!\left(\!\frac{P_{\rm R}\! +\! \bar{\sigma}_2^2}{\bar{\sigma}_2^2}\!\cdot\! \frac{\bar{\sigma}_4^2}{P_{\rm R} \!+\! \bar{\sigma}_4^2}\!\cdot\! \frac{P_{\rm R} \!+\! \bar{\sigma}_1^2}{\bar{\sigma}_1^2}\!\cdot\! \frac{\bar{\sigma}_3^2}{P_{\rm R} \!+\! \bar{\sigma}_3^2}
   	    \!\right)\\
   	     \label{D2.5_CASE3_13_3}
   	     \frac{1}{2}\log \left( 2\frac{p_{\rm{R}3} + p_{\rm{R}4} +  \bar{\sigma}_4^2}{ p_{\rm{R}4} +  \bar{\sigma}_4^2}\right)& \ge   \frac{1}{2}\log \left(\frac{P_{\rm R} + \bar{\sigma}_4^2}{\bar{\sigma}_4^2}\!\cdot\! \frac{\bar{\sigma}_1^2}{P_{\rm R} + \bar{\sigma}_1^2}\!\cdot\! \frac{P_{\rm R} + \bar{\sigma}_3^2}{\bar{\sigma}_3^2}
   	     	 \right),
  \end{flalign}
  \end{subequations}
  where the left hand side (LHS) of (\ref{D2.5_CASE3_13_1}) is equal to $R_1 + {1\over 2}$ with $R_1$ given by (\ref{Pro4.2_1}), the right hand side (RHS) of (\ref{D2.5_CASE3_13_1}) is equal to $R_1^{'}$ given by  (\ref{D2.5_R1'}), the LHS of (\ref{D2.5_CASE3_13_3}) is equal to $R_3 + {1\over 2}$ with $R_3$ given by (\ref{Pro4.2_3}), and the RHS of (\ref{D2.5_CASE3_13_3}) is equal to $R_3^{'}$ given by  (\ref{D2.5_R3'}). Note that (\ref{D2.5_CASE3_13_0}) can be rewritten as
   \begin{subequations}
    \label{D2.5_CASE3_13_0_1}
 \begin{align}
  2\frac{p_{\rm{R}4} + \bar{\sigma}_2^2}{\bar{\sigma}_2^2} \ge& \frac{(P_{\rm R} + \bar{\sigma}_2^2)\bar{\sigma}_4^2(P_{\rm R} + \bar{\sigma}_1^2)\bar{\sigma}_3^2}{\bar{\sigma}_2^2(P_{\rm R} + \bar{\sigma}_4^2)\bar{\sigma}_1^2(P_{\rm R} + \bar{\sigma}_3^2)} \\
 2\frac{ P_{\rm R} + \bar{\sigma}_4^2}{p_{\rm{R}4} + \bar{\sigma}_4^2} \ge &\frac{(P_{\rm R} + \bar{\sigma}_4^2)\bar{\sigma}_1^2(P_{\rm R} + \bar{\sigma}_3^2)}{\bar{\sigma}_4^2(P_{\rm R} + \bar{\sigma}_1^2)\bar{\sigma}_3^2}.
\end{align}
\end{subequations}
We can further write (\ref{D2.5_CASE3_13_0_1}) as
 \begin{subequations}
 \label{D2.5_CASE3_maxmin}
 \begin{flalign}
    \quad \quad\quad \quad\quad \quad\quad \quad\quad \quad  p_{\rm{R}4} \ge {}& \frac{\bar{\sigma}_4^2(P_{\rm R} + \bar{\sigma}_1^2)\bar{\sigma}_3^2}{\bar{\sigma}_1^2(P_{\rm R} + \bar{\sigma}_3^2)}\cdot \frac{P_{\rm R} + \bar{\sigma}_2^2}{2(P_{\rm R} + \bar{\sigma}_4^2)} 
      - \bar{\sigma}_2^2& \\
    \stackrel{\vartriangle }{=}{}& p_{\rm{R}4,\textrm{min}} &
 \end{flalign}
\end{subequations}
  \vspace{-6mm} 
 \begin{subequations}
 	\label{D2.5_CASE3_maxmin_1}
 	\begin{flalign}
    \quad \quad\quad \quad\quad \quad\quad \quad\quad \quad  p_{\rm{R}4}\le {}& \frac{\bar{\sigma}_4^2(P_{\rm R} + \bar{\sigma}_1^2)\bar{\sigma}_3^2}{\bar{\sigma}_1^2(P_{\rm R} + \bar{\sigma}_3^2)}\cdot 2 - \bar{\sigma}_4^2&\\
    \stackrel{\vartriangle }{=}{}& p_{\rm{R}4,\textrm{max}}.&
 \end{flalign}
 \end{subequations}
 To prove the existence of $p_{\rm{R}4}$ satisfying (\ref{D2.5_CASE3_maxmin}) and (\ref{D2.5_CASE3_maxmin_1}), we need to show that the following inequalities hold:  
 \begin{subequations}
 \label{D2.5_CASE3_max&min}
 \begin{align}
 \label{D2.5_CASE3_max&min_1}
 {}& p_{\rm{R}4,\textrm{max}}- p_{\rm{R}4,\textrm{min}}\ge 0\\
 \label{D2.5_CASE3_max&min_2}
 {}& p_{\rm{R}4,\textrm{max}}\ge  0\\
 \label{D2.5_CASE3_max&min_3}
 {}& p_{\rm{R}4,\textrm{min}}\le p_{\rm{R}3} + p_{\rm{R}4}.
 \end{align}
 \end{subequations}
We have the following results:  
 \begin{subequations}
 \begin{flalign}
     \quad \quad \quad \quad p_{\rm{R}4,\textrm{max}}& - p_{\rm{R}4,\textrm{min}}=  \frac{\bar{\sigma}_4^2(P_{\rm R} + \bar{\sigma}_1^2)\bar{\sigma}_3^2}{\bar{\sigma}_1^2(P_{\rm R} + \bar{\sigma}_3^2)}\cdot \left(2-\frac{P_{\rm R} + \bar{\sigma}_2^2}{2(P_{\rm R} + \bar{\sigma}_4^2)}\right) + \bar{\sigma}_2^2 -  \bar{\sigma}_4^2&\\ 
    &  = \frac{\bar{\sigma}_4^2(P_{\rm R} + \bar{\sigma}_1^2)\bar{\sigma}_3^2}{\bar{\sigma}_1^2(P_{\rm R} + \bar{\sigma}_3^2)}\!\cdot\! \left(\frac{3P_{\rm R} + 4\bar{\sigma}_4^2- \bar{\sigma}_2^2}{2(P_{\rm R} + \bar{\sigma}_4^2)}\right) + \bar{\sigma}_2^2 -  \bar{\sigma}_4^2&\\
      \label{D2.5_CASE3_max-min_1}
     & \ge\left(\frac{(P_{\rm R} + \bar{\sigma}_1^2)\bar{\sigma}_3^2}{\bar{\sigma}_1^2(P_{\rm R} + \bar{\sigma}_3^2)}\cdot \frac{3}{2} -1\right)\bar{\sigma}_4^2  + \bar{\sigma}_2^2 &\\
       \label{D2.5_CASE3_max-min_2}
     &  \ge  \frac{\bar{\sigma}_4^2}{2} + \bar{\sigma}_2^2 &\\
     & \ge  0 &
 \end{flalign}
 \end{subequations}
 \vspace{-6mm} 
 \begin{subequations}
 \begin{flalign}
     \quad \quad \quad \quad p_{\rm{R}4,\textrm{max}}& =  2\frac{\bar{\sigma}_4^2(P_{\rm R} + \bar{\sigma}_1^2)\bar{\sigma}_3^2}{\bar{\sigma}_1^2(P_{\rm R} + \bar{\sigma}_3^2)} - \bar{\sigma}_4^2 &\\
    & = \left(2\frac{(P_{\rm R} + \bar{\sigma}_1^2)\bar{\sigma}_3^2}{\bar{\sigma}_1^2(P_{\rm R} + \bar{\sigma}_3^2)} -1\right)\bar{\sigma}_4^2&\\
     \label{D2.5_CASE3_max_1}   
    &  \ge  0&
 \end{flalign}
 \end{subequations}
 \vspace{-6mm} 
 \begin{subequations}
 \begin{flalign}
     \quad \quad \quad \quad p_{\rm{R}3}  &+  p_{\rm{R}4} - p_{\rm{R}4,\textrm{min}}\nonumber&\\&\quad\; \!=\! P_{\rm R} \!- \! \frac{\bar{\sigma}_4^2(P_{\rm R} \!+ \!\bar{\sigma}_1^2)\bar{\sigma}_3^2}{(P_{\rm R} \!+\! \bar{\sigma}_4^2)\bar{\sigma}_1^2(P_{\rm R} \!+\! \bar{\sigma}_3^2)}\!\cdot\! \frac{P_{\rm R} \!+\! \bar{\sigma}_2^2}{2} 
         \!+\! \bar{\sigma}_2^2\!&\\
      &\quad\;\!=\!(P_{\rm R} \!+\! \bar{\sigma}_2^2)\!\left(\!1 \!-\! \frac{1}{2}\!\cdot\! \frac{\bar{\sigma}_4^2(P_{\rm R} \!+\! \bar{\sigma}_1^2)\bar{\sigma}_3^2}{(P_{\rm R} \!+\! \bar{\sigma}_4^2)\bar{\sigma}_1^2(P_{\rm R} \!+\! \bar{\sigma}_3^2)}\!\right)\!&\\
     \label{D2.5_CASE3_min_1}
     &\quad\;\ge (P_{\rm R} + \bar{\sigma}_2^2)(1- \frac{1}{2})&\\
     & \quad\;\ge 0. &
 \end{flalign}
 \end{subequations}
 In the above, step (\ref{D2.5_CASE3_max-min_1}) follows from $4\bar{\sigma}_4^2 - \bar{\sigma}_2^2 \ge 3\bar{\sigma}_4^2 $ for $\bar{\sigma}_4^2 \ge \bar{\sigma}_2^2 $; and step (\ref{D2.5_CASE3_max-min_2}) from $\frac{(P_{\rm R} + \bar{\sigma}_1^2)\bar{\sigma}_3^2}{\bar{\sigma}_1^2(P_{\rm R} + \bar{\sigma}_3^2)} \ge 1$ for  $\bar{\sigma}_3^2 \ge \bar{\sigma}_1^2$.
Step (\ref{D2.5_CASE3_max_1}) follows from $\frac{(P_{\rm R} + \bar{\sigma}_1^2)\bar{\sigma}_3^2}{\bar{\sigma}_1^2(P_{\rm R} + \bar{\sigma}_3^2)} \ge 1$ for  $\bar{\sigma}_3^2 \ge \bar{\sigma}_1^2$.
Step (\ref{D2.5_CASE3_min_1}) from $\frac{\bar{\sigma}_4^2(P_{\rm R} + \bar{\sigma}_1^2)}{(P_{\rm R} + \bar{\sigma}_4^2)\bar{\sigma}_1^2} \le 1$ for $\bar{\sigma}_4^2 \le \bar{\sigma}_1^2$ and $\frac{\bar{\sigma}_3^2}{P_{\rm R} + \bar{\sigma}_3^2} \le 1$.  This proves the existence of $p_{\rm{R}3}$ and $p_{\rm{R}4}$ satisfying (\ref{D2.5_CASE3_13_0}).

$\rm {(iv)} \; 2\bar{\sigma}_1^2 \le \bar{\sigma}_3^2 < 3\bar{\sigma}_1^2 \; \textrm{and} \; P_{\rm R}\ge  \frac{\bar{\sigma}_1^2\bar{\sigma}_3^2}{\bar{\sigma}_3^2 - 2 \bar{\sigma}_1^2}$: 
 Using the achievability rates in Proposition 4.2,  we set $p_{{\rm R1}} = P_{\rm R} - \bar{\sigma}_3^2,\, p_{\rm{R}2
} + p_{\rm{R}3} + p_{\rm{R}4}= \bar{\sigma}_3^2$ and $p_{\rm{R}3} + p_{\rm{R}4} = \frac{\bar{\sigma}_1^2(P_{\rm R} + \bar{\sigma}_3 ^2) }{(P_{\rm R} + \bar{\sigma}_1 ^2)\bar{\sigma}_3^2}\cdot 2(\bar{\sigma}_3^2 + \bar{\sigma}_1^2) - \bar{\sigma}_1^2$ in (\ref{Pro4.2}). Note that 
\begin{subequations}
 \begin{align}
  p_{\rm{R}3} + p_{\rm{R}4} &= \frac{\bar{\sigma}_1^2(P_{\rm R} + \bar{\sigma}_3 ^2) }{(P_{\rm R} + \bar{\sigma}_1 ^2)\bar{\sigma}_3^2}\cdot 2(\bar{\sigma}_3^2 + \bar{\sigma}_1^2) - \bar{\sigma}_1^2 \\
  \label{D2.5_CASE4_0_1}
  &\le \frac{\bar{\sigma}_3}{2\bar{\sigma}_1}\cdot  \frac{\bar{\sigma}_1}{\bar{\sigma}_3}\cdot 2(\bar{\sigma}_3^2 + \bar{\sigma}_1^2) - \bar{\sigma}_1^2 \\
  &= \bar{\sigma}_3^2,
  \end{align}
  \end{subequations}
 where (\ref{D2.5_CASE4_0_1}) follows from  $\frac{P_{\rm R} + \bar{\sigma}_3 ^2 }{P_{\rm R} + \bar{\sigma}_1 ^2} \le \frac{\bar{\sigma}_3}{2\bar{\sigma}_1} $ for $P_{\rm R} \ge \frac{\bar{\sigma}_1^2\bar{\sigma}_3^2}{\bar{\sigma}_3^2 - 2 \bar{\sigma}_1^2}$, by noting  the fact that for $\bar{\sigma}_3^2\ge \bar{\sigma}_1^2$, $\frac{x + \bar{\sigma}_3^2}{ x + \bar{\sigma}_1^2}$ is monotonically decreasing in $x$. 
 Thus, $p_{\rm{R}2} = \bar{\sigma}_3^2 - p_{\rm{R}3} - p_{\rm{R}4}\ge 0 $. The remaining proof for this case is strictly follows the proof for (i)
$\left\{\bar{\sigma}_3^2 \ge 3\bar{\sigma}_1^2,\, P_{\rm R}\ge  \bar{\sigma}_3^2\right\}$, and thus omit here for brevity.

 (v) $2\bar{\sigma}_1^2 \le \bar{\sigma}_3^2 < 3\bar{\sigma}_1^2 \; \textrm{and} \; P_{\rm R} \in \!\left(\!\bar{\sigma}_3^2, \frac{\bar{\sigma}_1^2\bar{\sigma}_3^2}{\bar{\sigma}_3^2 - 2 \bar{\sigma}_1^2}\!\right)\!$ :
  Using the achievable rates in Proposition 4.2, we set $p_{\rm{R}2} = 0$ and $p_{\rm{R}3} + p_{\rm{R}4} = 2\bar{\sigma}_3^2\frac{P_{\rm R} + \bar{\sigma}_1^2}{P_{\rm R} + \bar{\sigma}_3^2} -\bar{\sigma}_1^2$ in (\ref{Pro4.2}). Note that 
 $p_{\rm{R}3} + p_{\rm{R}4} = \frac{P_{\rm R}(2\bar{\sigma}_3^2- \bar{\sigma}_1^2) + \bar{\sigma}_1^2\bar{\sigma}_3^2}{P_{\rm R} + \bar{\sigma}_3^2} \ge 0$ for $\bar{\sigma}_3^2 \ge \bar{\sigma}_1^2$, and 
$p_{\rm{R}3} + p_{\rm{R}4} - P_{\rm R} = \frac{(\bar{\sigma}_3^2 - P_{\rm R})(P_{\rm R}+  \bar{\sigma}_1^2)}{P_{\rm R} +\bar{\sigma}_3^2} < 0 $ for $P_{\rm R} > \bar{\sigma}_3^2$. Thus, $p_{\rm{R}1} = P_{\rm R}- p_{\rm{R}3}- p_{\rm{R}4} >0$.
 Then 
 \begin{subequations}
 \begin{flalign}
 \label{D2.5_CASE5_2}
 R_2 + \frac{1}{2}           
    ={}&\frac{1}{2}\log \left(\frac{p_{\rm{R}2} + p_{\rm{R}3} + p_{\rm{R}4} + \bar{\sigma}_1^2 }{p_{\rm{R}3} + p_{\rm{R}4} +  \bar{\sigma}_1^2}\right) + \frac{1}{2}\\
     \label{D2.5_CASE5_2_1}
     = {}&\frac{1}{2}\\ 
     \label{D2.5_CASE5_2_2}
     \ge {}&\frac{1}{2}\log \left(1 + \frac{P_{\rm R}}{\bar{\sigma}_1^2}\right) - \frac{1}{2}\log \left(1 + \frac{P_{\rm R}}{\bar{\sigma}_3^2}\right)\\
     \label{D2.5_CASE5_2_3}
     = {}& R_2 ^{'},
 \end{flalign}
 \end{subequations}
 where step (\ref{D2.5_CASE5_2}) follows from (\ref{Pro4.2_2}); step (\ref{D2.5_CASE5_2_2}) from
 \begin{align}
 \frac{P_{\rm R} + \bar{\sigma}_1^2}{\bar{\sigma}_1^2} \frac{\bar{\sigma}_3^2}{P_{\rm R} +\bar{\sigma}_3^2} \le 2,
 \end{align} or effectively, 
 \begin{align}
 (2\bar{\sigma}_1^2 - \bar{\sigma}_3^2)P_{\rm R} + \bar{\sigma}_1^2\bar{\sigma}_3^2 \ge 0
 \end{align} for $\left\{2\bar{\sigma}_1^2\le \bar{\sigma}_3^2 \le 3\bar{\sigma}_1^2,\,P_{\rm R} \in (\bar{\sigma}_3^2, \frac{\bar{\sigma}_1^2\bar{\sigma}_3^2}{\bar{\sigma}_3^2 - 2 \bar{\sigma}_1^2})\right\}$;  and step (\ref{D2.5_CASE5_2_3}) from
 (\ref{D2.5_R2'}).

 Then we need to show  that there exists $p_{\rm{R}3}$ and $p_{\rm{R}4}$ satisfying the following inequalities:
\begin{subequations}
\label{D2.5_CASE5_134_0}
\begin{align}
\label{D2.5_CASE5_134_1}
{}& R_1 + \frac{1}{2} \ge R_1 ^{'}\\
\label{D2.5_CASE5_134_3} 
{}& R_3 + \frac{1}{2} \ge R_3 ^{'}\\
\label{D2.5_CASE5_134_4}
{}& R_4 + \frac{1}{2} \ge R_4 ^{'}.
\end{align}
\end{subequations}
Note that (\ref{D2.5_CASE5_134_1}) and (\ref{D2.5_CASE5_134_3}) can be written as
 \begin{subequations}
  \label{D2.5_CASE5_13_0}
  \begin{flalign}
  \label{D2.5_CASE5_13_1}
   	   &\frac{1}{2}\log \left( 2\frac{p_{\rm{R}4} + \bar{\sigma}_2^2}{\bar{\sigma}_2^2}\right)   \ge \frac{1}{2}\log \!\left(\!\frac{P_{\rm R}\! +\! \bar{\sigma}_2^2}{\bar{\sigma}_2^2}\!\cdot\! \frac{\bar{\sigma}_4^2}{P_{\rm R} \!+\! \bar{\sigma}_4^2}\!\cdot\! \frac{P_{\rm R} \!+\! \bar{\sigma}_1^2}{\bar{\sigma}_1^2}\!\cdot\! \frac{\bar{\sigma}_3^2}{P_{\rm R} \!+\! \bar{\sigma}_3^2}
   	    \!\right)\\
   	    \label{D2.5_CASE5_13_3}
   	     &\frac{1}{2}\log \left( 2\frac{P_{\rm R} + \bar{\sigma}_1^2}{p_{\rm{R}3} + p_{\rm{R}4} + \bar{\sigma}_1^2}\cdot\frac{p_{\rm{R}3} + p_{\rm{R}4} +  \bar{\sigma}_4^2}{ p_{\rm{R}4} +  \bar{\sigma}_4^2}\right)\ge   \frac{1}{2}\log \left(\frac{P_{\rm R} + \bar{\sigma}_4^2}{\bar{\sigma}_4^2}\!\cdot\! \frac{\bar{\sigma}_1^2}{P_{\rm R} + \bar{\sigma}_1^2}\!\cdot\! \frac{P_{\rm R} + \bar{\sigma}_3^2}{\bar{\sigma}_3^2}
   	     	 \right),
  \end{flalign}
  \end{subequations}
  where the left hand side (LHS) of (\ref{D2.5_CASE5_13_1}) is equal to $R_1 + {1\over 2}$ with $R_1$ given by (\ref{Pro4.2_1}), the right hand side (RHS) of (\ref{D2.5_CASE5_13_1}) is equal to $R_1^{'}$ given by  (\ref{D2.5_R1'}), the LHS of (\ref{D2.5_CASE5_13_3}) is equal to $R_3 + {1\over 2}$ with $R_3$ given by (\ref{Pro4.2_3}), and the RHS of (\ref{D2.5_CASE5_13_3}) is equal to $R_3^{'}$ given by  (\ref{D2.5_R3'}). Equivalently, (\ref{D2.5_CASE5_13_0}) can be rewritten as
    \begin{subequations}
    \label{D2.5_CASE5_13_0_1}
 \begin{align}
       & 2\frac{p_{\rm {R4}} + \bar{\sigma}_2^2}{ \bar{\sigma}_2^2 } \ge   \frac{(P_{\rm R} + \bar{\sigma}_2^2)\bar{\sigma}_4^2(P_{\rm R} + \bar{\sigma}_1^2)\bar{\sigma}_3^2}{\bar{\sigma}_2^2(P_{\rm R} + \bar{\sigma}_4^2)\bar{\sigma}_1^2(P_{\rm R} + \bar{\sigma}_3^2)}\\
         &2\frac{(P_{\rm{R}} + \bar{\sigma}_1^2)(p_{\rm{R}3} + p_{\rm{R}4} +  \bar{\sigma}_4^2)}{(p_{\rm{R}3} + p_{\rm{R}4} +  \bar{\sigma}_1^2)( p_{\rm{R}4} +  \bar{\sigma}_4^2)}  \ge \frac{(P_{\rm R} + \bar{\sigma}_4^2)\bar{\sigma}_1^2}{\bar{\sigma}_4^2(P_{\rm R} + \bar{\sigma}_1^2)}
        \cdot \frac{(P_{\rm R} + \bar{\sigma}_3^2)}{\bar{\sigma}_3^2}.
 \end{align}
 \end{subequations}
Together with $p_{\rm{R}3} + p_{\rm{R}4} = 2\bar{\sigma}_3^2\frac{P_{\rm R} + \bar{\sigma}_1^2}{P_{\rm R} + \bar{\sigma}_3^2} -\bar{\sigma}_1^2$, we can further write (\ref{D2.5_CASE5_13_0_1}) as
 \begin{subequations}
 \label{D2.5_CASE5_maxmin}
 \begin{flalign}
     \quad \quad \quad \quad \quad p_{\rm{R}4} \ge{}& \frac{\bar{\sigma}_4^2(P_{\rm R} + \bar{\sigma}_1^2)\bar{\sigma}_3^2}{(P_{\rm R} +  \bar{\sigma}_4^2)\bar{\sigma}_1^2(P_{\rm R} + \bar{\sigma}_3^2)} \cdot \frac{P_{\rm R} + \bar{\sigma}_2^2}{2} - \bar{\sigma}_2^2&\\
     \stackrel{\vartriangle }{=}{}& p_{\rm{R}4,\textrm{min}} &
 \end{flalign}
\end{subequations}
 \vspace{-6mm} 
 \begin{subequations}
 	\label{D2.5_CASE5_maxmin_1}
 	\begin{flalign}
     \quad \quad \quad \quad \quad p_{\rm{R}4}\le{} & \frac{2(P_{\rm R} + \bar{\sigma}_1^2)\bar{\sigma}_4^2(P_{\rm R} + \bar{\sigma}_1^2)\bar{\sigma}_3^2}{(P_{\rm R} +  \bar{\sigma}_4^2)\bar{\sigma}_1^2(P_{\rm R} + \bar{\sigma}_3^2)}\cdot \frac{p_{\rm{R}3} + p_{\rm{R}4} +\bar{\sigma}_4^2}{p_{\rm{R}3} + p_{\rm{R}4} +\bar{\sigma}_1^2}  - \bar{\sigma}_4^2 &\\
     ={}& \frac{2(P_{\rm R} + \bar{\sigma}_1^2)\bar{\sigma}_4^2(P_{\rm R} + \bar{\sigma}_1^2)\bar{\sigma}_3^2}{(P_{\rm R} +  \bar{\sigma}_4^2)\bar{\sigma}_1^2(P_{\rm R} + \bar{\sigma}_3^2)}\cdot \frac{2\bar{\sigma}_3^2\frac{P_{\rm R} + \bar{\sigma}_1^2}{P_{\rm R} + \bar{\sigma}_3^2} -\bar{\sigma}_1^2 +\bar{\sigma}_4^2}{2\bar{\sigma}_3^2\frac{P_{\rm R} + \bar{\sigma}_1^2}{P_{\rm R} + \bar{\sigma}_3^2} }- \bar{\sigma}_4^2 &\\
     ={} & \frac{\bar{\sigma}_4^2(P_{\rm R} + \bar{\sigma}_1^2)\bar{\sigma}_3^2}{(P_{\rm R} +  \bar{\sigma}_4^2)\bar{\sigma}_1^2(P_{\rm R} + \bar{\sigma}_3^2)} \cdot \left((2 +   \frac{\bar{\sigma}_4^2 - \bar{\sigma}_1^2}{\bar{\sigma}_3^2})P_{\rm R} + \bar{\sigma}_1^2 + \bar{\sigma}_4^2\right)  - \bar{\sigma}_4^2 &\\
    \stackrel{\vartriangle }{=}{}& p_{\rm{R}4,\textrm{max}}.&
 \end{flalign}
 \end{subequations}
Assume $p_{\rm{R}3} + \bar{\sigma}_1^2 \ge \bar{\sigma}_3^2$ for $p_{\rm{R}3} + p_{\rm{R}4} = 2\bar{\sigma}_3^2\frac{P_{\rm R} + \bar{\sigma}_1^2}{P_{\rm R} + \bar{\sigma}_3^2} -\bar{\sigma}_1^2$, that is, 
\begin{subequations}
\label{D2.5_CASE5_*}
\begin{align} 
	  Ê p_{\rm{R}4} {}& \le 2 \bar{\sigma}_3^2\frac{P_{\rm R} + \bar{\sigma}_1^2}{P_{\rm R} + \bar{\sigma}_3^2} -  \bar{\sigma}_3^2 \\
	  Ê {}& = \frac{P_{\rm R} + 2\bar{\sigma}_1^2 - \bar{\sigma}_3^2}{P_{\rm R} + \bar{\sigma}_3^2} \bar{\sigma}_3^2 \\
	  Ê {}& \stackrel{\vartriangle }{=}p_{\rm{R}4}^*.
\end{align}
\end{subequations} 
Then  for (\ref{D2.5_CASE5_134_4}), we have
 \begin{subequations}
  \begin{align}  
  \label{D2.5_CASE5_4} 
      R_4 \!+\! \frac{1}{2}
      ={}&\frac{1}{2} + \rm{min}\left \{\frac{1}{2}\log \left( 1+  \frac{p_{{\rm R1}}}{p_{\rm{R}2} + p_{\rm{R}4} + \bar{\sigma}_3^2} \right),\right .\nonumber\\
      {}& \ \ \ \left .\frac{1}{2}\log \left(1 + \frac{p_{{\rm R1}}}{p_{\rm{R}2} + p_{\rm{R}3} + p_{\rm{R}4} + \bar{\sigma}_1^2} \right)\right\}  \\
      \label{D2.5_CASE5_4_1}  
      ={} &  \frac{1}{2}\log \!\left(\!1 + \frac{p_{{\rm R1}}}{p_{\rm{R}2} + p_{\rm{R}3} + p_{\rm{R}4} + \bar{\sigma}_1^2} \!\right)\! \!+ \!\frac{1}{2}\\          
      ={}&\frac{1}{2}\log \left(2\frac{P_{\rm R} + \bar{\sigma}_1^2}{p_{\rm{R}2} + p_{\rm{R}3} + p_{\rm{R}4} + \bar{\sigma}_1^2} \right)\\
      ={} &\frac{1}{2}\log \left(2\frac{P_{\rm R} + \bar{\sigma}_1^2}{2\bar{\sigma}_3^2\frac{P_{\rm R} + \bar{\sigma}_1^2}{P_{\rm R} + \bar{\sigma}_3^2} }\right)\\
      ={}&\frac{1}{2}\log \left(1 + \frac{P_{\rm R}}{\bar{\sigma}_3^2}\right)\\
      \label{D2.5_CASE5_4_2}
       = {}& R_4 ^{'},
  \end{align}
  \end{subequations}
where step (\ref{D2.5_CASE5_4}) follows from (\ref{Pro4.2_4}),  step (\ref{D2.5_CASE5_4_1}) from $p_{\rm{R}3} + \bar{\sigma}_1^2 \ge \bar{\sigma}_3^2$; and step  (\ref{D2.5_CASE5_4_2}) from (\ref{D2.5_R4'}).

Combining the above discussions, we see that to prove the   existence of $p_{\rm{R}3}$ and  $p_{\rm{R}4}$ satisfying (\ref{D2.5_CASE5_maxmin}),  (\ref{D2.5_CASE5_maxmin_1}) and (\ref{D2.5_CASE5_*}), we need to prove 
 \begin{subequations}
 \label{D2.5_CASE5_max&min}
 \begin{align}
 \label{D2.5_CASE5_max&min_1}
 {}& p_{\rm{R}4,\textrm{max}}- p_{\rm{R}4,\textrm{min}}\ge 0\\
 \label{D2.5_CASE5_max&min_2}
 {}& p_{\rm{R}4,\textrm{max}}\ge  0\\
 \label{D2.5_CASE5_max&min_3}
 {}& p_{\rm{R}4,\textrm{min}}\le p_{\rm{R}3} + p_{\rm{R}4}\\
 \label{D2.5_CASE5_max&min_*_1}
 {}&  p_{\rm{R}4}^*- p_{\rm{R}4,\textrm{min}}\ge 0\\
 \label{D2.5_CASE5_max&min_*_2}
  {}& p_{\rm{R}4}^*\ge  0.
 \end{align}
 \end{subequations}
We have the following results:   
 \begin{subequations}
 \begin{flalign}
     \quad \; p_{\rm{R}4,\textrm{max}}&-p_{\rm{R}4,\textrm{min}}= \frac{\bar{\sigma}_4^2(P_{\rm R} + \bar{\sigma}_1^2)\bar{\sigma}_3^2}{(P_{\rm R} +  \bar{\sigma}_4^2)\bar{\sigma}_1^2(P_{\rm R} + \bar{\sigma}_3^2)}  \cdot \left((\frac{3}{2} +  \frac{\bar{\sigma}_4^2 - \bar{\sigma}_1^2}{\bar{\sigma}_3^2})P_{\rm R} + \bar{\sigma}_4^2 + \bar{\sigma}_1^2 - \frac{\bar{\sigma}_2^2}{2}\right)\nonumber&\\
     &\ \ \ \ \ \  + \bar{\sigma}_2^2 - \bar{\sigma}_4^2 &\\
     &\ge\! \left(\! \frac{(P_{\rm R} + \bar{\sigma}_1^2)\bar{\sigma}_3^2}{(P_{\rm R} +  \bar{\sigma}_4^2)\bar{\sigma}_1^2(P_{\rm R} + \bar{\sigma}_3^2)}
    \cdot\!\left(\!\left(\!\frac{3}{2} \!+\!  \frac{\bar{\sigma}_4^2 \!-\! \bar{\sigma}_1^2}{\bar{\sigma}_3^2}\right)P_{\rm R} \!+\! \bar{\sigma}_4^2 \!\right)\!-\!1 \!\right)\!\cdot\! \bar{\sigma}_4^2&\\
     \label{D2.5_CASE5_max-min_1}
     &\ge 0 &
 \end{flalign}
 \end{subequations}
  \vspace{-6mm} 
 \begin{subequations}
 \begin{flalign}
      \quad \;p_{\rm{R}4,\textrm{max}}&=  \frac{\bar{\sigma}_4^2(P_{\rm R} + \bar{\sigma}_1^2)\bar{\sigma}_3^2}{(P_{\rm R} +  \bar{\sigma}_4^2)\bar{\sigma}_1^2(P_{\rm R} + \bar{\sigma}_3^2)} \cdot\!\left(\!(2 +  \frac{\bar{\sigma}_4^2 \!-\! \bar{\sigma}_1^2}{\bar{\sigma}_3^2})P_{\rm R} \!+\! \bar{\sigma}_4^2 \!+\! \bar{\sigma}_1^2\!\right)\! \! -\! \bar{\sigma}_4^2 &\\
     \label{D2.5_CASE5_max_1}
     &\ge \!\!\left(\!\!\frac{1}{P_{\rm R} +  \bar{\sigma}_4^2}\!\cdot\! \!\left(\!(\!2 \!-\! \frac{\bar{\sigma}_1^2 \!-\! \bar{\sigma}_4^2}{\bar{\sigma}_3^2}\!)P_{\rm R} + \bar{\sigma}_4^2\!\right)\! \!-\! 1\!\right)\!\bar{\sigma}_4^2\!\!&\\
     \label{D2.5_CASE5_max_2}
     &\ge  0 &
 \end{flalign}
 \end{subequations}
  \vspace{-6mm} 
 \begin{subequations}
 \begin{flalign}
       \quad \;p_{\rm{R}3} &+ p_{\rm{R}4} -  p_{\rm{R}4,\textrm{min}}= \left(2 - \frac{(P_{\rm R} + \bar{\sigma}_2^2)\bar{\sigma}_4^2}{2(P_{\rm R} + \bar{\sigma}_4^2)\bar{\sigma}_1^2}\right)  \cdot \bar{\sigma}_3^2 \frac{P_{\rm R} + \bar{\sigma}_1^2}{P_{\rm R} + \bar{\sigma}_3^2}
      -\bar{\sigma}_1^2 + \bar{\sigma}_2^2\\
     & \quad \;=  \frac{P_{\rm R}(4\bar{\sigma}_1^2 - \bar{\sigma}_4^2) + \bar{\sigma}_4^2(4\bar{\sigma}_1^2 - \bar{\sigma}_2^2)} {2(P_{\rm R} + \bar{\sigma}_4^2)\bar{\sigma}_1^2} \cdot \bar{\sigma}_3^2 \frac{P_{\rm R} + \bar{\sigma}_1^2}{P_{\rm R} + \bar{\sigma}_3^2} -\bar{\sigma}_1^2 + \bar{\sigma}_2^2\\
      \label{D2.5_CASE5_min_1}
      &\quad \; \ge \frac{3\bar{\sigma}_1^2P_{\rm R} + 3\bar{\sigma}_1^2\bar{\sigma}_4^2}{2(P_{\rm R} + \bar{\sigma}_4^2)\bar{\sigma}_1^2}  
      \!\cdot\! \bar{\sigma}_3^2 \frac{P_{\rm R} + \bar{\sigma}_1^2}{P_{\rm R} + \bar{\sigma}_3^2} \!-\!\bar{\sigma}_1^2 \!+\! \bar{\sigma}_2^2&\\
      &\quad \; = \frac{3}{2}\bar{\sigma}_3^2\frac{P_{\rm R} + \bar{\sigma}_1^2}{P_{\rm R} + \bar{\sigma}_3^2} - \bar{\sigma}_1^2  + \bar{\sigma}_2^2&\\
      &\quad \; =  \frac{(3\bar{\sigma}_3^2 - 2\bar{\sigma}_1^2)P_{\rm R} + \bar{\sigma}_1^2\bar{\sigma}_3^2}{2(P_{\rm R} + \bar{\sigma}_3^2)} + \bar{\sigma}_2^2&\\
      \label{D2.5_CASE5_min_2}
      &\quad \; \ge  0.  & 
 \end{flalign}
 \end{subequations}
   \vspace{-6mm} 
 \begin{subequations}
  \begin{flalign} 
    Ê  \quad \;p_{\rm{R}4}^* -& p_{\rm{R}4,\textrm{min}}= \!\left(\!P_{\rm R}\! +\! 2\bar{\sigma}_1^2\! -\! \bar{\sigma}_3^2 \!-\!\frac{(P_{\rm R} \!+\! \bar{\sigma}_2^2)\bar{\sigma}_4^2(P_{\rm R} +\! \bar{\sigma}_1^2)}{2(P_{\rm R} \!+\! \bar{\sigma}_4^2)\bar{\sigma}_1^2}\!\right)  \cdot \frac{\bar{\sigma}_3^2}{P_{\rm R} + \bar{\sigma}_3^2}  +  \bar{\sigma}_2^2 \\
    Ê &\ \,  = \left(\frac{a P_{\rm R}^2 + bP_{\rm R} + c}{2(P_{\rm R} + \bar{\sigma}_4^2)\bar{\sigma}_1^2}\right) \cdot \frac{\bar{\sigma}_3^2}{P_{\rm R} + \bar{\sigma}_3^2}  +  \bar{\sigma}_2^2&\\
    Ê \label{D2.5_CASE5_4_10}
    Ê &\ \,\ge 0.  &
  \end{flalign}
  \end{subequations}
    \vspace{-6mm} 
\begin{flalign}
\label{D2.5_CASE5_4_11}
\quad \; p_{\rm{R}4}^* {}& = \frac{P_{\rm R} + 2\bar{\sigma}_1^2 - \bar{\sigma}_3^2}{P_{\rm R} + \bar{\sigma}_3^2} \bar{\sigma}_3^2 > 0. & Ê    
\end{flalign}
In the above, step (\ref{D2.5_CASE5_max-min_1}) follows from $ 2\bar{\sigma}_1^2 \le \bar{\sigma}_3^2$ and $\bar{\sigma}_3^2 \ge \bar{\sigma}_1^2$.
Step (\ref{D2.5_CASE5_max_1}) follows from $ \frac{(P_{\rm R} + \bar{\sigma}_1^2)\bar{\sigma}_3^2}{\bar{\sigma}_1^2(P_{\rm R} + \bar{\sigma}_3^2)} \ge 1$ for $\bar{\sigma}_3^2 \ge \bar{\sigma}_1^2$; and step (\ref{D2.5_CASE5_max_2}) from $\bar{\sigma}_3^2 \ge \bar{\sigma}_1^2$.
Step (\ref{D2.5_CASE5_min_1}) follows from  $4\bar{\sigma}_1^2 - \bar{\sigma}_4^2 \ge 3\bar{\sigma}_1^2$ and $4\bar{\sigma}_1^2-\bar{\sigma}_2^2 \ge 3\bar{\sigma}_1^2$ for $\bar{\sigma}_1^2\ge \bar{\sigma}_4^2,\,\bar{\sigma}_1^2\ge \bar{\sigma}_2^2$; and  step (\ref{D2.5_CASE5_min_2}) from $\bar{\sigma}_3^2 \ge \bar{\sigma}_1^2$. Step (\ref{D2.5_CASE5_4_10}) follows from $a P_{\rm R}^2 + bP_{\rm R} + c \ge 0$ for $P_{\rm R} > \bar{\sigma}_3^2$, which will be proved in the following discussion, 
where $a = 2\bar{\sigma}_1^2 - \bar{\sigma}_4^2, \,b = 4\bar{\sigma}_1^4-2 \bar{\sigma}_1^2\bar{\sigma}_3^2 + \bar{\sigma}_1^2\bar{\sigma}_4^2 -\bar{\sigma}_2^2\bar{\sigma}_4^2$ and $c = 4\bar{\sigma}_1^4\bar{\sigma}_4^2 - 2\bar{\sigma}_1^2\bar{\sigma}_3^2\bar{\sigma}_4^2 - \bar{\sigma}_1^2\bar{\sigma}_2^2\bar{\sigma}_4^2$. The last step of (\ref{D2.5_CASE5_4_11}) is from $P_{\rm R} >  \bar{\sigma}_3^2$.

What remains is to show that $a P_{\rm R}^2 + bP_{\rm R} + c \ge 0$ for $P_{\rm R} > \bar{\sigma}_3^2$.  Let  $f(x)\stackrel{\vartriangle }{=} a x^2 + bx + c$. Since 
   \begin{equation}
   \label{D2.5_CASE5_4_12}
   a = 2\bar{\sigma}_1^2 - \bar{\sigma}_4^2  \ge 0,
  \end{equation}
   the quadratic function $f(x)$ achieves the minimum at
 \begin{subequations}
 \label{D2.5_CASE5_4_8}
 \begin{align}
    x = {}& -\frac{b}{2a}\\
      = {}& \frac{2\bar{\sigma}_1^2(\bar{\sigma}_3^2 - 2\bar{\sigma}_1^2) + \bar{\sigma}_4^2(\bar{\sigma}_2^2 - \bar{\sigma}_1^2)}{2(2\bar{\sigma}_1^2 - \bar{\sigma}_4^2)} \\
       \label{D2.5_CASE5_4_3}
      \le{}&\frac{\bar{\sigma}_1^2(\bar{\sigma}_3^2-2\bar{\sigma}_1^2)}{2\bar{\sigma}_1^2 - \bar{\sigma}_4^2}\\
       \label{D2.5_CASE5_4_4}
     \le  {}& \bar{\sigma}_3^2,
 \end{align}
 \end{subequations}
  where step (\ref{D2.5_CASE5_4_3}) follows from $\bar{\sigma}_2^2 \le \bar{\sigma}_1^2$; and step(\ref{D2.5_CASE5_4_4}) from
  \begin{align}
  \bar{\sigma}_1^2(\bar{\sigma}_3^2-2\bar{\sigma}_1^2) \le \bar{\sigma}_3^2(2\bar{\sigma}_1^2 - \bar{\sigma}_4^2)
  \end{align}  
  by noting 
  $\bar{\sigma}_1^2\bar{\sigma}_3^2 -\bar{\sigma}_3^2\bar{\sigma}_4^2 + 2\bar{\sigma}_1^4 = \bar{\sigma}_3^2 (\bar{\sigma}_1^2 - \bar{\sigma}_4^2) + 2\bar{\sigma}_1^4 \ge 0
  $ for $\bar{\sigma}_1^2 \ge \bar{\sigma}_4^2$. Thus, $f(x)$ is monotonically increasing for $x > \bar{\sigma}_3^2$. Then, we only need to show $f(\bar{\sigma}_3^2) \ge 0$. To this end, we have 
\begin{subequations}
 \label{D2.5_CASE5_4_9}
 \begin{align}
     f(\bar{\sigma}_3^2) ={}& a \bar{\sigma}_3^4 + b\bar{\sigma}_3^2 + c\\ 
     ={}& 4\bar{\sigma}_1^4(\bar{\sigma}_3^2 + \bar{\sigma}_4^2) - \bar{\sigma}_4^2  (\bar{\sigma}_3^2 + \bar{\sigma}_2^2)  (\bar{\sigma}_1^2 + \bar{\sigma}_3^2)  \\
     \label{D2.5_CASE5_4_5}
     \ge {}&(\bar{\sigma}_3^2 + \bar{\sigma}_4^2)(4\bar{\sigma}_1^4 - \bar{\sigma}_4^2(\bar{\sigma}_3^2 + \bar{\sigma}_1^2))\\
     \label{D2.5_CASE5_4_6}
     \ge {}& (\bar{\sigma}_3^2 + \bar{\sigma}_4^2)(3\bar{\sigma}_1^4 - \bar{\sigma}_4^2\bar{\sigma}_3^2)\\
     \label{D2.5_CASE5_4_7}
     \ge {}& 0, 
 \end{align}
 \end{subequations} 
 where step (\ref{D2.5_CASE5_4_5}) follows from $\bar{\sigma}_3^2 + \bar{\sigma}_4^2 \ge \bar{\sigma}_3^2 + \bar{\sigma}_2^2$  for $\bar{\sigma}_4^2 \ge \bar{\sigma}_2^2$; step (\ref{D2.5_CASE5_4_6}) from $\bar{\sigma}_1^4 \ge \bar{\sigma}_4^2\bar{\sigma}_1^2$ for $\bar{\sigma}_1^2 \ge \bar{\sigma}_4^2$, and step (\ref{D2.5_CASE5_4_7}) from  $3\bar{\sigma}_1^4 \ge \bar{\sigma}_3^2\bar{\sigma}_1^2 \ge \bar{\sigma}_3^2\bar{\sigma}_4^2$ for $2\bar{\sigma}_1^2 \le \bar{\sigma}_3^2 \le 3\bar{\sigma}_1^2$. Combine (\ref{D2.5_CASE5_4_12})-(\ref{D2.5_CASE5_4_9}), we have $f(P_{\rm R}) \ge f(\bar{\sigma}_3^2)  \ge 0$ for $P_{\rm R} > \bar{\sigma}_3^2$.
 
 $\rm {(vi)} \; 2\bar{\sigma}_1^2 \le \bar{\sigma}_3^2 < 3\bar{\sigma}_1^2 \; \textrm{and} \; P_{\rm R}\le  \bar{\sigma}_3^2 $: The proof for this case strictly follows the proof for (iii)
 $\left\{\bar{\sigma}_3^2 \ge 3\bar{\sigma}_1^2,\,P_{\rm R}\le  \frac{\bar{\sigma}_1^2\bar{\sigma}_3^2}{\bar{\sigma}_3^2 - 2 \bar{\sigma}_1^2} \right\}$,
 except that step (\ref{D2.5_CASE3_2_0_1}) should be replaced by 
  \begin{align}
  \frac{P_{\rm R} + \bar{\sigma}_1^2}{\bar{\sigma}_1^2}\cdot\frac{\bar{\sigma}_3^2}{P_{\rm R} + \bar{\sigma}_3^2} \le \frac{\bar{\sigma}_1^2 + \bar{\sigma}_3^2}{2\bar{\sigma}_1^2}\le 2,
  \end{align}
  where the first step is from $P_{\rm R} \le \bar{\sigma}_3^2$, and the fact that for $\bar{\sigma}_3^2\ge \bar{\sigma}_1^2$, $\frac{x + \bar{\sigma}_1^2}{ x + \bar{\sigma}_3^2}$ is monotonically increasing in $x$.

 $\rm {(vii)} \;   \bar{\sigma}_3^2 < 2\bar{\sigma}_1^2$ and $P_{\rm R} \ge  \bar{\sigma}_4^2$:  
 Using the achievable rates in  Proposition 4.3, 
 we set $p_{{\rm R1}} = P_{\rm R} - \bar{\sigma}_4^2,\,p_{{\rm R2}} = \bar{\sigma}_4^2$ in (\ref{Pro4.3}). Then
	\begin{subequations}
		\label{D2.5_CASE7_1_0}
		\begin{flalign}
		\label{D2.5_CASE7_1}
		\quad \quad \quad \quad \quad \quad \quad  R_1 + \frac{1}{2}           
		= & \frac{1}{2}\log \left( 2\frac{p_{\rm R2} + \bar{\sigma}_2^2}{ \bar{\sigma}_2^2}\right)&\\
		\label{D2.5_CASE7_1_1}
		=& \frac{1}{2}\log \left(2 \frac{\bar{\sigma}_2^2 + \bar{\sigma}_4^2}{ \bar{\sigma}_2^2}\right)&\\
		\label{D2.5_CASE7_1_2}
		\ge &\frac{1}{2}\log \left(\frac{P_{\rm R} + \bar{\sigma}_2^2}{\bar{\sigma}_2^2}\cdot \frac{\bar{\sigma}_4^2}{P_{\rm R} + \bar{\sigma}_4^2}\cdot \frac{P_{\rm R} + \bar{\sigma}_1^2}{\bar{\sigma}_1^2}\cdot \frac{\bar{\sigma}_3^2}{P_{\rm R} + \bar{\sigma}_3^2}\right)  &\\
		\label{D2.5_CASE7_1_3}
		= & R_1 ^{'}&
		\end{flalign}
	\end{subequations}
	\vspace{-6mm} 
	\begin{subequations}
		\begin{flalign}
		\label{D2.5_CASE7_2}   
			\quad \quad \quad \quad \quad \quad \quad R_2 + \frac{1}{2} =& \frac{1}{2}\log \left(2\frac{P_{\rm R} + \bar{\sigma}_1^2}{p_{\rm{R}1} + \bar{\sigma}_1^2}   \right) &\\ 
			\label{D2.5_CASE7_2_1}
		= &\frac{1}{2}\log \left(2\frac{P_{\rm R} + \bar{\sigma}_1^2}{P_{\rm R} - \bar{\sigma}_4^2 + \bar{\sigma}_1^2}  \right) &\\
		\label{D2.5_CASE7_2_2} 
		\ge  &\frac{1}{2}\log \left(\frac{P_{\rm R} + \bar{\sigma}_1^2}{\bar{\sigma}_1^2}\cdot\frac{\bar{\sigma}_3^2}{P_{\rm R} + \bar{\sigma}_3^2} \right)&\\
		\label{D2.5_CASE7_2_3} 
		=& R_2 ^{'}&
		\end{flalign}
	\end{subequations}
	\vspace{-6mm} 
	\begin{subequations}
		\label{D2.5_CASE7_3_0}
		\begin{flalign}
		\label{D2.5_CASE7_3}
		\quad \quad \quad \quad \quad \quad \quad	R_3 + \frac{1}{2}           
		= & \frac{1}{2}\log \left( 2\frac{P_{\rm R}  + \bar{\sigma}_4^2}{p_{\rm{R}2} + \bar{\sigma}_4^2}\right) &\\
		\label{D2.5_CASE7_3_1}
		= &\frac{1}{2}\log \left(\frac{P_{\rm R} + \bar{\sigma}_4^2}{\bar{\sigma}_4^2}\right)&\\
		\label{D2.5_CASE7_3_2}
		\ge &\frac{1}{2}\log \left(\frac{P_{\rm R} + \bar{\sigma}_4^2}{\bar{\sigma}_4^2}\cdot \frac{\bar{\sigma}_1^2}{P_{\rm R} + \bar{\sigma}_1^2}\cdot \frac{P_{\rm R} + \bar{\sigma}_3^2}{\bar{\sigma}_3^2}
		\right)&\\
		\label{D2.5_CASE7_3_3}
		= & R_3 ^{'} &  
		\end{flalign}
	\end{subequations}
	\vspace{-6mm} 
	\begin{subequations}
		\begin{flalign}
		\label{D2.5_CASE7_4}
		\quad \quad \quad \quad \quad \quad \quad	R_4 + \frac{1}{2} = & \frac{1}{2}\log \left( 2\frac{P_{\rm R}  + \bar{\sigma}_3^2}{p_{\rm{R}2} + \bar{\sigma}_3^2}\right) &\\
		\label{D2.5_CASE7_4_1}
		= &\frac{1}{2}\log \left(2\frac{P_{\rm R} + \bar{\sigma}_3^2}{\bar{\sigma}_3^2 + \bar{\sigma}_4^2}\right)&\\
		\label{D2.5_CASE7_4_2}
		\ge&\frac{1}{2}\log \left( \frac{P_{\rm R} + \bar{\sigma}_3^2}{\bar{\sigma}_3^2}
		\right)&\\
		\label{D2.5_CASE7_4_3}
		= & R_4 ^{'}.&
		\end{flalign}
	\end{subequations}
 In the above, step (\ref{D2.5_CASE7_1}) follows from (\ref{Pro4.3_1}); step (\ref{D2.5_CASE7_1_2}) from $2\bar{\sigma}_1^2 > \bar{\sigma}_3^2,\,\bar{\sigma}_4^2 \ge \bar{\sigma}_2^2 $ and $\bar{\sigma}_3^2 \ge \bar{\sigma}_1^2$; and step (\ref{D2.5_CASE7_1_3}) from
 (\ref{D2.5_R1'}). Step (\ref{D2.5_CASE7_2}) follows from (\ref{Pro4.3_2}); step (\ref{D2.5_CASE7_2_2}) follows from $2\bar{\sigma}_1^2 > \bar{\sigma}_3^2$ and $\bar{\sigma}_3^2 \ge \bar{\sigma}_1^2$; and step (\ref{D2.5_CASE7_2_3}) from (\ref{D2.5_R2'}). Step (\ref{D2.5_CASE7_3}) follows from (\ref{Pro4.3_3}); (\ref{D2.5_CASE7_3_2})  from $\bar{\sigma}_1^2 \le \bar{\sigma}_3^2$; and step (\ref{D2.5_CASE7_3_3}) from (\ref{D2.5_R3'}). Step (\ref{D2.5_CASE7_4}) follows from (\ref{Pro4.3_4}); step (\ref{D2.5_CASE7_4_2}) from $\bar{\sigma}_4^2 \le \bar{\sigma}_3^2$; and step (\ref{D2.5_CASE7_4_3}) from (\ref{D2.5_R4'}).

 	 $\rm {(viii)} \;   \bar{\sigma}_3^2 < 2\bar{\sigma}_1^2$ and $P_{\rm R} <\bar{\sigma}_4^2$:
 	 Using the achievable rates in  Proposition 4.3, 
 	 we set $p_{{\rm R1}} = 0$ and $p_{{\rm R2}} = P_{\rm R}$ in (\ref{Pro4.3}). Then
 		\begin{subequations}
 			\label{D2.5_CASE8_1_0}
 			\begin{flalign}
 			\label{D2.5_CASE8_1}
 		\quad\quad \quad 	\quad \quad 	\quad \quad 	R_1 + \frac{1}{2}           
 			= & \frac{1}{2}\log \left( 2\frac{p_{\rm R2} + \bar{\sigma}_2^2}{ \bar{\sigma}_2^2}\right)&\\
 			\label{D2.5_CASE8_1_1}
 			= & \frac{1}{2}\log \left( 2\frac{P_{\rm R} + \bar{\sigma}_2^2}{ \bar{\sigma}_2^2}\right)&\\
 			\label{D2.5_CASE8_1_2}
 			\ge &\frac{1}{2}\log \left(\frac{P_{\rm R} + \bar{\sigma}_2^2}{\bar{\sigma}_2^2}\cdot \frac{\bar{\sigma}_4^2}{P_{\rm R} + \bar{\sigma}_4^2}\cdot \frac{P_{\rm R} + \bar{\sigma}_1^2}{\bar{\sigma}_1^2}\cdot \frac{\bar{\sigma}_3^2}{P_{\rm R} + \bar{\sigma}_3^2}\right)  &\\
 			\label{D2.5_CASE8_1_3}
 			= & R_1 ^{'}&
 			\end{flalign}
 		\end{subequations}
 		\vspace{-6mm} 
 		\begin{subequations}
 			\begin{flalign}
 			\label{D2.5_CASE8_2}   
 		\quad\quad \quad 	\quad \quad 	\quad \quad 	R_2 + \frac{1}{2} =& \frac{1}{2}\log \left(2\frac{P_{\rm R} + \bar{\sigma}_1^2}{ \bar{\sigma}_1^2}   \right) &\\ 
 			\label{D2.5_CASE8_2_1} 
 			\ge  &\frac{1}{2}\log \left(\frac{P_{\rm R} + \bar{\sigma}_1^2}{\bar{\sigma}_1^2}\cdot\frac{\bar{\sigma}_3^2}{P_{\rm R} + \bar{\sigma}_3^2} \right)&\\
 			\label{D2.5_CASE8_2_2} 
 			=& R_2 ^{'}&
 			\end{flalign}
 		\end{subequations}
 		\vspace{-6mm} 
 		\begin{subequations}
 			\label{D2.5_CASE8_3_0}
 			\begin{flalign}
 			\label{D2.5_CASE8_3}
 		\quad\quad \quad 	\quad \quad 	\quad \quad 	R_3 + \frac{1}{2}           
 			= & \frac{1}{2} &\\
 			\label{D2.5_CASE8_3_1}
 			\ge &\frac{1}{2}\log \left(\frac{P_{\rm R} + \bar{\sigma}_4^2}{\bar{\sigma}_4^2}\cdot \frac{\bar{\sigma}_1^2}{P_{\rm R} + \bar{\sigma}_1^2}\cdot \frac{P_{\rm R} + \bar{\sigma}_3^2}{\bar{\sigma}_3^2}
 			\right)&\\
 			\label{D2.5_CASE8_3_2}
 			= & R_3 ^{'} &  
 			\end{flalign}
 		\end{subequations}
 		\vspace{-6mm} 
 		\begin{subequations}
 			\begin{flalign}
 			\label{D2.5_CASE8_4}
 	\quad	\quad \quad 	\quad \quad 	\quad \quad 	R_4 + \frac{1}{2} = & \frac{1}{2}&\\
 			\label{D2.5_CASE8_4_1}
 			\ge&\frac{1}{2}\log \left( \frac{P_{\rm R} + \bar{\sigma}_3^2}{\bar{\sigma}_3^2}
 			\right)&\\
 			\label{D2.5_CASE8_4_2}
 			= & R_4 ^{'}.&
 			\end{flalign}
 		\end{subequations}
 	 In the above, step (\ref{D2.5_CASE8_1}) follows from (\ref{Pro4.3_1}); step (\ref{D2.5_CASE8_1_2}) from $\bar{\sigma}_4^2 \le \bar{\sigma}_1^2 $; and step (\ref{D2.5_CASE8_1_3}) from
 	 (\ref{D2.5_R1'}). Step (\ref{D2.5_CASE8_2}) follows from (\ref{Pro4.3_2});  and step (\ref{D2.5_CASE8_2_2}) from (\ref{D2.5_R2'}). Step (\ref{D2.5_CASE8_3}) follows from (\ref{Pro4.3_3}); (\ref{D2.5_CASE8_3_1})  from $\bar{\sigma}_1^2 \le \bar{\sigma}_3^2$ and $P_{\rm R} < \bar{\sigma}_4^2$; and step (\ref{D2.5_CASE8_3_2}) from (\ref{D2.5_R3'}). Step (\ref{D2.5_CASE8_4}) follows from (\ref{Pro4.3_4}); step (\ref{D2.5_CASE8_4_1}) from $\bar{\sigma}_4^2 \le \bar{\sigma}_3^2$ and $P_{\rm R} < \bar{\sigma}_4^2$; and step (\ref{D2.5_CASE8_4_2}) from (\ref{D2.5_R4'}).

Combining the above eight subcases,  we see that (D2.5) is achievable to within ${1\over 2}$ bit. This concludes the proof of Case II for the relay-to-user link.

\subsubsection{Case III}
With (\ref{case3}), the outer bound of the relay-to-user link in (\ref{outer}) can be written as
\begin{subequations} \label{c3}
\begin{align}
R_1 +R_3& \leq  D_2\\
R_1 +R_4 &\leq  D_2\\
R_2 +R_3 &\leq  D_4 \\
R_2 +R_4 &\leq  D_3\\
R_2 &  \leq   D_1   \\
R_i & \geq 0, \ i\in\mathcal{I}.
\end{align}\end{subequations}
For  the polytope defined by (\ref{c3}), we have the following five maximal vertices with the rate tuples $(R_1', R_2', R_3' R_4')$ given by 
\begin{subequations} \label{98}
\begin{flalign}
\label{D3.1_R1'}
{\rm (D3.1)}  \quad \quad \quad \quad \quad\quad \quad \quad \quad \quad  \quad  \quad 	
R_1'  & = D_2  &&\\
\label{D3.1_R2'}
R_2' & = D_1  &&\\
\label{D3.1_R3'}
R_3' & = 0 &&\\
\label{D3.1_R4'}
R_4' & = 0,&&
\end{flalign}\end{subequations}  \vspace{-6mm}
\begin{subequations} \label{99}
\begin{flalign}
\label{D3.2_R1'}
{\rm (D3.2)}  \quad\quad \quad \quad \quad \quad \quad \quad \quad \quad  \quad  \quad 	R_1' &=D_2 -D_4 && \\
\label{D3.2_R2'}
R_2' &= 0 &&\\
\label{D3.2_R3'}
R_3' &=D_4 &&\\
\label{D3.2_R4'}
R_4' &= D_3, &&
\end{flalign}\end{subequations}  \vspace{-6mm}
\begin{subequations} \label{100}
\begin{flalign}
\label{D3.3_R1'}
{\rm (D3.3)}  \quad \quad \quad \quad \quad \quad\quad \quad \quad \quad  \quad  \quad 	
R_1' & =  D_2 -D_3      &&    \\
\label{D3.3_R2'}
R_2' & =       0   && \\
\label{D3.3_R3'}
R_3' & =     D_3      &&\\
\label{D3.3_R4'}
R_4' & =          D_3,&&
\end{flalign}\end{subequations}  \vspace{-6mm}
\begin{subequations} \label{101}
\begin{flalign} 
\label{D3.4_R1'}
{\rm (D3.4)}  \quad \quad \quad \quad \quad \quad\quad \quad \quad \quad  \quad  \quad 	
R_1' & = D_2 - D_4 +D_1  &&\\
\label{D3.4_R2'}
R_2' & =D_1 &&\\
\label{D3.4_R3'}
R_3' &=D_4-D_1 &&\\
\label{D3.4_R4'}
R_4' &= D_3 -D_1, &&
\end{flalign}			\end{subequations}  \vspace{-6mm}
\begin{subequations} \label{102}
\begin{flalign}
\label{D3.5_R1'}
{\rm (D3.5)}  \quad \quad \quad \quad \quad \quad \quad\quad \quad \quad  \quad  \quad 	
R_1' & = D_2 - D_3 +D_1 &&\\
\label{D3.5_R2'}
R_2' & =D_1 &&\\
\label{D3.5_R3'}
R_3' &=D_3-D_1 &&\\
\label{D3.5_R4'}
R_4' &= D_3-D_1.&&
\end{flalign}			\end{subequations}
We    need to prove that these five maximal vertices (D3.1)-(D3.5) are achievable to within   $1\over 2$ bit. We use the achievable rates in  Proposition 4.4 for the proof.

For (D3.1), we   set $p_{{\rm R}1}=[P_{\rm R} - \bar \sigma_1^2]^+$, $  p_{{\rm R2}}=0$, and $p_{{\rm R3}}= \min (P_{\rm R}, \bar \sigma_1^2)$   in (\ref{Pro4.4}). 
For  $P_{\rm R} \geq \bar \sigma_1^2$, we obtain $p_{{\rm R3}}= \bar \sigma_1^2$. Then 
\begin{subequations}
\begin{flalign}\label{D3.1_CASE1_1}
		\quad\quad	\quad\quad	\quad\quad\quad\quad\quad R_1+{1\over 2} &=  {1\over 2} \log \left( 2  {\bar \sigma_1^2 + \bar \sigma_2^2 \over \bar \sigma_2^2} \cdot  {P_{\rm R}+ \bar \sigma_3^2 \over \bar \sigma_1^2+\bar \sigma_3^2}\right)  &\\
	\label{D3.1_CASE1_1_1} 
	&     \geq   {1\over 2} \log \left(   {\bar \sigma_1^2 + \bar \sigma_2^2 \over \bar \sigma_2^2} \cdot {P_{\rm R}+ \bar \sigma_1^2 \over \bar \sigma_1^2 }\right)  &\\ 
	\label{D3.1_CASE1_1_2}
  &   \geq  {1\over 2} \log \left(  { {P_{\rm R} + \bar \sigma_2^2}\over \bar \sigma_2^2 }   \right) & \\  
  \label{D3.1_CASE1_1_3}
  &= R_1' 
\end{flalign}
\end{subequations}
\vspace{-6mm}
\begin{subequations}
\begin{flalign}\label{D3.1_CASE1_2}   
  \quad\quad	\quad\quad	\quad\quad\quad\quad\quad R_2+{1\over 2} & \ge \frac{1}{2}\log\left(1+\frac{p_{\rm R1}}{p_{\rm R2}+p_{\rm R3}+\bar\sigma_1^2}\right) + {1\over 2} &\\
  \label{D3.1_CASE1_2_1}
   &=     {1\over 2} \log \left(   {P_{\rm R}+ \bar \sigma_1^2 \over \bar \sigma_1^2 }\right)  & \\  
  \label{D3.1_CASE1_2_2}
  &  = R_2'. &  
\end{flalign}
\end{subequations}
In the above, step (\ref{D3.1_CASE1_1}) follows from (\ref{Pro4.4_1}); step (\ref{D3.1_CASE1_1_1}) from
\begin{align}
2{P_{\rm R}+ \bar \sigma_3^2 \over \bar \sigma_1^2+\bar \sigma_3^2} - {P_{\rm R}+ \bar \sigma_1^2 \over \bar \sigma_1^2 } = {(\bar \sigma_1^2-\bar \sigma_3^2)(P_{\rm R}- \bar \sigma_1^2) \over \bar \sigma_1^2(\bar \sigma_1^2+\bar \sigma_3^2)} \geq 0 
\end{align}
 for $\bar \sigma_1^2 \geq \bar \sigma_3^2$ and $P_{\rm R} \geq \bar \sigma_1^2$; step (\ref{D3.1_CASE1_1_2}) from $\bar \sigma_1^2 \geq \bar \sigma_2^2$; and step (\ref{D3.1_CASE1_1_3}) from (\ref{D3.1_R1'}). Step (\ref{D3.1_CASE1_2}) follows from (\ref{Pro4.4_2});  and step (\ref{D3.1_CASE1_2_2}) from (\ref{D3.1_R2'}).

For $P_{\rm R} < \bar \sigma_1^2$, we have $p_{{\rm R3}}= P_{\rm R}$. Then 
\begin{subequations}
\begin{align}
R_1 + {1\over 2} & = {1\over 2}\log \left(1+  {P_{\rm R}\over \bar \sigma_2^2}\right) + {1\over 2}  > R_1' \\
R_2 + {1\over 2} & = {1\over 2} > {1\over 2}\log \left(1+  {P_{\rm R}\over \bar \sigma_1^2}\right)  = R_2'  \\
R_3 + {1\over 2} & = {1\over 2} > 0 =  R_3' \\
R_4 + {1\over 2} &={1\over 2}  > 0 = R_4'. 
\end{align}
\end{subequations}
 Therefore, (D3.1) is achievable to within ${1\over 2}$ bit.

We now consider ({D3}.2). We set $  p_{{\rm R1}}=0$, $  p_{\rm R2} =  (P_{\rm R} - \bar \sigma_4^2)^+$, and $  p_{{\rm R}3 }= \min   (\bar \sigma_4^2, P_{\rm R} )$  in (\ref{Pro4.4}). For  $P_{\rm R}\geq \bar \sigma_4^2$, we obtain $p_{{\rm R}3 }= \bar \sigma_4^2$. Then
\begin{subequations}
\begin{flalign}\label{D3.2_CASE1_1}
\quad\quad\quad\quad\quad\quad\quad\quad\quad\quad R_1 + {1\over 2} & =   {1\over 2} \log \left(   2 { \bar \sigma_4^2 + \bar \sigma_2^2 \over \bar \sigma_2^2}   \right) &\\
 \label{D3.2_CASE1_1_1}
&    \geq  {1\over 2} \log \left(    { {P_{\rm R} + \bar \sigma_2^2}\over \bar \sigma_2^2 } \cdot  {{  \bar \sigma_4^2}\over P_{\rm R}+  \bar \sigma_4^2 }    \right)  &\\
\label{D3.2_CASE1_1_2}
 &=R_1'&
\end{flalign}
\end{subequations}
\vspace{-6mm}
\begin{flalign}
\quad\quad\quad\quad\quad\quad\quad\quad\quad\quad R_2 + {1\over 2} & = {1\over 2} > 0 =  R_2' &
\end{flalign}
\vspace{-6mm}
\begin{subequations}
\begin{flalign}\label{D3.2_CASE1_3} 
 \quad\quad\quad\quad\quad\quad\quad\quad\quad\quad R_3+ {1\over 2} &=     {1\over 2} \log \left(    {P_{\rm R}   + \bar \sigma_4^2 \over    \bar \sigma_4^2}   \right)    &\\ 
 \label{D3.2_CASE1_3_1}
 & =R_3'& 
 \end{flalign}
 \end{subequations}
 \vspace{-6mm}
 \begin{subequations}
 \begin{flalign}\label{D3.2_CASE1_4}
 \quad\quad\quad\quad\quad\quad\quad\quad\quad\quad R_4+{1\over 2}  &=   {1\over 2} \log \left(  2  {P_{\rm R}   + \bar \sigma_3^2 \over    \bar \sigma_4^2 + \bar \sigma_3^2}   \right)   &\\
 \label{D3.2_CASE1_4_1}
 &    \geq   {1\over 2} \log \left(   {{P_{\rm R} + \bar \sigma_3^2}\over \bar \sigma_3^2 }   \right)  & \\  
 \label{D3.2_CASE1_4_2}
 &  = R_4'.&
\end{flalign}
\end{subequations}
In the above, step (\ref{D3.2_CASE1_1}) follows from (\ref{Pro4.4_1}); step (\ref{D3.2_CASE1_1_1}) from
$\bar \sigma_4^2 \geq \bar \sigma_2^2$; and step (\ref{D3.2_CASE1_1_2}) from (\ref{D3.2_R1'}). Step (\ref{D3.2_CASE1_3}) follows from (\ref{Pro4.4_3});  and step (\ref{D3.2_CASE1_3_1}) from (\ref{D3.2_R3'}). Step (\ref{D3.2_CASE1_4}) follows from (\ref{Pro4.4_4}); step (\ref{D3.2_CASE1_4_1}) from $\bar \sigma_3^2 \geq \bar \sigma_4^2$; and step (\ref{D3.2_CASE1_4_2}) from (\ref{D3.2_R4'}).

   For  $P_{\rm R} < \bar \sigma_4^2$, we have $p_{{\rm R}3 }=  P_{\rm R}$. Then 
   \begin{subequations}
   	\begin{align}
   	R_1+{1\over 2} &  = {1\over 2}  \log \left( 1 + { {P_{\rm R}}\over \bar \sigma_2^2 }  \right) + {1\over 2} > D_2 >  R_1' \\
  	R_2 + {1\over 2} & >0 =   R_2'\\
   	R_3+{1\over 2} & = {1\over 2}> {1\over 2}  \log \left( 1 + { {P_{\rm R}}\over \bar \sigma_4^2 }  \right) = R_3' \\
   	R_4+{1\over 2} & = {1\over 2}> {1\over 2}  \log \left( 1 + { {P_{\rm R}}\over \bar \sigma_3^2 }  \right) = R_4'.
	 \end{align}\end{subequations}
 Therefore, (D3.2) is achievable to within ${1\over 2}$ bit.

We next consider ({D3}.3).   We     set $p_{{\rm R}1}=  0$, $p_{\rm R2} =  (P_{\rm R}- \bar \sigma_3^2)^+$, and $p_{{\rm R}3 }=  \min (P_{\rm R}, \bar \sigma_3^2) $   in (\ref{Pro4.4}). For $P_{\rm R}\geq \bar \sigma_3^2$, we obtain $p_{{\rm R}3 }=   \bar \sigma_3^2$. Then
\begin{subequations}
\begin{flalign}\label{D3.3_CASE1_1}
\quad\quad\quad\quad\quad\quad\quad\quad\quad\quad
R_1 + {1\over 2} & =  {1\over 2}  \log \left(   2 { \bar \sigma_3^2 + \bar \sigma_2^2 \over \bar \sigma_2^2} \right)        & \\ 
\label{D3.3_CASE1_1_1}
&   \geq    {1\over 2}  \log \left( { {P_{\rm R} + \bar \sigma_2^2}\over \bar \sigma_2^2 }\cdot   {{  \bar \sigma_3^2}\over P_{\rm R}+  \bar \sigma_3^2 }  \right)    & \\
\label{D3.3_CASE1_1_2}
 & = R_1' 
\end{flalign}
\end{subequations}
\vspace{-6mm}
\begin{flalign}
\quad\quad\quad\quad\quad\quad\quad\quad\quad\quad R_2 + {1\over 2} & = {1\over 2} > 0 =  R_2' &
\end{flalign}
\vspace{-6mm}
\begin{subequations}
\begin{flalign}\label{D3.3_CASE1_3}
\quad\quad\quad\quad\quad\quad\quad\quad\quad\quad R_3 +{1\over 2} & =  {1\over 2} \log \left( 2   {P_{\rm R}   + \bar \sigma_4^2 \over    \bar \sigma_3^2 +  \bar \sigma_4^2}   \right)   &\\
\label{D3.3_CASE1_3_1}
&  \geq      {1\over 2} \log \left(    {P_{\rm R}   + \bar \sigma_3^2 \over      \bar \sigma_3^2}   \right)    &\\ 
\label{D3.3_CASE1_3_2}
 & =R_3' 
\end{flalign}
\end{subequations}
\vspace{-6mm}
\begin{subequations}
\begin{flalign}\label{D3.3_CASE1_4}
\quad\quad\quad\quad\quad\quad\quad\quad\quad\quad R_4+{1\over 2} & =   {1\over 2} \log \left(   {{P_{\rm R} + \bar \sigma_3^2}\over \bar \sigma_3^2 }   \right)     &\\  
\label{D3.3_CASE1_4_1}
&  =R_4'.
\end{flalign}
\end{subequations}
In the above, step (\ref{D3.3_CASE1_1}) follows from (\ref{Pro4.4_1}); step (\ref{D3.3_CASE1_1_1}) from $\bar \sigma_3^2 \geq \bar \sigma_2^2$; and step (\ref{D3.3_CASE1_1_2}) from (\ref{D3.3_R1'}).  Step (\ref{D3.3_CASE1_3}) follows from (\ref{Pro4.4_3});  step (\ref{D3.3_CASE1_3_1}) from
\begin{align}
2{P_{\rm R}+ \bar \sigma_4^2 \over \bar \sigma_3^2+\bar \sigma_4^2} - {P_{\rm R}+ \bar \sigma_3^2 \over \bar \sigma_3^2 } = {(\bar \sigma_3^2-\bar \sigma_4^2)(P_{\rm R}- \bar \sigma_3^2) \over \bar \sigma_3^2(\bar \sigma_3^2+\bar \sigma_4^2)} \geq 0 
\end{align}
 for $\bar \sigma_3^2 \geq \bar \sigma_4^2$ and $P_{\rm R} \geq \bar \sigma_3^2$; 
 and step (\ref{D3.3_CASE1_3_2}) from (\ref{D3.3_R3'}).
 Step (\ref{D3.3_CASE1_4}) follows from (\ref{Pro4.4_4});  and step (\ref{D3.3_CASE1_4_1}) from (\ref{D3.3_R4'}).

For  $P_{\rm R} < \bar \sigma_3^2$, we have $p_{{\rm R}3 }=  P_{\rm R}$. Then 
   	\begin{subequations}
   	   	\begin{align}
   	   	R_1+{1\over 2} &  = {1\over 2}  \log \left( 1 + { {P_{\rm R}}\over \bar \sigma_2^2 }  \right) + {1\over 2} > D_2 >  R_1' \\
   	   	R_2 + {1\over 2} & >0 =   R_2'\\
   	   	R_3+{1\over 2} & = {1\over 2}> {1\over 2}  \log \left( 1 + { {P_{\rm R}}\over \bar \sigma_3^2 }  \right) = R_3' \\
   	   	R_4+{1\over 2} & = {1\over 2}> {1\over 2}  \log \left( 1 + { {P_{\rm R}}\over \bar \sigma_3^2 }  \right) = R_4'.
   		 \end{align}\end{subequations}
Therefore, (D3.3) is achievable to within ${1\over 2}$ bit.

We next consider ({D3}.4).   If  $P_{\rm R} \geq \bar \sigma_1^2$, we set  $p_{{\rm R}1}= P_{\rm R}-\bar \sigma_1^2$, $p_{\rm R2} =  \bar \sigma_1^2 - \bar \sigma_4^2$, and $p_{{\rm R}3 }=  \bar \sigma_4^2$   in (\ref{Pro4.4}). Then
\begin{subequations}
\begin{flalign}\label{D3.4_CASE1_1}
\quad\quad\quad\quad\quad\quad\quad\quad R_1 \!+\! {1\over 2} & \!=\!		 {1\over 2}  \log \left(    	2 {\bar \sigma_4^2+ \bar \sigma_2^2 \over \bar \sigma_2^2} \cdot  {P_{\rm R} + \bar \sigma_3^2 \over \bar \sigma_1^2 + \bar \sigma_3^2} \right) &\\ 
\label{D3.4_CASE1_1_1}
&  \!\geq\!  {1\over 2}  \log \left(      {\bar \sigma_4^2+ \bar \sigma_2^2 \over \bar \sigma_2^2}\cdot  {P_{\rm R} + \bar \sigma_1^2 \over \bar \sigma_1^2  } \right) &\\ 
\label{D3.4_CASE1_1_2}
& \!\geq\!   {1\over 2}  \log \left(   { {P_{\rm R} + \bar \sigma_2^2}\over \bar \sigma_2^2 } \! \cdot\! {{  \bar \sigma_4^2}\over P_{\rm R}+  \bar \sigma_4^2 } \! \cdot\!  {{P_{\rm R} + \bar \sigma_1^2}\over \bar \sigma_1^2 }     \right)  &\\
\label{D3.4_CASE1_1_3}
& \!=\! R_1'&
\end{flalign}
\end{subequations}
\vspace{-6mm}
\begin{subequations}
\begin{flalign}\label{D3.4_CASE1_2}
\quad\quad\quad\quad\quad\quad\quad\quad R_2 \!+\!{1\over 2} &  \!\geq\! \frac{1}{2}\log\left(1+\frac{p_{\rm R1}}{p_{\rm R2}+p_{\rm R3}+\bar\sigma_1^2}\right) + {1\over 2}& \\ 
\label{D3.4_CASE1_2_1} 
&\!=\! {1\over 2}  \log \left(  {{P_{\rm R} + \bar \sigma_1^2}\over \bar \sigma_1^2 }     \right)   &\\ 
\label{D3.4_CASE1_2_2}
 & \!=\! R_2'  &
\end{flalign}
\end{subequations}
\vspace{-6mm}
\begin{subequations}
\begin{flalign}\label{D3.4_CASE1_3}
\quad\quad\quad\quad\quad\quad\quad\quad R_3 \!+\! {1\over 2} &   \!=\!     {1\over 2}  \log \left(  { \bar \sigma_1^2 + \bar \sigma_4^2 \over   \bar \sigma_4 ^2}   \right)  &\\
\label{D3.4_CASE1_3_1}
&   \!\geq\!     {1\over 2}  \log \left(   {{P_{\rm R} + \bar \sigma_4^2}\over \bar \sigma_4^2 }  \cdot  {{  \bar \sigma_1^2}\over P_{\rm R}+ \bar \sigma_1^2 }    \right)  &\\ 
\label{D3.4_CASE1_3_2}
 & \!=\! R_3' &
\end{flalign}
\end{subequations}
\vspace{-6mm}
\begin{subequations}
\begin{flalign}\label{D3.4_CASE1_4}
\quad\quad\quad\quad\quad\quad\quad\quad R_4 \!+\! {1\over 2} & \!=\!   {1\over 2}  \log \left(  	2{ \bar \sigma_1^2  + \bar \sigma_3^2 \over   \bar \sigma_4^2+ \bar \sigma_3 ^2}   \right) & \\
\label{D3.4_CASE1_4_1}
&\!\geq\! {1\over 2}  \log \left(  	{ \bar \sigma_1^2  + \bar \sigma_3^2 \over   \bar \sigma_3 ^2}   \right) &\\
\label{D3.4_CASE1_4_2}
&  \!\geq\!     {1\over 2}  \log \left(  {{P_{\rm R} + \bar \sigma_3^2}\over \bar \sigma_3^2 }  \cdot  {{  \bar \sigma_1^2}\over P_{\rm R}+ \bar \sigma_1^2 } \right)
  &\\  
\label{D3.4_CASE1_4_3}
&  \!=\! R_4'.&
\end{flalign}
\end{subequations}
In the above, step (\ref{D3.4_CASE1_1}) follows from (\ref{Pro4.4_1}); step (\ref{D3.4_CASE1_1_1}) from
\begin{align}
2{P_{\rm R}+ \bar \sigma_3^2 \over \bar \sigma_1^2+\bar \sigma_3^2} - {P_{\rm R}+ \bar \sigma_1^2 \over \bar \sigma_1^2 } = {(\bar \sigma_1^2-\bar \sigma_3^2)(P_{\rm R}- \bar \sigma_1^2) \over \bar \sigma_1^2(\bar \sigma_1^2+\bar \sigma_3^2)} \geq 0 
\end{align}
 for $\bar \sigma_1^2 \geq \bar \sigma_3^2$ and $P_{\rm R} \geq \bar \sigma_1^2$; step (\ref{D3.4_CASE1_1_2}) from $\bar \sigma_4^2 \geq \bar \sigma_2^2$; and step (\ref{D3.4_CASE1_1_3}) from (\ref{D3.4_R1'}). Step (\ref{D3.4_CASE1_2}) follows from (\ref{Pro4.4_2});  and step (\ref{D3.4_CASE1_2_2}) from (\ref{D3.4_R2'}).  Step (\ref{D3.4_CASE1_3}) follows from (\ref{Pro4.4_3});  step (\ref{D3.4_CASE1_3_1}) from
  $\bar \sigma_4^2 \geq \bar \sigma_1^2$; 
  and step (\ref{D3.4_CASE1_3_2}) from (\ref{D3.4_R3'}).
  Step (\ref{D3.4_CASE1_4}) follows from (\ref{Pro4.4_4}); 
  step (\ref{D3.4_CASE1_4_1}) from
  $\bar \sigma_3^2 \geq \bar \sigma_4^2$; step (\ref{D3.4_CASE1_4_2}) from
  $\bar \sigma_3^2 \leq \bar \sigma_1^2$; and step (\ref{D3.4_CASE1_4_3}) from (\ref{D3.4_R4'}).

 For $P_{\rm R} < \bar \sigma_1^2$,  we set  $p_{{\rm R}1}= 0$, $p_{{\rm R}2} +  p_{{\rm R}3} = P_{\rm R} $, and $p_{\rm R3} = {P_{\rm R}(\bar \sigma_4^2 - \bar \sigma_2^2) \over P_{\rm R} + \bar \sigma_4^2}$   in (\ref{Pro4.4}). Then 
 \begin{subequations}
 \begin{flalign}\label{D3.4_CASE2_1}
 \quad\quad\quad\quad\quad\quad\quad\quad R_1 \!+\! {1\over 2} & \!=\!		 {1\over 2}  \log \left(     1 + {p_{{\rm R}3} \over \bar \sigma_2^2}   \right) + {1\over 2} &\\ 
 \label{D3.4_CASE2_1_1}
 & \!=\! {1\over 2}  \log \left(  2 { {P_{\rm R} + \bar \sigma_2^2}\over \bar \sigma_2^2 }  \cdot {{  \bar \sigma_4^2}\over P_{\rm R}+  \bar \sigma_4^2 }     \right)  &\\
 \label{D3.4_CASE2_1_2} 
 & \!\geq\!   {1\over 2}  \log \left(   { {P_{\rm R} + \bar \sigma_2^2}\over \bar \sigma_2^2 } \! \cdot\! {{  \bar \sigma_4^2}\over P_{\rm R}+  \bar \sigma_4^2 } \! \cdot\!  {{P_{\rm R} + \bar \sigma_1^2}\over \bar \sigma_1^2 }     \right)  &\\
 \label{D3.4_CASE2_1_3} 
 & \!=\! R_1' &
 \end{flalign}
 \end{subequations}
 \vspace{-6mm}
 \begin{subequations}
 \begin{flalign}\label{D3.4_CASE2_2}
 \quad\quad\quad\quad\quad\quad\quad\quad R_2 \!+\! {1\over 2} &    \!=\!    {1\over 2} &\\ 
 \label{D3.4_CASE2_2_1}  
 & \!>\!{1\over 2}  \log \left(  {{P_{\rm R} + \bar \sigma_1^2}\over \bar \sigma_1^2 }     \right)   &\\  
 \label{D3.4_CASE2_2_2}
 & \!=\! R_2'  &
 \end{flalign}
 \end{subequations}
 \vspace{-6mm}
 \begin{subequations}
 \begin{flalign}\label{D3.4_CASE2_3}
 \quad\quad\quad\quad\quad\quad\quad\quad R_3 \!+\! {1\over 2} & \!=\! {1\over 2}  \log \left( 2{P_{\rm R} +  \bar \sigma_4^2  \over  p_{{\rm R}3} + \bar \sigma_4^2}   \right)  &\\
 \label{D3.4_CASE2_3_1}
 &   \!=\!    {1\over 2}  \log \left( 2{P_{\rm R} +  \bar \sigma_4^2  \over  {P_{\rm R}(\bar \sigma_4^2 - \bar \sigma_2^2) \over P_{\rm R} + \bar \sigma_4^2} + \bar \sigma_4^2}   \right)  &\\
 \label{D3.4_CASE2_3_2}
 &   \!\geq\!      {1\over 2}  \log \left(   {{P_{\rm R} + \bar \sigma_4^2}\over \bar \sigma_4^2 }  \cdot  {{  \bar \sigma_1^2}\over P_{\rm R}+ \bar \sigma_1^2 }    \right)  &\\
 \label{D3.4_CASE2_3_3}
  & \!=\!R_3' &
 \end{flalign}
 \end{subequations}
 \vspace{-6mm}
 \begin{subequations}
 \begin{flalign}\label{D3.4_CASE2_4}
 \quad\quad\quad\quad\quad\quad\quad\quad R_4 \!+\! {1\over 2} & \!=\!  {1\over 2}  \log \left( 2{P_{\rm R} +  \bar \sigma_3^2  \over  p_{{\rm R}3} + \bar \sigma_3^2}   \right)  &\\
 \label{D3.4_CASE2_4_1}
  &   \!=\!    {1\over 2}  \log \left( 2{P_{\rm R} +  \bar \sigma_3^2  \over  {P_{\rm R}(\bar \sigma_4^2 - \bar \sigma_2^2) \over P_{\rm R} + \bar \sigma_4^2} + \bar \sigma_3^2}   \right)  &\\ 
  \label{D3.4_CASE2_4_2}
 &  \!\geq\!     {1\over 2}  \log \left(  {{P_{\rm R} + \bar \sigma_3^2}\over \bar \sigma_3^2 }  \cdot  {{  \bar \sigma_1^2}\over P_{\rm R}+ \bar \sigma_1^2 } \right)  &\\ 
 \label{D3.4_CASE2_4_3} &  \!=\!R_4'.&
  \end{flalign}
  \end{subequations}
In the above, step (\ref{D3.4_CASE2_1}) follows from (\ref{Pro4.4_1}); step (\ref{D3.4_CASE2_1_2}) from
${P_{\rm R} + \bar \sigma_1^2 \over \bar \sigma_1^2} < 2$ for $ P_{\rm R} < \bar \sigma_1^2$; step (\ref{D3.4_CASE2_1_3}) from (\ref{D3.4_R1'}). Step (\ref{D3.4_CASE2_2}) follows from (\ref{Pro4.4_2}); step (\ref{D3.4_CASE2_2_1}) from $ P_{\rm R} < \bar \sigma_1^2$; and step (\ref{D3.4_CASE2_2_2}) from (\ref{D3.4_R2'}).  Step (\ref{D3.4_CASE2_3}) follows from (\ref{Pro4.4_3});  step (\ref{D3.4_CASE2_3_2}) from
\begin{subequations}
\begin{align}
2\bar \sigma_4^2(P_{\rm R}+ \bar \sigma_1^2)&\geq 2\bar \sigma_1^2\bar \sigma_4^2\\
& \geq {P_{\rm R}\bar \sigma_1^2\bar \sigma_4^2 \over P_{\rm R}+ \bar \sigma_4^2} + \bar \sigma_1^2\bar \sigma_4^2\\
& \geq \left({P_{\rm R}(\bar \sigma_4^2- \bar \sigma_2^2 ) \over P_{\rm R}+ \bar \sigma_4^2} +\bar \sigma_4^2\right)\bar \sigma_1^2;
\end{align}
\end{subequations} 
and step (\ref{D3.4_CASE2_3_3}) from (\ref{D3.4_R3'}).
Step (\ref{D3.4_CASE2_4}) follows from (\ref{Pro4.4_4}); 
 step (\ref{D3.4_CASE2_4_2}) from 
\begin{subequations}
\begin{align}
2\bar \sigma_3^2(P_{\rm R}+ \bar \sigma_1^2)&\geq 2\bar \sigma_1^2\bar \sigma_3^2\\
\label{D3.4_CASE2_4_5}
& \geq {P_{\rm R}\bar \sigma_1^2\bar \sigma_4^2 \over P_{\rm R}+ \bar \sigma_4^2} + \bar \sigma_1^2\bar \sigma_3^2\\
& \geq \left({P_{\rm R}(\bar \sigma_4^2- \bar \sigma_2^2 ) \over P_{\rm R}+ \bar \sigma_4^2} + \bar \sigma_3^2\right)\bar \sigma_1^2,
\end{align}
\end{subequations}
where (\ref{D3.4_CASE2_4_5}) from 
$\bar \sigma_3^2 \geq \bar \sigma_4^2$; and step (\ref{D3.4_CASE2_4_3}) from (\ref{D3.4_R4'}).     
 Therefore, (D3.4) is achievable to within ${1\over 2}$ bit.

We next consider ({D3}.5). For  $P_{\rm R} \geq \bar \sigma_1^2$, we  set $p_{{\rm R1}}=P_{\rm R} - \bar \sigma_1^2$, $p_{{\rm R3} }=  \bar \sigma_3^2 $, and $p_{\rm R2} =  \bar \sigma_1^2 - \bar \sigma_3^2$   in (\ref{Pro4.4}). Then   
\begin{subequations}
 \begin{flalign}\label{D3.5_CASE1_1}
		\quad\quad\quad\quad\quad\quad\quad\quad R_1 \!+\! {1\over 2} & \!=\!		 {1\over 2}  \log \left(    	2 {\bar \sigma_3^2+ \bar \sigma_2^2 \over \bar \sigma_2^2}   \cdot {P_{\rm R} + \bar \sigma_3^2 \over \bar \sigma_1^2 + \bar \sigma_3^2} \right) &\\
		\label{D3.5_CASE1_1_1}
		& \!\geq\! {1\over 2}  \log \left(  {\bar \sigma_3^2+ \bar \sigma_2^2 \over \bar \sigma_2^2}   \cdot {P_{\rm R} + \bar \sigma_1^2 \over \bar \sigma_1^2} \right)&\\
		\label{D3.5_CASE1_1_2}
		&  \!\geq\!   {1\over 2}  \log \left(   { {P_{\rm R} + \bar \sigma_2^2}\over \bar \sigma_2^2 }  \!\cdot\! {{  \bar \sigma_3^2}\over P_{\rm R}+  \bar \sigma_3^2 } \!\cdot\!  {{P_{\rm R} + \bar \sigma_1^2}\over \bar \sigma_1^2 }     \right)  &\\ 
		\label{D3.5_CASE1_1_3}
		& \!=\! R_1'&
 \end{flalign}
 \end{subequations}
 \vspace{-6mm}
 \begin{subequations}
 \begin{flalign}\label{D3.5_CASE1_2}		
		\quad\quad\quad\quad\quad\quad\quad\quad R_2 \!+\! {1\over 2} &   \!\geq\! \frac{1}{2}\log\left(1+\frac{p_{\rm R1}}{p_{\rm R2}+p_{\rm R3}+\bar\sigma_1^2}\right) + {1\over 2}& \\ 
		\label{D3.5_CASE1_2_1}
		 & \!=\!      {1\over 2}  \log \left(  {{P_{\rm R} + \bar \sigma_1^2}\over \bar \sigma_1^2 }     \right)   &\\ 
		 \label{D3.5_CASE1_2_2} 
		 & \!=\!R_2' & 
 \end{flalign}
 \end{subequations}
 \vspace{-6mm}
 \begin{subequations}
 \begin{flalign}\label{D3.5_CASE1_3}		
		\quad\quad\quad\quad\quad\quad\quad\quad R_3 \!+\! {1\over 2} &   \!=\!    {1\over 2}  \log \left(2  { \bar \sigma_1^2 + \bar \sigma_4^2 \over    \bar \sigma_3^2 + \bar \sigma_4^2}   \right)  &\\
		\label{D3.5_CASE1_3_1}
		& \!\geq\! {1\over 2}  \log \left(  { \bar \sigma_1^2 + \bar \sigma_4^2 \over    \bar \sigma_3^2 }   \right)  &\\
		\label{D3.5_CASE1_3_2}
		&   \!\geq\!      {1\over 2}  \log \left(   {{P_{\rm R} + \bar \sigma_3^2}\over \bar \sigma_3^2 } \cdot {{  \bar \sigma_1^2}\over P_{\rm R}+ \bar \sigma_1^2 }    \right)  &\\  
		\label{D3.5_CASE1_3_3}
		& \!=\!R_3' &
 \end{flalign}
 \end{subequations}
 \vspace{-6mm}
 \begin{subequations}
 \begin{flalign}\label{D3.5_CASE1_4}		
		\quad\quad\quad\quad\quad\quad\quad\quad R_4 \!+\! {1\over 2} & \!=\!  {1\over 2}  \log \left(   { \bar \sigma_1^2  + \bar \sigma_3^2 \over    \bar \sigma_3 ^2}   \right)  &\\
		\label{D3.5_CASE1_4_1} 
		&  \!\geq\!     {1\over 2}  \log \left(  {{P_{\rm R} + \bar \sigma_3^2}\over \bar \sigma_3^2 } \cdot {{  \bar \sigma_1^2}\over P_{\rm R}+ \bar \sigma_1^2 } \right)   &\\ 
		\label{D3.5_CASE1_4_2}
		 &\!=\!R_4'.&
 \end{flalign}
 \end{subequations}
 In the above, step (\ref{D3.5_CASE1_1}) follows from (\ref{Pro4.4_1}); step (\ref{D3.5_CASE1_1_1}) from
 \begin{align}
 2{P_{\rm R}+ \bar \sigma_3^2 \over \bar \sigma_1^2+\bar \sigma_3^2} - {P_{\rm R}+ \bar \sigma_1^2 \over \bar \sigma_1^2 } = {(\bar \sigma_1^2-\bar \sigma_3^2)(P_{\rm R}- \bar \sigma_1^2) \over \bar \sigma_1^2(\bar \sigma_1^2+\bar \sigma_3^2)} \geq 0 
 \end{align}
  for $\bar \sigma_1^2 \geq \bar \sigma_3^2$ and $P_{\rm R} \geq \bar \sigma_1^2$; step (\ref{D3.5_CASE1_1_2}) from $\bar \sigma_3^2 \geq \bar \sigma_2^2$; and step (\ref{D3.5_CASE1_1_3}) from (\ref{D3.5_R1'}). Step (\ref{D3.5_CASE1_2}) follows from (\ref{Pro4.4_2});  and step (\ref{D3.5_CASE1_2_2}) from (\ref{D3.5_R2'}).  Step (\ref{D3.5_CASE1_3}) follows from (\ref{Pro4.4_3});  step (\ref{D3.5_CASE1_3_1}) from
   $\bar \sigma_3^2 \geq \bar \sigma_4^2$; 
   step (\ref{D3.5_CASE1_3_2}) from
      $\bar \sigma_3^2 \leq \bar \sigma_1^2$;
   and step (\ref{D3.5_CASE1_3_3}) from (\ref{D3.5_R3'}).
   Step (\ref{D3.5_CASE1_4}) follows from (\ref{Pro4.4_4}); 
   step (\ref{D3.5_CASE1_4_1}) from
   $\bar \sigma_3^2 \leq \bar \sigma_1^2$; step (\ref{D3.5_CASE1_4_2}) from
    (\ref{D3.5_R4'}).

For  $P_{\rm R} < \bar \sigma_1^2$,  we set  $p_{{\rm R}1}= 0$, $p_{{\rm R}2} +  p_{{\rm R}3} = P_{\rm R} $, and $p_{\rm R3} = {P_{\rm R}(\bar \sigma_3^2 - \bar \sigma_2^2) \over P_{\rm R} + \bar \sigma_3^2}$   in (\ref{Pro4.4}). Then 
  \begin{subequations}
  \begin{flalign}\label{D3.5_CASE2_1}
 \quad\quad\quad\quad\quad\quad\quad\quad R_1\! +\! {1\over 2} &\! =\!	 {1\over 2}  \log \left( 1 + {p_{{\rm R}3} \over \bar \sigma_2^2}   \right) + {1\over 2} &\\
 \label{D3.5_CASE2_1_1} 
 & \! =\!	 {1\over 2}  \log \left(  2 { {P_{\rm R} + \bar \sigma_2^2}\over \bar \sigma_2^2 }  \cdot {{  \bar \sigma_3^2}\over P_{\rm R}+  \bar \sigma_3^2 }     \right)   &\\
 \label{D3.5_CASE2_1_2}
 &  \!\geq\!   {1\over 2}  \log \left(   { {P_{\rm R} + \bar \sigma_2^2}\over \bar \sigma_2^2 }  \!\cdot\! {{  \bar \sigma_3^2}\over P_{\rm R}+  \bar \sigma_3^2 } \!\cdot\!  {{P_{\rm R} + \bar \sigma_1^2}\over \bar \sigma_1^2 }     \right)   &\\ 
 \label{D3.5_CASE2_1_3}
 & \! =\!	 R_1' &
 \end{flalign}
  \end{subequations}
  \vspace{-6mm}
  \begin{subequations}
  \begin{flalign}\label{D3.5_CASE2_2}
 \quad\quad\quad\quad\quad\quad\quad\quad  R_2 \! +\! {1\over 2} &   \! =\!	    {1\over 2}  &\\ 
 \label{D3.5_CASE2_2_1}  
 & \!>\! {1\over 2}  \log \left(  {{P_{\rm R} + \bar \sigma_1^2}\over \bar \sigma_1^2 }     \right)     &\\ \label{D3.5_CASE2_2_2} 
 & \! =\!	R_2'   &
 \end{flalign}
  \end{subequations}
  \vspace{-6mm}
  \begin{subequations}
  \begin{flalign}\label{D3.5_CASE2_3}
 \quad\quad\quad\quad\quad\quad\quad\quad  R_3 \! +\! {1\over 2} & \! =\!	 {1\over 2}  \log \left( 2{P_{\rm R} +  \bar \sigma_4^2  \over  p_{{\rm R}3} + \bar \sigma_4^2}   \right)   &\\
 \label{D3.5_CASE2_3_1}
 &   \! =\!	    {1\over 2}  \log \left( 2{P_{\rm R} +  \bar \sigma_4^2  \over  {P_{\rm R}(\bar \sigma_3^2 - \bar \sigma_2^2) \over P_{\rm R} + \bar \sigma_3^2} + \bar \sigma_4^2}   \right)   &\\
 \label{D3.5_CASE2_3_2}
 &   \!\geq\!     {1\over 2}  \log \left(   {{P_{\rm R} + \bar \sigma_3^2}\over \bar \sigma_3^2 }  \cdot  {{  \bar \sigma_1^2}\over P_{\rm R}+ \bar \sigma_1^2 }    \right)   &\\ \label{D3.5_CASE2_3_3}
  & \! =\!	R_3' &
 \end{flalign}
  \end{subequations}
  \vspace{-6mm}
  \begin{subequations}
  \begin{flalign}\label{D3.5_CASE2_4}
 \quad\quad\quad\quad\quad\quad\quad\quad  R_4 \! +\!{1\over 2} & \! =\!	 {1\over 2}  \log \left( 2{P_{\rm R} +  \bar \sigma_3^2  \over  p_{{\rm R}3} + \bar \sigma_3^2}   \right)   &\\
 \label{D3.5_CASE2_4_1}
  &   \! =\!	   {1\over 2}  \log \left( 2{P_{\rm R} +  \bar \sigma_3^2  \over  {P_{\rm R}(\bar \sigma_3^2 - \bar \sigma_2^2) \over P_{\rm R} + \bar \sigma_3^2} + \bar \sigma_3^2}   \right)   &\\
  \label{D3.5_CASE2_4_2} 
 &  \!\geq\!     {1\over 2}  \log \left(  {{P_{\rm R} + \bar \sigma_3^2}\over \bar \sigma_3^2 }  \cdot  {{  \bar \sigma_1^2}\over P_{\rm R}+ \bar \sigma_1^2 } \right)   &\\ \label{D3.5_CASE2_4_3}
 &  \! =\!	R_4'.&
 \end{flalign}
 \end{subequations}
 In the above, step (\ref{D3.5_CASE2_1}) follows from (\ref{Pro4.4_1}); step (\ref{D3.5_CASE2_1_2}) from
${P_{\rm R} + \bar \sigma_1^2 \over \bar \sigma_1^2} < 2$ for $ P_{\rm R} < \bar \sigma_1^2$; step (\ref{D3.5_CASE2_1_3}) from (\ref{D3.5_R1'}). Step (\ref{D3.5_CASE2_2}) follows from (\ref{Pro4.4_2}); step (\ref{D3.5_CASE2_2_1}) from $ P_{\rm R} < \bar \sigma_1^2$; and step (\ref{D3.5_CASE2_2_2}) from (\ref{D3.5_R2'}).  Step (\ref{D3.5_CASE2_3}) follows from (\ref{Pro4.4_3});  step (\ref{D3.5_CASE2_3_2}) from
 \begin{subequations}
 \begin{align}
2\bar \sigma_3^2&(P_{\rm R}+ \bar \sigma_1^2)(P_{\rm R}+ \bar \sigma_4^2)\geq 2\bar \sigma_1^2\bar \sigma_3^2(P_{\rm R}+ \bar \sigma_4^2)\\
\label{D3.5_CASE2_3_5}
& \geq P_{\rm R}\bar \sigma_1^2(\bar \sigma_4^2- \bar \sigma_2^2)  + \bar \sigma_1^2\bar \sigma_3^2(P_{\rm R}+ \bar \sigma_4^2)\\
&\! =\! \bar \sigma_1^2(P_{\rm R}+ \bar \sigma_3^2)\left({P_{\rm R}(\bar \sigma_3^2- \bar \sigma_2^2 ) \over P_{\rm R}+ \bar \sigma_3^2} + \bar  \sigma_4^2\right),
 \end{align}
 \end{subequations}
 where  (\ref{D3.5_CASE2_3_5}) from  $\bar \sigma_3^2 \geq \bar \sigma_4^2$; 
 and step (\ref{D3.5_CASE2_3_3}) from (\ref{D3.5_R3'}).
Step (\ref{D3.5_CASE2_4}) follows from (\ref{Pro4.4_4}); 
 step (\ref{D3.5_CASE2_4_2}) from 
 \begin{subequations}
 \begin{align}
 2\bar \sigma_3^2(P_{\rm R}+ \bar \sigma_1^2)&\geq 2\bar \sigma_1^2\bar \sigma_3^2\\
 \label{D3.5_CASE2_4_5}
& \geq {P_{\rm R}\bar \sigma_1^2\bar \sigma_3^2 \over P_{\rm R}+ \bar \sigma_3^2} + \bar \sigma_1^2\bar \sigma_3^2\\
 & \geq {P_{\rm R}\bar \sigma_1^2(\bar \sigma_3^2- \bar \sigma_2^2 ) \over P_{\rm R}+ \bar \sigma_3^2} + \bar \sigma_1^2\bar \sigma_3^2;
 \end{align}
 \end{subequations}
  and step (\ref{D3.5_CASE2_4_3}) from (\ref{D3.5_R4'}). 
  Therefore, (D3.5) is achievable to within ${1\over 2}$ bit, which concludes the proof of Theorem 1.	
  \hfill $\blacksquare$

\section{Conclusion}

In this paper, we put forth a novel transmission scheme and  an  efficient rate analysis method to study the capacity of the two-pair two-way relay channel. Based on the message-reassembling strategy, we  decoupled the coding design for the user-to-relay and relay-to-user transmissions, so as to fully exploit the channel randomness. We employed Gaussian random coding, nested lattice coding, and superposition coding for signal encoding, as well as successive interference cancellation for signal decoding. Careful power allocation was also used in the capacity analysis. Our approach improved the achievability of the capacity region from the existing result of within ${3\over 2}$ bits per user to within ${1\over 2}$ bit per user.

\appendices
\renewcommand{\theequation}{\thesection.\arabic{equation}}

\section{Proof of Proposition 1}
	It suffices to prove (\ref{outer_a}) and (\ref{outer_e}) of Proposition 1, since the other inequalities are similar. Besides, (\ref{outer_e}) is obtained straightforwardly by the cut-set theorem. Thus, we only need to prove (\ref{outer_a}). 
	
\begin{figure}[htbp]	
		\centering
		\subfigure[]{
			\label{fig:genie_aided_1} 
			\includegraphics[width=2.8in]{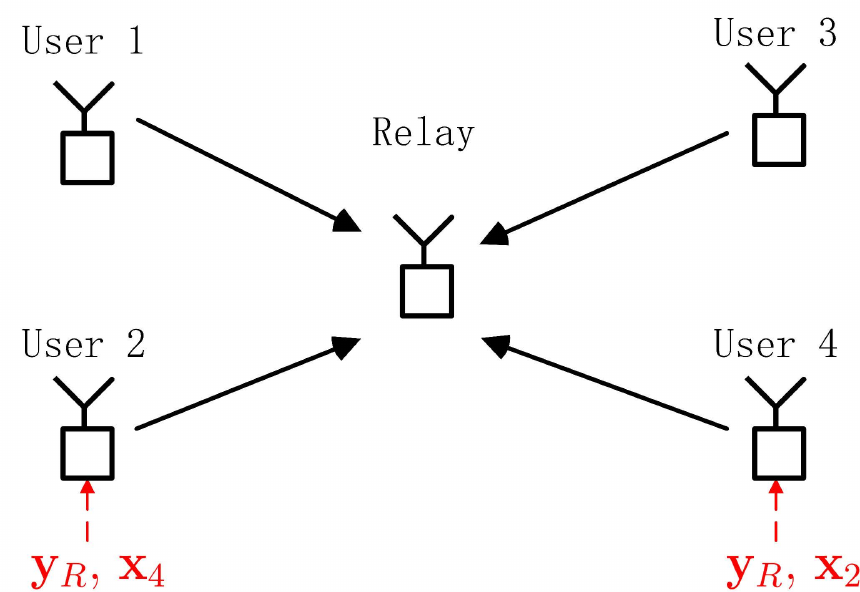}}
		\hspace{0.3in}
		\subfigure[]{
			\label{fig:genie_aided_2} 
			\includegraphics[width=2.8in]{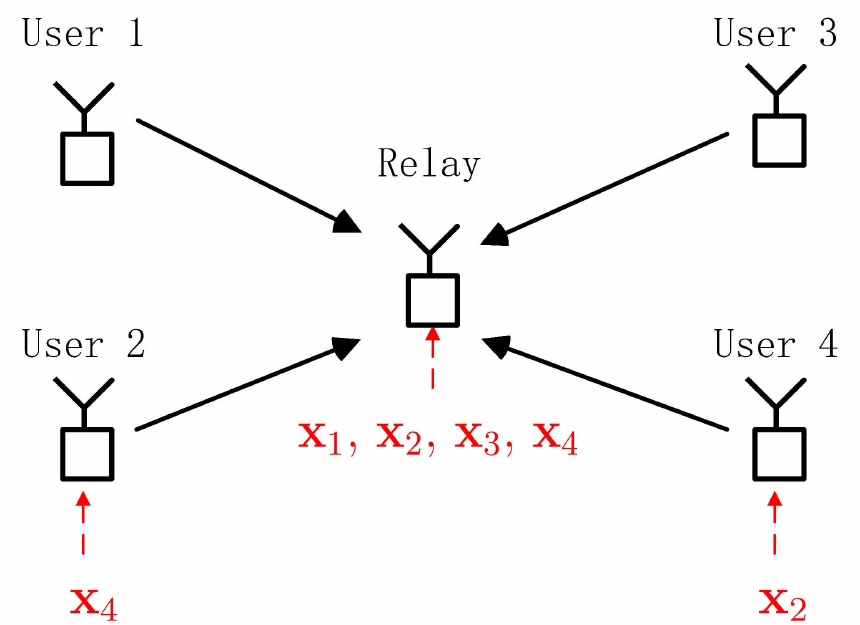}}
		\caption{An illustration of the genie information provided to the two-pair TWRC channel in proving (\ref{outer_a}).}
		\label{fig:genie_aided} 
	\end{figure}

	We use the genie-aided approach to prove inequality (\ref{outer_a}). Suppose that a genie provides side information $\{\mb{y}_R,\,\mb{x}_4 \}$  to user 2 and $ \{\mb{y}_R,\,\mb{x}_2\}$ to user 4,  as shown in Fig. \ref{fig:genie_aided_1}. 
	Then, from the definition of the two-pair TWRC model, the genie-aided system requires user 2 to decode $\mb{x}_1$ from $\{\mb{x}_2, \,\mb{y}_2,\,\mb{y}_R, \mb{x}_4\}$, and user 4 to decode $\mb{x}_3$ from $\{\mb{x}_4, \,\mb{y}_4,\,\mb{y}_R, \mb{x}_2\}$. Since $\{\mb{x}_2, \,\mb{y}_2,\,\mb{y}_R, \mb{x}_4\} \rightarrow \{\mb{x}_2,\,\mb{y}_R, \mb{x}_4\} \rightarrow \mb{x}_1$ forms a Markov chain, it is equivalent to say that the genie-aided system requires user 2 to decode $\mb{x}_1$ from $\{\mb{x}_2, \,\mb{y}_R, \mb{x}_4\}$. Similarly, $\{\mb{x}_4, \,\mb{y}_4,\,\mb{y}_R, \mb{x}_2\} \rightarrow \{\mb{x}_4, \,\mb{y}_R, \mb{x}_2\}\rightarrow \mb{x}_3$ forms a Markov chain, implying that user 4 is required  to decode $\mb{x}_3$ from $\{\mb{x}_4, \,\mb{y}_R, \mb{x}_2\}$. Therefore, considering users 2 and 4 together, the system needs  to decode both $\mb{x}_1$ and $\mb{x}_3$ from a common message set $\{\mb{x}_4, \,\mb{y}_R, \mb{x}_2\}$. From $\mb{y}_R$ in (\ref{up_v}), we further see that this is equivalent to decoding $\mb{x}_1$ and $\mb{x}_3$ from $\mb{y}_R^{'} = h_1\mb{x}_1 + h_3\mb{x}_3 + \mb{n}_R$. Since $\mb{y}_R^{'} = h_1\mb{x}_1 + h_3\mb{x}_3 + \mb{n}_R$ is a standard two-user multiple access channel, we obtain the capacity constraint of $R_1$ and $R_3$ as
	\begin{align}
	\label{R_1+ R_3 1}
	R_1 + R_3 \leq C_{13}. 
	\end{align}
	Since the capaicty of the genie-aided system generally serves as an outer bound of the original two-pair TWRC, we obtain the first half of (\ref{outer_a}).
	
\begin{figure}[htbp]	
		\centering
		\subfigure[]{
			\label{fig:genie_aided_3} 
			\includegraphics[width=2.8in]{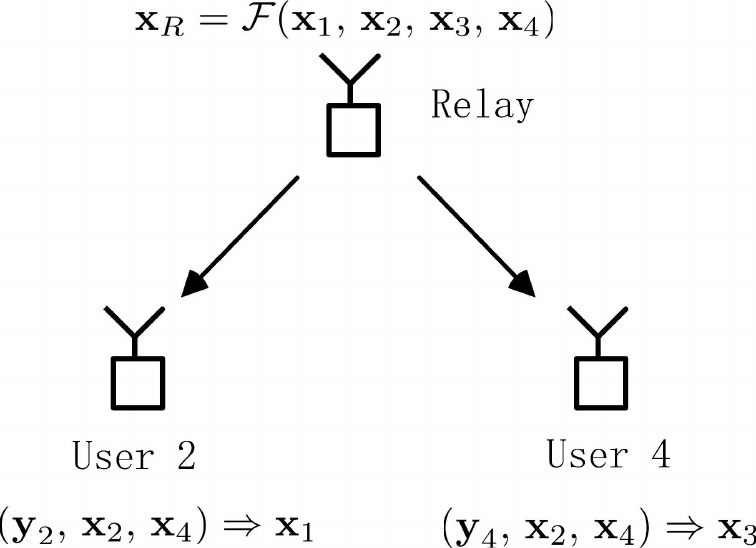}}
		\hspace{0.5in}
		\subfigure[]{
			\label{fig:genie_aided_4} 
			\includegraphics[width=2.42in]{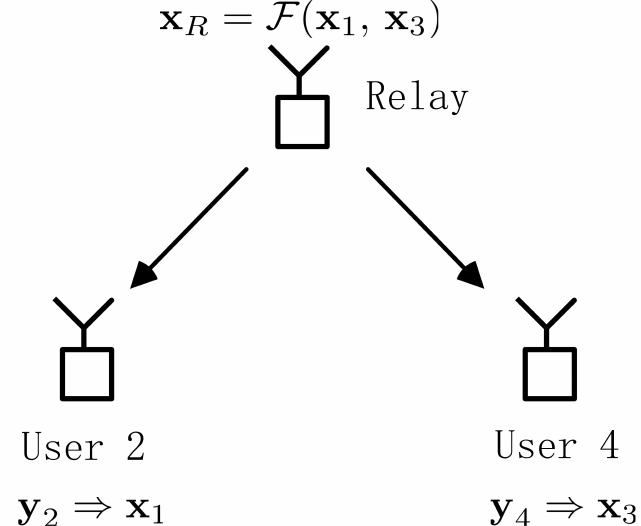}}
		\caption{The reduced systems of the genie-aided system in  Fig. \ref{fig:genie_aided_2}}
		\label{fig:genie_aided_induced} 
	\end{figure}

	Now suppose that a genie provides side information $\{\mb{x}_1, \,\mb{x}_2,\,\mb{x}_3,\,\mb{x}_4\}$ to the relay,   $\mb{x}_4$ to user 2, and $\mb{x}_2$ to user 4, as shown in Fig.  \ref{fig:genie_aided_2}.
	The genie-aided system requires user 2 to decode $\mb{x}_1$ from $\{\mb{y}_2,\, \mb{x}_2,\,\mb{x}_4\}$,  and  user 4 to decode $\mb{x}_3$ from $\{\mb{y}_4,\, \mb{x}_2,\,\mb{x}_4\}$, where
	\begin{align}
	\label{apdix_a_down}
	\mb y_2 =  g_2   \mb x_\mathrm{R}  +  \mb n_2, \quad \mb y_4 =  g_4   \mb x_\mathrm{R}  +  \mb n_4,
	\end{align} 	
	and $\mb{x}_R$ is generally a function of message set $\{\mb{y}_R,\,\mb{x}_1,\,\mb{x}_2,\,\mb{x}_3,\,\mb{x_4}\}$. 
	Since $\{\mb{y}_R,\,\mb{x}_1,\,\mb{x}_2,\,\mb{x}_3,\,\mb{x_4}\}\rightarrow \{\mb{x}_1, \,\mb{x}_2,\,\mb{x}_3,\,\mb{x}_4\} \rightarrow \{\mb{y}_2,\, \mb{x}_2,\,\mb{x}_4\}$ and $\{\mb{y}_R,\,\mb{x}_1,\,\mb{x}_2,\,\mb{x}_3,\,\mb{x_4}\}\rightarrow \{\mb{x}_1, \,\mb{x}_2,\,\mb{x}_3,\,\mb{x}_4\} \rightarrow \{\mb{y}_4,\, \mb{x}_2,\,\mb{x}_4\}$ form  Markov chains,
	$\mb{x}_R$ is reduced to  a function of message set $\{\mb{x}_1,\,\mb{x}_2,\,\mb{x}_3,\,\mb{x_4}\}$ without affecting the capacity of the genie-aided system, yielding an equivalent system in Fig. \ref{fig:genie_aided_3}. Further, since $\mb{x}_2$ and $\mb{x}_4$ are both known by the relay and the two users, the system in Fig. \ref{fig:genie_aided_3} is equivalent to the standard broadcast channel in Fig. \ref{fig:genie_aided_4}, where $\mb{x}_R$ is a function of $\mb{x}_1$ and $\mb{x}_3$, user 2 is required to decode $\mb{x}_1$ from $\mb{y}_2$ and user 4 is required to decode $\mb{x}_3$ from $\mb{y}_4$.
	
	 From \cite{Gamal12}, if ${g_2^2 P_R\over \sigma_2^2 } \geq {g_4^2 P_R\over \sigma_4^2 }$ ,  then the capacity region of the broadcast channel in Fig. \ref{fig:genie_aided_4} is given by 
	\begin{subequations}
	\begin{align}
	R_1 &\leq {1\over 2} \log\left(1 + {\alpha g_2^2 P_R \over \sigma_2^2}\right)\\
	R_3 &\leq {1\over 2}\log \left(1 + {(1-\alpha)g_4^2P_R \over \alpha g_4^2 P_R + \sigma_4^2}\right)
	\end{align}		
	\end{subequations}
	where $\alpha \in [0,1]$.
	Then
	\begin{subequations}
		\begin{align}
		R_1 + R_3 &\leq {1\over 2} \log\left(\left(1 + {\alpha g_2^2 P_R \over \sigma_2^2}\right) \left( {1 + {g_4^2P_R\over \sigma_4^2} \over {\alpha g_4^2 P_R \over \sigma_4^2} + 1}\right)\right)\\
		&	=  {1\over 2} \log\left({{1 + {\alpha g_2^2 P_R \over \sigma_2^2} + {g_4^2P_R\over \sigma_4^2} + {\alpha g_2^2 P_R \over \sigma_2^2}\cdot {g_4^2P_R\over \sigma_4^2} }\over {\alpha g_4^2 P_R \over \sigma_4^2} + 1}	\right)	\\
		\label{apdix_a_r_1_r_3}
		& \leq  {1\over 2} \log\left({{1 + { g_2^2 P_R \over \sigma_2^2} + {\alpha g_4^2P_R\over \sigma_4^2} + {\alpha g_2^2 P_R \over \sigma_2^2}\cdot {g_4^2P_R\over \sigma_4^2} }\over {\alpha g_4^2 P_R \over \sigma_4^2} + 1}	\right)\\
		& =  {1\over 2} \log\left({\left(1 + { g_2^2 P_R \over \sigma_2^2}\right)\left( 1 + {\alpha g_4^2P_R\over \sigma_4^2}  \right)\over {\alpha g_4^2 P_R \over \sigma_4^2} + 1}	\right)\\
		& = {1\over 2} \log\left(1 + { g_2^2 P_R \over \sigma_2^2}\right)\\
		& = D_2
		\end{align}
	\end{subequations}
	where (\ref{apdix_a_r_1_r_3}) follows from  ${g_2^2 P_R\over \sigma_2^2 } \geq {g_4^2 P_R\over \sigma_4^2 }$.
	Similarly, if ${g_2^2 P_R\over \sigma_2^2 } \leq {g_4^2 P_R\over \sigma_4^2 }$, we have $R_1 + R_3 \leq D_4$.	
	Thus, the rates of $\mb{x}_1$  and $\mb{x}_3$ are bounded by	
	\begin{align}
	\label{R_1 + R_3_2}
	R_1 + R_3 \leq \max(D_2, D_4).
	\end{align}
	Combining (\ref{R_1+ R_3 1}) and (\ref{R_1 + R_3_2}), we obtain (\ref{outer_a}), which concludes the proof.

\section{Proof of  of (18)}
We use the following nested lattice scheme for the user-to-relay transmission.

\emph{Encoding:} Recall from subsection III-B that $h_1   \mb x_{10}$ and $h_2 \mb x_2$  use the same codebook $\mathcal{C}_{\rm 10}= \Lambda_{\rm 1}^f\bigcap \mathcal{V}{(\Lambda_{\rm 1}^c)}$ with the power  $p_{{\rm 1}0}   = h_1^2\alpha_{10} P_1    =  h_2^2\alpha_{2} P_2 $, where ${\bf  x}_{10}$  is power-constrained  by $\alpha_{10}P_1$, and $\mb x_2$ is power-constrained by $\alpha_2 P_2$. The codewords transmitted by users 1 and 2 satisfy
\begin{align*}
h_1   \mb x_{10} = (\mb{w}_1 + \mb{u}_1)\,\rm {mod} \, \Lambda_{\rm 1}^{c} \\   h_2 \mb{x}_2 = (\mb{w}_2 + \mb{u}_2)\,\rm {mod} \, \Lambda_{\rm 1}^{c}
\end{align*} 
where  $\mb{w}_i, \, i\in \{1,2\}$ are the nested lattice codewords in $\mathcal{C}_{\rm 10}$,  and 
$\mb{u}_i,\,i\in \{1,2\}$ are  random dither vectors with $\mb{u}_i \sim \rm{Unif}(\mathcal {V}(\Lambda_1^c))$. The dither vectors $\{\mb{u}_i\}$ are independent of each other and also independent of $\{\mb{w}_i\}$ and the noise $\mb{n}_R$. Also, $\{\mb{u}_i\}$ are known to the source nodes and the relay. Note that, from the crypto-lemma \cite{Forney03}, $h_1\mb{x}_{10}$ and $h_2\mb{x}_2$ are respectively uniformly distributed over $\mathcal {V}(\Lambda_1^c)$ and independent of $\mb{w}_1$ and $\mb{w}_2$. The average power of $h_1\mb{x}_{10}$ (or $h_2\mb{x}_2$)  approaches $p_{10}$ as $n$ tends to infinity. 

\emph{Decoding:} From (\ref{network_decoding_y_r}), the received vector at the relay is given by
\begin{align*}
\mb y_R =  h_1 \mb{x}_{10} + h_2 \mb{x}_2 +\mb z_{\rm R}
\end{align*}	
where $\mb{z}_R =\mb s' +\mb n_{\rm R}$. As  $\mb s'$ and $\mb n_{\rm R}$ are independent of each other, the power of $\mb{z}_R $ is given by 
\begin{align}
\label{apdix_b_sigma_r}
\tilde{\sigma}^2_R = p_{s'} + \sigma^2_R.
\end{align}
Upon receiving $\mb{y_R}$, the relay computes
\begin{align*}
\tilde{\mb{y}}_R &=  \alpha \mb{y}_R - \sum_{j = 1}^{2} \mb{u}_j \\
& = \alpha(h_1 \mb{x}_{10} + h_2 \mb{x}_2 +\mb z_{\rm R})- \sum_{j = 1}^{2} \mb{u}_j\\
& =  (h_1 \mb{x}_{10} + h_2 \mb{x}_2 - \mb{u}_1 - \mb{u}_2) + (\alpha -1)(h_1 \mb{x}_{10} + h_2 \mb{x}_2) + \alpha \mb{z}_R\\
& =  \sum_{j = 1}^{2}[(\mb{w}_j + \mb{u}_j)\rm {mod} \; \Lambda_{\rm 1}^{c} - \mb{u}_j ] + (\alpha -1)(h_1 \mb{x}_{10} + h_2 \mb{x}_2) + \alpha \mb{z}_R\\
& =  \sum_{j = 1}^{2}(\mb{w}_j - Q_{\Lambda_1^c}(\mb{w}_j + \mb{u}_j) ) + (\alpha -1)(h_1 \mb{x}_{10} + h_2 \mb{x}_2) + \alpha \mb{z}_R\\
& = \mb{t} + \tilde{\mb{z}}_R
\end{align*}  
where 
\begin{align*}
\mb{t} &= \sum_{j = 1}^{2}(\mb{w}_j - Q_{\Lambda_1^c}(\mb{w}_j + \mb{u}_j) ) =  h_1 \mb{x}_{10} + h_2 \mb{x}_2 - \mb{u}_1 - \mb{u}_2\\
\tilde{\mb{z}}_R & = (\alpha -1)(h_1 \mb{x}_{10} + h_2 \mb{x}_2) + \alpha \mb{z}_R.
\end{align*}
In the above, $\alpha\in [0,1]$ is a scaling factor and $Q_{\Lambda_1^c}(\cdot)$ represents the nearest neighbor lattice quantizer associated with $\Lambda_1^c$. Also,  the property of $\mb{x}\, \rm {mod}\;\Lambda_1^c = \mb{x} - Q_{\Lambda_1^c}(\mb{x})$ is used in the above derivation. Let $\alpha$ be the minimum mean-square error (MMSE) coefficient: 
\begin{align*}
\alpha = \frac{2 p_{10}}{2 p_{10} + \tilde{\sigma}^2_R}.
\end{align*}
Then, we obtain the variance of the effective noise as
\begin{align}
\label{apdix_b_sigma}
\sigma^2(\tilde{\mb{z}}_R) = \frac{1}{n} E ||\tilde{\mb{z}}_R||^2 \leq \frac{2p_{10}\tilde{\sigma}^2_R}{2p_{10} + \tilde{\sigma}^2_R}.
\end{align}
Clearly, $\mb{t} \in \Lambda_1^f$ is a valid lattice point of $\Lambda_1^f$.

The relay aims to recover $\mb{t}$ from $\tilde{\mb{y}}_R$. We employ minimum Euclidean distance lattice decoding to find the closest point to $\tilde{\mb{y}}_R$ in $\Lambda_1^f$. Thus, an estimate of $\mb{t} $ is given by 
\begin{align*}
\tilde{\mb{t}} &= Q_{\Lambda_1^f}(\tilde{\mb{y}}_R)\\
&= Q_{\Lambda_1^f}( \mb{t} + \tilde{\mb{z}}_R)
\end{align*}
where $Q_{\Lambda_1^f}(\cdot)$ is the nearest neighbor lattice quantizer associated with $\Lambda_1^f$.
Then, from the lattice symmetry and the independence between $\mb{ t}$ and $\tilde{\mb{z}}_R$, the probability of decoding error is 
\begin{align*}
p_e &=\rm {Pr}\{\tilde{\mb{t}} \neq \mb{t}\}\\
&={\rm Pr} \{\tilde{\mb{z}}_R \notin \mathcal {V}(\Lambda_1^f)\}.
\end{align*}
From \cite{Erez04} and \cite{Poltyrev94}, for $\Lambda_1^f$ good for coding, we have $p_e \rightarrow 0$,  when
\begin{align*}
\left((\sigma^2(\tilde{\mb{z}}_R)\large)^{\frac{1}{2}}\right)^n < {\rm{Vol}}(\Lambda_1^f), {\textrm {as}}  \,n \rightarrow \infty.
\end{align*}
Thus, the rate of the nested lattice code is given by 
\begin{align}
\label{apdix_b_r_10}
R_{10} &= \frac{1}{n}\log|\mathcal{C}_{10}| = \frac{1}{n}\log \frac{{\rm{Vol}}(\Lambda_1^c)}{{\rm{Vol}}(\Lambda_1^f)} =\frac{1}{n}\log_2\frac{(p_{10})^\frac{n}{2}}{{\rm{Vol}}(\Lambda_1^f)} \leq \frac{1}{n}\log_2\frac{(p_{10})^\frac{n}{2}}{\left((\sigma^2(\tilde{\mb{z}}_R))^{\frac{1}{2}}\right)^n}
\end{align}	
Substituting (\ref{apdix_b_sigma_r}) and (\ref{apdix_b_sigma}) into (\ref{apdix_b_r_10}), we obtain (\ref{network_decoding_r_10}). 
Given $\tilde{\mb{t}} = \mb{t}$, the relay  compute $\mb{t} + \mb{u}_1 + \mb{u}_2  = h_1 \mb{x}_{10} + h_2 \mb{x}_2$, and then cancels it from the received signal $\mb{y}_R$. 

We note that the above arguments basically follows the proof of Theorem 3 in \cite{Nam10}. One major difference is that here we do not take modulo of $\tilde{\mb{y}}_R$ over $\Lambda_1^c$ since $h_1\mb{x}_{10} + h_2\mb{x}_2$ needs to be directly computed for interference cancellation. We also note that similar interference cancellation techniques have been used in \cite{Nazer16} for successive computation at a multi-antenna receiver.

\end{document}